\documentclass[12pt, oneside, final]{huthesis}
\usepackage{amsmath}
\usepackage{graphicx}
\usepackage{amsfonts}
\usepackage{amssymb}
\input{epsf}
\author{Nathan Salwen}
\title{Non-perturbative Methods in Modal Field Theory}
\degreemonth{October} 
\degreeyear{2001}
\degree{Doctor of Philosophy}
\field{Physics}
\department{Physics}
\advisor{Sidney Coleman} 

\begin{document}

\maketitle 

\copyrightpage

\begin{abstract}
Several issues in the modal approach to quantum field
  theory are discussed.  Within the formalism of spherical field
  theory, differential renormalization is presented and shown to
  result in a finite number of renormalization parameters.
  Computations of the massless Thirring model in 1+1 dimensions are
  presented using this approach.
  
  Diagonalization techniques in periodic field theory are
  demonstrated.  Issues of very large Hilbert spaces are considered
  and several approaches are presented.  The quasi sparse eigenvector
  (QSE) approach takes advantage of the relatively small number of
  basis states that typically contribute significantly to any
  particular eigenvector.  Stochastic correction methods use Monte
  Carlo calculations to calculate higher order corrections to the
  quasi sparse result.
  
  The quasi sparse eigenvector method and stochastic error correction
  are applied to the Hubbard model.  With $\frac{U}{t}=4$, the shift
  in the ground energy below the $U=0$ value is found to within 1\%
  for the 8x8 Hubbard model with $\frac{25}{64}$ filling.
\end{abstract}

\newpage
\addcontentsline{toc}{section}{Table of Contents}
\tableofcontents

\begin{acknowledgments}
  First I would like to acknowledge my thesis advisor, Sidney Coleman,
  who took responsibility for me despite his better judgement.  His
  beautiful introduction to the subject of quantum field theory has
  made Physics 253 a legendary course at Harvard.  The joy and pain of
  this course and Daniel Fisher's ``Statistical Mechanics'' were the
  deciding factors in my decision to study physics in graduate school.

  I would also like to thank Dean Lee, my mentor and partner.  I have
  often thought about the fortuitous meeting in Amherst the day you
  showed me your work on spherical field theory.  And I wonder if you
  were a guardian angel sent from above to keep me focused and healthy
  as I attempt to learn a little something about physics. 
  
  Two other people helped particularly with this document.  Efthinio
  Kaxiras agreed to serve on my committee at the last minute and
  generously found the time to provide corrections and comments.  His
  interest has encouraged me to continue my research.  And Alex
  Barnett generously shared his experience dealing with Harvard's
  formatting requirements and has provided a Latex file to make it
  easy for the rest of us.

  My path to this degree has been a long one with some twists and
  turns.  I would like to thank some people who helped me along the
  way.  Raoul Bott took me under his wing and introduced the beautiful
  subject of differential geometry to me.  Andrew Lesniewski worked as
  my advisor and introduced me to the subject of quantization of
  discrete maps.  Rick Heller generously agreed to advise me when I
  found myself orphaned.  I still hope to revisit some of the
  interesting problems he presented.  

  Ron Rubin deserves a special mention.  He showed that a strong
  vision can overcome technical difficulties, even if those
  difficulties seem insurmountable.  He saw the path to quantization
  of the Baker's map and he sees the path to peace in the Middle
  East.  The community of graduate students at Harvard was very
  generous with me explaining many points I found confusing and
  sharing the joy of their explorations.

  I would like to thank my parents who have been patient with me as I
  find my path.  My father, a physicist himself, has always been
  available to help clarify new subjects, especially quantum
  mechanics and perturbation theory.  I am proud to be entering his
  community of scholars.
  
\end{acknowledgments}

\newpage

\startarabicpagination

\chapter{Introduction}

I first encountered quantum field theory as an undergraduate.  I was
drawn by the beauty of its symmetric and seemingly simple equations.  But 
I was also enticed by the opacity of those same equations which so
stubbornly resist calculations.

Free theories can be calculated exactly and others can
be calculated in perturbation theory, but one quickly runs into
problems of infinities.  Even after regularization and renormalization
the perturbation series is at best asymptotic and for many physical
systems is virtually useless.

Lattice regularization is another approach which avoids the
constraints on coupling constant.  From a constructivist viewpoint, it
is reassuring that the field theory can be put on a lattice and a
finite answer extracted.  The fermion derivative term presents
problems in discretized space.  Although these can be handled, 
the complications thus engendered invite a new approach.  

Thus, when I was introduced to spherical field theory, I was initially
attracted by two main features.  The first was that although the
answer would still be expressed as an infinite sum, just as in
perturbation theory, the series would in principle converge.  The
second was that, since space is still treated as a continuous
variable, fermions could be treated in a naive manner.  

On the other hand, since, space was still continuous, spherical field
theory faced the problem of renormalization.  Because the Hamiltonian
and the natural regulators are functions of "t", the radial
coordinate, and thus, it is conceivable that arbitrary functions might
be required to renormalize a theory.  Diagrammatic renormalization is
only useful for super-renormalizable theories with a finite set of
such diagrams.  The second chapter, "Renormalization in spherical field
theory", addresses this problem.  It shows that a small set of local
counterterms is sufficient,in general, to remove all ultraviolet
divergences in a manner such that the renormalized theory is finite
and translationally invariant.

The Thirring model in 1+1 dimensions has a four fermion interaction
and is not super-renormalizable.  The next chapter, "The massless
Thirring model in spherical field theory", serves as concrete test of
the regularization scheme.  It also served as a laboratory to test the
efficacy of different techniques for handling fermions.  The Hilbert
space of the system was small enough to fit in memory and a direct
Runge-Kutta approach was used to integrate the equations of motion.  

A Monte Carlo integration was also attempted at this time but the
results were mediocre and were not included.  Some difficulties of the
spherical approach became apparent during this investigation.  The
time dependence of the Hamiltonian made small steps necessary near
$t=0$ but at large $t$, the Hamiltonian is small and a long time is
necessary for the state to evolve to the ground state.  This problem
is mainly technical and was solved with adaptive time steps.

The time dependence in the regularator also created a moving target
for the number of modes necessary for the calculation.  At small $t$,
where $Mt$ was small only a few modes would be necessary but at larger
$t$, more were required.  The majority of computer time was spent
calculating the time evolution for small $t$.  The requirement to
include the extra modes for their effect at large $t$ resulted in a
waste of computer resources.  Finally, the time dependence of the
Hamiltonian prevented the precalculation of certain constants and
other optimizations that a time independent $H$ would have allowed.

In quantum mechanics, the greatest optimization for calculating time
evolution is expression of the system in terms of energy eigenstates.
The next chapter tackles this problem head-on for $\phi^4$ theory in
1+1 dimensions.  Space is now a periodic box of length $2L$.  The
advantage of a time independent $H$ has been gained at the cost of a
new parameter, $L$ which must also be taken to infinity before we can
use our results.  

The work in this chapter was exciting to me for two reasons.  One was
the potential to work directly in Minkowski space.  Even more exciting
was the potential for explicit examination of particular eigenstates of
the system.  Figure 8 in chapter 5 is a perfect example of the types
of work that could be done within this approach.  The eigenstates are
tracked across a phase transition and the symmetry relationships are
exposed.  

At this point we were ready to try our new approaches on systems in
2+1 or 3+1 dimensions.  The exponentially higher dimensionality of the
Hilbert space became the dominant issue we faced.  I considered simply
waiting.  It is a ``rule'' in the computer industry that computer
power doubles about every 2 years.  If the rate sped up a little, in
thirty years we might be able to attack some 2+1 dimensional systems
and in another thirty years we could try problems in 3+1 dimensions.

The following question then presents itself.  ``Are these systems in
2+1 dimensions so complicated that they cannot be described without
reference to $10^{18}$ states or more, or after the diagonalization was
completed would we find that the results could be described in a
simpler way?''  In particular, it is possible that any particular
eigenstate of the Hilbert space may require only a relatively small
number of basis states to accurately describe it.

Chapter 5 argues for the prevalence of this condition, known as
``quasi-sparsity''.  A careful analysis shows that this is as much a
statement about the types of bases we are likely to use as it is a
statement about the Hamiltonians we will encounter.  In our work we
use either a Fock state basis or a position state basis.  Different
problems turn out to fit into the quasi-sparse model to different
extents.  As pointed out in chapter 8, the Hubbard model with a Fock
state basis seems particularly ill-suited for this approach.

The power of lattice field theory comes from its ability to tap into
Monte Carlo computational methods.  The dimensionalities of the relevant
spaces are even larger than those considered in this thesis but because
of the smoothness of contributions as a function on the  configuration
space, it is possible to do a good job of sampling the important
configurations.  I am convinced that efficient use of Monte Carlo is
essential to solving most physically relevant quantum field theory
problems.  

In lattice field theory, Monte Carlo is restricted to imaginary time
calculations and cannot handle unquenched fermions well.  A successful
integration of Monte Carlo computation into the approach described in
chapter 4 could address these limitations and give visibility to more
than one state in a symmetry sector.  Two potential methods of doing
this integration are presented in chapter 6.

The thesis ends with an attempt to apply the results of chapters 5 and
6 to the Hubbard model.  It was expected that the complexity of its
ground state would present a challenge for our methods and we were not
disappointed.  Adjustments to the stochastic methods of chapter 6 are
presented in chapter 8 and a reasonable estimate of the ground state
energy of the 8x8 Hubbard model is extracted.  Other questions such as
the value of particular correlators in the ground state could also be
computed with this framework.  Other information, such as the makeup
of excited states will require further modifications.  Some potential
directions for further improvement are mentioned in the chapter.

One area where our current state of the art seems particularly lacking is
in the sampling of configurations for Monte Carlo.  We run
trajectories so the choice at each step has no global knowledge of its
path.  While I was able to find a good ``distance'' based guiding
scheme for the Hubbard model it does not take the energy of the states
into account.  It may be that the Hilbert space of a fermion system
does not support the same notion of close paths that the lattice field
theory boson computations use.  But if it does, a sampling scheme
which uses it will almost certainly be an improvement.

The order of chapters in this thesis corresponds with the
chronological order of the papers they contain.  They also mirror the
logical progression of my approach to quantum field theory.  In trying
to do computations on more difficult systems I have been driven to
adopt features of the lattice field theory approach.  The goal for the
future will be to keep some of the advantages of modal field
theory while learning from the sampling techniques of lattice Monte
Carlo.

\chapter[Renormalization in spherical field theory]{Renormalization in spherical field theory \footnote{D. Lee, N. Salwen, Phys. Lett. B460 (1999) 107.}}

\section{Introduction}

Spherical field theory is a non-perturbative method which uses the spherical
partial wave expansion to reduce a general $d$-dimensional Euclidean field
theory into a set of coupled radial systems (\cite{a1, fermion, a3}). High
spin partial waves correspond with large tangential momenta and can be
neglected if the theory is properly renormalized. The remaining system can
then be converted into differential equations and solved using standard
numerical methods. $\phi^{4}$ theory in two dimensions was considered in
\cite{a1}. In that case there was only one divergent diagram, and it could be
completely removed by normal ordering. In general any super-renormalizable
theory can be renormalized by removing the divergent parts of divergent
diagrams. Using a high-spin cutoff $J_{\max}$ and discarding partial waves
with spin greater than $J_{\max}$, we simply compute the relevant counterterms
using spherical Feynman rules.

The $J_{\max}$ cutoff scheme however is not translationally invariant. It
preserves rotational invariance but regulates ultraviolet processes
differently depending on radial distance. In the two-dimensional $\phi^{4}$
example it was found that the mass counterterm had the form
\begin{equation}
\mathcal{L}_{c.t.}\propto\phi^{2}(\vec{t})\left[  K_{0}(\mu t)I_{0}(\mu t)+2%
{\textstyle\sum_{n=1,J_{\max}}}
K_{n}(\mu t)I_{n}(\mu t)\right]  ,
\end{equation}
where $I_{n}$, $K_{n}$ are $n^{\text{th}}$-order modified Bessel functions of
the first and second kinds, $\mu$ is the bare mass, and $t$ is the magnitude
of $\vec{t}$. As $J_{\max}\rightarrow\infty$, we find%
\begin{equation}
\mathcal{L}_{c.t.}\propto\phi^{2}(\vec{t})\left[  \log(\tfrac{2J_{\max}}{\mu
t})+O(J_{\max}^{-1})\right]  .
\end{equation}
Our regularization scheme varies with $t$, and we see that the counterterm
also depends on $t$. The physically relevant issue, however, is whether or not
the renormalized theory is independent of $t$. In this case the answer is yes.
Any $t$ dependence in renormalized amplitudes is suppressed by powers of
$J_{\max}^{-1}$, and translational invariance becomes exact as $J_{\max
}\rightarrow\infty$.

We now consider general renormalizable theories, in particular those which are
not super-renormalizable. In this case the number of divergent diagrams is
infinite. Since we are primarily interested in non-perturbative phenomena, a
diagram by diagram subtraction method is not useful. In the same manner
strictly perturbative methods such as dimensional regularization are not
relevant either. Our interest is in non-perturbative renormalization, where
coefficients of renormalization counterterms are determined by
non-perturbative computations.\footnote{We should mention that Pauli-Villars
regularization is compatible with non-perturbative renormalization. \ However
this introduces additional unphysical degrees of freedom and tends to be
computationally inefficient.} In this paper we analyze the general theory of
non-perturbative renormalization in the spherical field formalism. In the
course of our analysis we answer the following three questions: (i) Can
ultraviolet divergences be cancelled by a finite number of local counterterms?
(ii) Can the renormalized theory be made translationally invariant? (iii) What
is the general form of the counterterms?

The organization of the paper is as follows. We begin with a discussion of
differential renormalization, a regularization-independent method which will
allow us to construct local counterterms. Next we describe a regularization
procedure which is convenient for spherical field theory. In the large radius
limit $t\rightarrow\infty$ our regularization procedure (which we call angle
smearing) is anisotropic but locally invariant under translations. For general
$t$ we expand in powers of $t^{-1}$ to generate the general form of the
counterterms. We conclude with two examples of one-loop divergent diagrams. We
show by direct calculation that the predicted counterterms render these
processes finite and translationally invariant.

\section{Differential renormalization}

Differential renormalization is the coordinate space version of the BPHZ
method.\footnote{Paraphrase of private communication with Jose Latorre.} It is
framed entirely in coordinate space, and renormalized amplitudes can be
defined as distributions without reference to any specific regularization
procedure. Differential renormalization was introduced in \cite{b}, and a
systematic analysis of differential renormalization to all orders in
perturbation theory using Bogoliubov's recursion formula was first described
in \cite{f}. The usual implementation of differential renormalization is
carried out using singular Poisson equations and their explicit solutions. In
our discussion, however, we find it more convenient to operate directly on the
distributions.\footnote{Our approach is similar to the natural renormalization
scheme described in \cite{d}. In contrast with \cite{d}, however, we do not a
priori specify the finite parts of amplitudes.} We describe the details of our
approach in the following. We should stress that the two approaches are
equivalent, differing only at the level of formalism.

We assume that we are working with a renormalizable theory. For indices
$i_{1},\cdots i_{j}$ let us define%
\begin{align}
t^{i_{1},\cdots i_{j}}  &  =t^{i_{1}}t^{i_{2}}\cdots t^{i_{j}},\\
\nabla_{i_{1},\cdots i_{j}}  &  =\nabla_{i_{1}}\nabla_{i_{2}}\cdots
\nabla_{i_{j}}.
\end{align}
Let $f(\vec{t})$ be a smooth test function, and let $I(\vec{t}-\vec{t}%
^{\prime};\mu^{2})$ be a smooth function with support on a region of scale
$\mu^{-1}$. We define $S_{\vec{t}^{\prime}}^{j}\left[  f\right]  (\vec{t})$ as
$I(\vec{t}-\vec{t}^{\prime};\mu^{2})$ multiplied by the $j^{th}$ order term in
the Taylor series of $f(\vec{t})$ about the point $\vec{t}^{\prime}$.
Inserting delta functions, we have
\begin{align}
S_{\vec{t}^{\prime}}^{j}f(\vec{t})  &  =I(\vec{t}-\vec{t}^{\prime};\mu^{2})%
{\textstyle\sum\limits_{i_{1},\cdots i_{j}}}
\left[  \tfrac{(t-t^{\prime})^{i_{1},\cdots i_{j}}}{j!}\nabla_{i_{1},\cdots
i_{j}}f(\vec{t}^{\prime})\right] \\
&  =I(\vec{t}-\vec{t}^{\prime};\mu^{2})%
{\textstyle\sum\limits_{i_{1},\cdots i_{j}}}
\tfrac{(t-t^{\prime})^{i_{1},\cdots i_{j}}}{j!}%
{\textstyle\int}
d^{4}\vec{z}\,\,\nabla_{i_{1},\cdots i_{j}}^{\vec{t}^{\prime}}\delta^{4}%
(\vec{t}^{\prime}-\vec{z})\,f(\vec{z}).\nonumber
\end{align}
For the purposes of this discussion we will require%
\begin{equation}
I(\vec{t}-\vec{t}^{\prime};\mu^{2})=1+O^{N}(\vec{t}-\vec{t}^{\prime
})\text{\quad as }\vec{t}^{\prime}\rightarrow\vec{t}\text{,}%
\end{equation}
where $N$ is some positive integer greater than the superficial degree of
divergence of any subdiagram\footnote{In our discussion a subdiagram is a
subset of vertices together with all lines contained in those vertices.} in
the theory we are considering. \ For any renormalizable theory $N>2$ will
suffice. In our formalism, $I(\vec{t}-\vec{t}^{\prime};\mu^{2})$ determines
how finite parts of renormalized amplitudes are assigned, and $\mu$ is the
renormalization mass scale.

We now consider a particular diagram, $G$, with $n$ vertices. We define
$K(\vec{t}_{1},\cdots\vec{t}_{n})$ to be the kernel of the amputated diagram,
i.e., the value of the diagram with vertices fixed at points $\vec{t}%
_{1},\cdots\vec{t}_{n}$. The amplitude is obtained by integrating $K(\vec
{t}_{1},\cdots\vec{t}_{n})$\ with respect to all internal vertices. We will
regard $K$ as a distribution acting on $n$ smooth test functions $f_{1},\cdots
f_{n}.\ $(For external vertices containing more than one external line and/or
derivatives, $f_{ext}(\vec{t}_{ext})$ should be regarded as a product of test
functions, with possible derivatives, at $\vec{t}_{ext}$.)%
\begin{equation}
K:f_{1},\cdots f_{n}\rightarrow%
{\textstyle\int}
d^{4}\vec{t}_{1}\cdots d^{4}\vec{t}_{n}\,K(\vec{t}_{1},\cdots\vec{t}_{n}%
)f_{1}(\vec{t}_{1})\cdots f_{n}(\vec{t}_{n}).
\end{equation}
Let us assume that our diagram is primitively divergent with superficial
degree of divergence $j$. We now define another distribution $T_{G}K$, which
extracts the divergent part of $K$. We start with the case when $G$ has more
than one vertex. Let us define $T_{G}K:f_{1},\cdots f_{n}\rightarrow$%

\begin{equation}%
{\textstyle\sum\limits_{_{j_{1}+\cdots+j_{n}\leq j}}}
{\textstyle\int}
d^{4}\vec{t}_{1}\cdots d^{4}\vec{t}_{n}\,K(\vec{t}_{1},\cdots\vec{t}%
_{n})S_{\vec{t}_{ave}}^{j_{1}}f_{1}(\vec{t}_{1})\cdots S_{\vec{t}_{ave}%
}^{j_{n}}f_{n}(\vec{t}_{n}),
\end{equation}
where $\vec{t}_{ave}=\tfrac{1}{n}(\vec{t}_{1}+\cdots+\vec{t}_{n}).$ We note
that the subtracted distribution $K-T_{G}K$ is finite and well-defined for all
$f_{1},\cdots f_{n}$. Let us define
\begin{align}
&  F_{K}^{i_{1,1},i_{2,1}\cdots i_{j_{n},n}}(\vec{t})\label{bw}\\
&  =%
{\textstyle\int}
d^{4}\vec{t}_{1}\cdots d^{4}\vec{t}_{n}\,\delta^{4}(\tfrac{\vec{t}_{1}%
+\cdots+\vec{t}_{n}}{n}-\vec{t})K(\vec{t}_{1},\cdots\vec{t}_{n})\left[
{\textstyle\prod\limits_{k=1,\cdots n}}
\tfrac{I(\vec{t}_{k}-\vec{t};\mu^{2})(t_{k}-t)^{i_{1,k},\cdots i_{j_{k},k}}%
}{j_{k}!}\right]  .\nonumber
\end{align}
We can then rewrite $T_{G}K:f_{1},\cdots f_{n}\rightarrow$%
\begin{equation}%
{\textstyle\sum\limits_{_{j_{1}+\cdots+j_{n}\leq j}}}
{\textstyle\sum\limits_{_{\substack{i_{1,1},i_{2,1}\cdots\\ i_{1,n}\cdots
i_{j_{n},n}}}}}
\left[
\begin{array}
[c]{c}%
{\textstyle\int}
d^{4}\vec{t}F_{K}^{i_{1,1},i_{2,1}\cdots i_{j_{n},n}}(\vec{t})\int d^{4}%
\vec{z}_{1}\cdots d^{4}\vec{z}_{n}\\
\left(  \prod_{k=1,\cdots n}\,\,\nabla_{i_{1,k},\cdots i_{j_{k},k}}^{\vec{t}%
}\delta^{4}(\vec{t}-\vec{z}_{k})\right)  f_{1}(\vec{z}_{1})\cdots f_{n}%
(\vec{z}_{n})
\end{array}
\right]  . \label{cou}%
\end{equation}
The delta functions make this kernel completely local. We can read off the
corresponding counterterm interaction by functional differentiation with
respect to each of the component functions of $f_{ext}(\vec{t}_{ext})$ for the
external vertices and setting $f_{int}(\vec{t}_{int})=1$ for the internal
vertices. We now turn to the case when $G$ has only one vertex. For this case
we set $T_{G}K=K$, which is equivalent to normal ordering the interactions in
our theory.\ In this case $K$ is itself local and therefore $T_{G}K$ and our
counterterm interaction are again local.

We now extend the definition of $T_{G}$ in (\ref{cou}) to include the case of
subdiagrams. Let $G$ be a general 1PI diagram, and let $G^{\prime}$ be a
renormalization part\footnote{A renormalization part is a 1PI subdiagram with
degree of divergence $\geq0$.} of $G$ with superficial degree of divergence
$j^{\prime}$. For notational convenience we will label the vertices of $G$ so
that the first $n^{\prime}$ vertices lie in $G^{\prime}$. If $G^{\prime}$ has
only one vertex then again we normal order the interaction. Otherwise we
define $T_{G^{\prime}}K:f_{1},\cdots f_{n}\rightarrow$
\begin{equation}%
{\textstyle\sum\limits_{j_{1}^{\prime}+\cdots+j_{n^{\prime}}^{\prime}\leq
j^{\prime}}}
{\textstyle\int}
d^{4}\vec{t}_{1}\cdots d^{4}\vec{t}_{n}\,K(\vec{t}_{1},\cdots\vec{t}%
_{n})\left[
\begin{array}
[c]{c}%
S_{\vec{t}_{ave}}^{j_{1}^{\prime}}f_{1}(\vec{t}_{1})\cdots S_{\vec{t}_{ave}%
}^{j_{n}^{\prime}}f_{n^{\prime}}(\vec{t}_{n^{\prime}})\\
\cdot f_{n^{\prime}+1}(\vec{t}_{n^{\prime}+1})\cdots f_{n}(\vec{t}_{n})
\end{array}
\right]  ,
\end{equation}
where $\vec{t}_{ave}=\tfrac{1}{n^{\prime}}(\vec{t}_{1}+\cdots+\vec
{t}_{n^{\prime}})$.\footnote{After applying $T_{G^{\prime}}$, we regard
$G^{\prime}$ as being contracted to single vertex at $\vec{t}_{ave}$.} This
definition can be used recursively to define products of $T_{G_{1}^{\prime}%
}T_{G_{2}^{\prime}}$ for disjoint subdiagrams $G_{1}^{\prime}\cap
G_{2}^{\prime}=\emptyset$ or nested subdiagrams $G_{1}^{\prime}\supset
G_{2}^{\prime}.$ For the case of nested subdiagrams we always order the
product so that larger diagrams are on the left.

It is straightforward to show that the $T$ operation acts as the identity on
local interactions and thus treats overlapping divergences in the same manner
as BPHZ. Following the standard BPHZ procedure (\cite{g, zimm1, zimm2}), we can
write Bogoliubov's $\bar{R}$ operation using Zimmerman's forest formula,%
\begin{equation}
\bar{R}=%
{\textstyle\sum\limits_{F}}
{\textstyle\prod\limits_{\gamma\in F}}
(-T_{\gamma}),
\end{equation}
where $F$ ranges over all forests\footnote{A forest is any set of
non-overlapping renormalization parts.} of $G$, and $\gamma$ ranges over all
renormalization parts of $F.$ In the product we have again ordered nested
subdiagrams so that larger diagrams are on the left. Let $j$ be the
superficial degree of divergence of $G$. The renormalized kernel$,$ $RK$, is
given by%
\begin{equation}
\left.
\begin{array}
[c]{c}%
RK=\bar{R}K\quad\\
RK=(1-T_{G})\bar{R}K\quad
\end{array}
\right.  \left.
\begin{array}
[c]{c}%
\text{for }j<0\\
\text{for }j\geq0.
\end{array}
\right.
\end{equation}
Our final result is that all required counterterms are local, and the form of
the counterterms is
\begin{equation}
\mathcal{L}_{c.t.}=%
{\textstyle\sum_{A(\phi,\nabla_{i}\phi)\text{ }}}
F_{A}(\vec{t})A(\phi(\vec{t}),\nabla_{i}\phi(\vec{t})), \label{be}%
\end{equation}
where the sum is over operators of renormalizable type. For the case of gauge
theories, our renormalization procedure is supplemented by the additional
requirement that the renormalized amplitudes satisfy Ward
identities.\footnote{See \cite{e1}, \cite{e2} for a discussion of gauge
theories using the method of differential renormalization.} If our
regularization procedure breaks gauge invariance these identities are not
automatic and the required local counterterms will in general be any operators
of renormalizable type (not merely gauge-invariant operators). This is,
however, a separate discussion, and the details of implementing Ward identity
constraints will be discussed in future work.

\section{Regularization by angle smearing}

\begin{figure}[htbp]
\begin{center}
\epsfbox{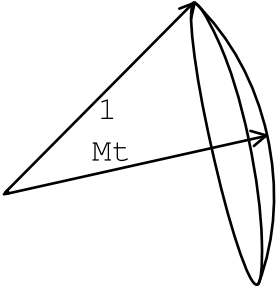}
\caption{Sketch of the angle-smearing region (three-dimensional rendering)}
\label{fig:ren1}
\end{center}
\end{figure}

In this section we determine the functional form of the coefficients
$F_{A}(\vec{t})$ in (\ref{be}). To make the discussion concrete, we will
illustrate using the example of massless $\phi^{4}$ theory in four dimensions
\begin{equation}
\mathcal{L}=\tfrac{1}{2}\phi\nabla^{2}\phi-\tfrac{\lambda}{4!}\phi
^{4}+\mathcal{L}_{c.t.}.
\end{equation}
From (\ref{be}) $\mathcal{L}_{c.t.}$ is given by%
\begin{equation}
F_{\phi^{2}}(\vec{t})\phi^{2}(\vec{t})+\text{ }%
{\textstyle\sum_{i,j}}
F_{\nabla\phi\nabla\phi}^{ij}(\vec{t})\nabla_{i}\phi(\vec{t})\nabla_{j}%
\phi(\vec{t})+F_{\phi^{4}}(\vec{t})\phi^{4}(\vec{t}).\text{ }%
\end{equation}
Let $G(\vec{t},\vec{t}^{\prime})$ be the free two-point correlator. We will
use a regularization scheme which preserves rotational invariance and is
convenient for spherical field theory, but one which breaks translational
invariance. We regulate the short distance behavior of $G$ by smearing the
endpoints over a radius $t$ spherical shell within a conical region $R_{M^{2}%
}(\vec{t})$, where $R_{M^{2}}(\vec{t})$ is the set of vectors $\vec{u}$ such
that the angle between $\vec{t}$ and $\vec{u}$ is between $-\frac{1}{Mt}$ and
$\frac{1}{Mt}$ (see Figure \ref{fig:ren1}).
 The result is a regulated correlator
\begin{equation}
G_{M^{2}}(\vec{t},\vec{t}^{\prime})=\tfrac{1}{\int_{\hat{u}\in R_{M^{2}}%
(\vec{t})}d^{3}\hat{u}\int_{\hat{u}^{\prime}\in R_{M^{2}}(\vec{t}^{\prime}%
)}d^{3}\hat{u}^{\prime}\,\,}\int_{\substack{\hat{u}\in R_{M^{2}}(\vec
{t})\\\hat{u}^{\prime}\in R_{M^{2}}(\vec{t}^{\prime})}}d^{3}\hat{u}d^{3}%
\hat{u}^{\prime}\,G(t\hat{u},t^{\prime}\hat{u}^{\prime}). \label{as}%
\end{equation}
We recall that our renormalized theory is determined by the translationally
invariant function $I(\vec{t}-\vec{t}_{ave};\mu^{2})$ described in the
previous section. Even though our regularization scheme breaks translational
invariance, the renormalized theory nevertheless remains invariant.

As the radius $t$ increases the curvature of the angle-smearing region becomes
negligible. In the limit $t\rightarrow\infty$ the region becomes a flat
three-dimensional ball with radius $\tfrac{1}{M}$ lying in the plane
perpendicular to the radial vector. In this limit our regularization is
invariant under local transformations and the counterterms converge to
constants independent of $\vec{t}$,%
\begin{align}
\lim_{t\rightarrow\infty}F_{\nabla\phi\nabla\phi}^{ij}(\vec{t})  &
=c_{\nabla\phi\nabla\phi}^{ij,(0)}(\tfrac{\mu^{2}}{M^{2}})\\
\lim_{t\rightarrow\infty}F_{\phi^{2}}(\vec{t})  &  =M^{2}c_{\phi^{2}}%
^{(0)}(\tfrac{\mu^{2}}{M^{2}})\\
\lim_{t\rightarrow\infty}F_{\phi^{4}}(\vec{t})  &  =c_{\phi^{4}}^{(0)}%
(\tfrac{\mu^{2}}{M^{2}}).
\end{align}
We have chosen our coefficients $c_{A}^{(0)}$ to be dimensionless. Although
our regularization scheme is invariant under rotations about the origin, the
radial vector has a special orientation which is normal to our
three-dimensional ball. Our regularization scheme is therefore not isotropic.
The result (as should be familiar from studies of anisotropic lattices) is
that the coefficient of the kinetic term has two independent components%
\begin{equation}
c_{\nabla\phi\nabla\phi}^{ij,(0)}(\tfrac{\mu^{2}}{M^{2}})=c_{\nabla\phi
\nabla\phi}^{\hat{t}\hat{t},(0)}(\tfrac{\mu^{2}}{M^{2}})+\delta^{ij}%
c_{\nabla\phi\nabla\phi}^{(0)}(\tfrac{\mu^{2}}{M^{2}}).
\end{equation}

Starting with the $t\rightarrow\infty$ result at lowest order, we now expand
our coefficient functions in powers of $\frac{1}{Mt}$,%
\begin{align}
F_{\nabla\phi\nabla\phi}^{ij}(\vec{t})  &  =c_{\nabla\phi\nabla\phi}%
^{ij,(0)}(\tfrac{\mu^{2}}{M^{2}})+\tfrac{1}{Mt}c_{\nabla\phi\nabla\phi
}^{ij,(1)}(\tfrac{\mu^{2}}{M^{2}})+\tfrac{1}{M^{2}t^{2}}c_{\nabla\phi
\nabla\phi}^{ij,(2)}(\tfrac{\mu^{2}}{M^{2}})+\cdots\\
F_{\phi^{2}}(\vec{t})  &  =M^{2}c_{\phi^{2}}^{(0)}(\tfrac{\mu^{2}}{M^{2}%
})+\tfrac{M}{t}c_{\phi^{2}}^{(1)}(\tfrac{\mu^{2}}{M^{2}})+\tfrac{1}{t^{2}%
}c_{\phi^{2}}^{(2)}(\tfrac{\mu^{2}}{M^{2}})+\cdots\\
F_{\phi^{4}}(\vec{t})  &  =c_{\phi^{4}}^{(0)}(\tfrac{\mu^{2}}{M^{2}}%
)+\tfrac{1}{Mt}c_{\phi^{4}}^{(1)}(\tfrac{\mu^{2}}{M^{2}})+\tfrac{1}{M^{2}%
t^{2}}c_{\phi^{4}}^{(2)}(\tfrac{\mu^{2}}{M^{2}})+\cdots.
\end{align}
For the moment let us assume\ $t\geq\Lambda^{-1}$ for%
\begin{equation}
\Lambda=m_{0}^{z}M^{1-z},
\end{equation}
for some fixed mass $m_{0}$ and constant $z$ such that $0<z<\frac{1}{2}$. In
this region our dimensionless expansion parameter $\frac{1}{Mt}$ is bounded by
$\tfrac{m_{0}^{z}}{M^{z}}$ and therefore diminishes uniformly as
$M\rightarrow\infty$.

In general the $\tfrac{\mu^{2}}{M^{2}}$ dependence in the functions
$c_{A}^{(k)}$ will contain analytic terms as $\mu^{2}\rightarrow0$ as well as
logarithmically divergent terms. There are, however, no inverse powers of
$\tfrac{\mu^{2}}{M^{2}}$. These would indicate severe infrared divergences not
present in the processes we are considering, as can be deduced from the long
distance behavior of the integral in (\ref{bw}).\footnote{If our theory
contained bare masses $m_{i}$, similar arguments would apply for the infrared
limit $\mu^{2},m_{i}^{2}\rightarrow0,$ with $\tfrac{m_{i}^{2}}{\mu^{2}}$
fixed.} With this we can neglect terms which vanish as $M\rightarrow\infty,$
\begin{align}
F_{\phi^{2}}(\vec{t})  &  =M^{2}c_{\phi^{2}}^{(0)}(\tfrac{\mu^{2}}{M^{2}%
})+\tfrac{1}{t^{2}}c_{\phi^{2}}^{(2)}(\tfrac{\mu^{2}}{M^{2}})\\
F_{\nabla\phi\nabla\phi}^{ij}(\vec{t})  &  =c_{\nabla\phi\nabla\phi}%
^{ij,(0)}(\tfrac{\mu^{2}}{M^{2}})\\
F_{\phi^{4}}(\vec{t})  &  =c_{\phi^{4}}^{(0)}(\tfrac{\mu^{2}}{M^{2}}).
\end{align}
Since our regularization scheme is invariant under $M\rightarrow-M$, we have
also omitted the term proportional$\ $to $c_{\phi^{2}}^{(1)}$ which is odd in
$M$.

We now consider what occurs in the small region near the origin, $t\leq
\Lambda^{-1}$. For the theory we are considering (and in fact for any
renormalizable theory) the highest ultraviolet divergence possible is
quadratic.\footnote{There may be additional logarithmic factors but this does
not matter for our purposes here.} In the limit $M\rightarrow\infty$ we deduce
that each $F_{A}$ scales no greater than $O(M^{2}).$ On the other hand the
volume of the region $t\leq\Lambda^{-1}$ diminishes as $O(M^{4z-4}).$ Thus the
total contribution from the region $t\leq\Lambda^{-1}$ scales as $O(M^{4z-2})$
and can be entirely neglected.

To summarize our results, the counterterm Lagrange density has the form
\begin{equation}
c_{\nabla\phi\nabla\phi}^{(0)}(\vec{\nabla}\phi(\vec{t}))^{2}+c_{\nabla
\phi\nabla\phi}^{\hat{t}\hat{t},(0)}(\hat{t}\cdot\vec{\nabla}\phi(\vec
{t}))^{2}+(M^{2}c_{\phi^{2}}^{(0)}+\tfrac{1}{t^{2}}c_{\phi^{2}}^{(2)})\phi
^{2}(\vec{t})+c_{\phi^{4}}^{(0)}\phi^{4}(\vec{t}). \label{an}%
\end{equation}

\section{Spherical fields}

We now examine the results of the previous section in the context of spherical
field theory. We start with the spherical partial wave expansion,%
\begin{equation}
\phi=%
{\textstyle\sum_{l=0,1,\cdots}}
{\textstyle\sum_{n=0,\cdots l}}
{\textstyle\sum_{m=-n,\cdots n}}
\phi_{l,n,m}(t)Y_{l,n,m}(\theta,\psi,\varphi),
\end{equation}
where $Y_{l,m,n}$ are four-dimensional spherical harmonics satisfying%
\begin{equation}%
{\textstyle\int}
d^{3}\Omega\,Y_{l^{\prime},n^{\prime},m^{\prime}}^{\ast}(\theta,\psi
,\varphi)Y_{l,n,m}(\theta,\psi,\varphi)=\delta_{l^{\prime},l}\delta
_{n^{\prime},n}\delta_{m^{\prime},m},
\end{equation}%
\begin{equation}
Y_{l,n,m}^{\ast}(\theta,\psi,\varphi)=(-1)^{m}Y_{l,n,-m}(\theta,\psi,\varphi).
\end{equation}
The explicit form of $Y_{l,m,n}$ can be found in \cite{fubini}.
\footnote{\cite{fubini}
deserves credit as the first discussion of radial (or covariant Euclidean)
quantization, an important part of the spherical field formalism.} The
integral of the free massless Lagrange density in terms of spherical fields is%
\begin{equation}%
{\textstyle\int}
d^{4}\vec{t}\,\mathcal{L}=%
{\textstyle\int_{0}^{\infty}}
dt\,\left\{  \sum_{l,m,n}\left[  (-1)^{m}\phi_{l,n,-m}\left[  \tfrac{\partial
}{\partial t}\tfrac{t^{3}}{2}\tfrac{\partial}{\partial t}-\tfrac{t}%
{2}l(l+2)\right]  \phi_{l,n,m}\right]  \right\}  .
\end{equation}
It can be shown that the process of angle smearing the field $\phi(\vec{t})$
is equivalent to multiplying $\phi_{l,n,m}(t)$ by an extra factor $s_{l}%
^{M}(t)$ where%
\begin{equation}
s_{l}^{M}(t)=\tfrac{2Mt\left[  (l+2)\sin(\frac{l}{Mt})-l\sin(\frac{l+2}%
{Mt})\right]  }{l(l+1)(l+2)\left[  2-Mt\sin(\frac{2}{Mt})\right]  }.
\end{equation}
For large $l$, $s_{l}^{M}(t)$ diminishes as $l^{-2}$, and so the correlator
receives an extra suppression of $l^{-4}$. We will later use this result to
estimate the contribution of high spin partial waves. The regularization of
our correlator can be implemented in our Lagrange density by dividing factors
of $s_{l}^{M}(t),$
\begin{align}
&  \phi_{l,n,-m}\left[  \tfrac{\partial}{\partial t}\tfrac{t^{3}}{2}%
\tfrac{\partial}{\partial t}-\tfrac{t}{2}l(l+2)\right]  \phi_{l,n,m}\\
&  \rightarrow\left[  (s_{l}^{M}(t))^{-1}\phi_{l,n,-m}\right]  \left[
\tfrac{\partial}{\partial t}\tfrac{t^{3}}{2}\tfrac{\partial}{\partial
t}-\tfrac{t}{2}l(l+2)\right]  \left[  (s_{l}^{M}(t))^{-1}\phi_{l,n,m}\right]
.\nonumber
\end{align}
We now include the interaction and counterterms. We first define%
\begin{align}
&  \left[
\genfrac{}{}{0pt}{1}{l_{1},n_{1},m_{1};l_{2},n_{2},m_{2}}{l_{3},n_{3}%
,m_{3};l_{4},n_{4},m_{4}}%
\right] \\
&  =%
{\textstyle\int}
d^{3}\Omega\,Y_{l_{1},n_{1},m_{1}}(\theta,\psi,\varphi)Y_{l_{2},n_{2},m_{2}%
}(\theta,\psi,\varphi)Y_{l_{3},n_{3},m_{3}}(\theta,\psi,\varphi)Y_{l_{4}%
,n_{4},m_{4}}(\theta,\psi,\varphi).\nonumber
\end{align}
We can write the full functional integral as%
\begin{equation}
\int\mathcal{D}\phi\exp\left[
{\textstyle\int}
d^{4}\vec{t}\,\mathcal{L}\right]  \propto\int\left(
{\textstyle\prod_{l,n,m}}
\mathcal{D}\phi_{l,n,m}^{\prime}\right)  \exp\left[
{\textstyle\int_{0}^{\infty}}
dt\,(L_{1}+L_{2}+L_{3})\right]  ,
\end{equation}
where%
\begin{equation}
L_{1}=\sum_{l,m,n}\left[  (-1)^{m}\left[  (s_{l}^{M}(t))^{-1}\phi
_{l,n,-m}^{\prime}\right]  \left[  \tfrac{\partial}{\partial t}\tfrac{t^{3}%
}{2}\tfrac{\partial}{\partial t}-\tfrac{t}{2}l(l+2)\right]  \left[  (s_{l}%
^{M}(t))^{-1}\phi_{l,n,m}^{\prime}\right]  \right]  , \label{l1}%
\end{equation}%
\begin{equation}
L_{2}=\sum_{l,m,n}\left[  (-1)^{m}\phi_{l,n,-m}^{\prime}\left[
\begin{array}
[c]{c}%
\left[  -c_{\nabla\phi\nabla\phi}^{(0)}-c_{\nabla\phi\nabla\phi}^{\hat{t}%
\hat{t},(0)}\right]  \tfrac{\partial}{\partial t}\tfrac{t^{3}}{2}%
\tfrac{\partial}{\partial t}\\
+c_{\nabla\phi\nabla\phi}^{(0)}\tfrac{t}{2}l(l+2)+t^{3}(M^{2}c_{\phi^{2}%
}^{(0)}+\tfrac{1}{t^{2}}c_{\phi^{2}}^{(2)})
\end{array}
\right]  \phi_{l,n,m}^{\prime}\right]  , \label{l2}%
\end{equation}%
\begin{equation}
L_{3}=-t^{3}(\tfrac{\lambda}{4!}-c_{\phi^{4}}^{(0)})\sum_{l_{i},m_{i},n_{i}%
}\left[
\genfrac{}{}{0pt}{1}{l_{1},m_{1},n_{1};l_{2},m_{2},n_{2}}{l_{3},m_{3}%
,n_{3};l_{4},m_{4},n_{4}}%
\right]  \phi_{l_{1},m_{1},n_{1}}^{\prime}\phi_{l_{2},m_{2},n_{2}}^{\prime
}\phi_{l_{3},m_{3},n_{3}}^{\prime}\phi_{l_{4},m_{4},n_{4}}^{\prime}.
\label{l3}%
\end{equation}
We have used primes in preparation for redefining the fields,%
\begin{equation}
(s_{l}^{M}(t))^{-1}\phi_{l,n,m}^{\prime}=\phi_{l,n,m}.
\end{equation}
The Jacobian of this transformation is a constant (although infinite) and can
be absorbed into the normalization of the functional integral. Now the
Lagrangian $L_{1}$ has the usual free-field form in terms of $\phi_{l,n,m}$
while $L_{2}$ and $L_{3}$ are now functions of $s_{l}^{M}(t)\phi_{l,n,m}$.

With $M$ serving as our ultraviolet regulator, the contribution of
high-spin\ partial waves decouples for sufficiently large spin $l$. We can
estimate the order of magnitude of this contribution in the following manner.
We first identify $t^{-1}l$ (where $t$ is the characteristic radius we are
considering) as an estimate of the magnitude of the tangential momentum,
$p_{T}$. For $p_{T}\gg M\gg t^{-1}$ our correlator scales as $\tfrac{M^{4}%
}{p_{T}^{6}}.$ By dimensional analysis, a diagram with $N_{L}$ loops and
$N_{I}$ internal lines will receive a contribution from partial waves with
spin $\geq l$ of order%
\begin{equation}
\left(  \tfrac{M^{4}}{p_{T}^{6}}\right)  ^{N_{I}}\left(  p_{T}\right)
^{4N_{L}}=\left(  \tfrac{M^{4}}{(t^{-1}l)^{6}}\right)  ^{N_{I}}\left(
t^{-1}l\right)  ^{4N_{L}}. \label{est}%
\end{equation}
\qquad

\section{One-loop examples}

\begin{figure}[htbp]
\begin{center}
\epsfbox{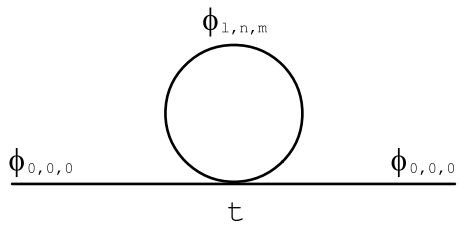}
\caption{One-loop two-point correlator for $\phi_{0,0,0}$}
\label{fig:ren2}
\end{center}
\end{figure}

\begin{figure}[htbp]
\begin{center}
\epsfbox{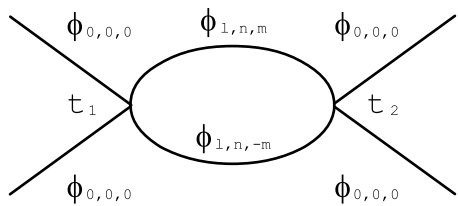}
\caption{One-loop four-point correlator for $\phi_{0,0,0}$}
\label{fig:ren3}
\end{center}
\end{figure}

We will devote the remainder of our discussion to computing one-loop spherical
Feynman diagrams as a check of our results. Our calculations are done both
numerically and analytically. The diagrams we will consider are shown in
Figures \ref{fig:ren2} and \ref{fig:ren3}.
 We start with the two-point function in Figure \ref{fig:ren2}. The
amplitude can be written as $t^{3}B(t)$ where%
\begin{equation}
B(t)\propto%
{\textstyle\sum_{l,n,m}}
\tfrac{1}{t^{2}(l+1)}(s_{l}^{M}(t))^{2}. \label{g}%
\end{equation}
Constants of proportionality are not important here and so we will define
$B(t)$ to be equal to the right side of (\ref{g}). Our results tell us that if
we choose our mass counterterms appropriately, the combination
\begin{equation}
B(t)+M^{2}c_{\phi^{2}}^{(0)}+\tfrac{1}{t^{2}}c_{\phi^{2}}^{(2)}%
\end{equation}
should be independent of $t$, or more succinctly,%
\begin{equation}
B(t)+\tfrac{1}{t^{2}}c_{\phi^{2}}^{(2)}%
\end{equation}
is independent of $t$. Let us first check this analytically. In the absence of
a high-spin cutoff, we can explicitly calculate the sum in (\ref{g}):%
\begin{equation}
B(t)=\tfrac{1}{t^{2}}+b(t)
\end{equation}
where%
\begin{equation}
b(t)=\tfrac{4M^{2}\sin^{4}(\frac{1}{Mt})}{(2-Mt\sin(\frac{2}{Mt}))^{2}}.
\end{equation}
In the limit $M\rightarrow\infty,$%
\begin{equation}
B(t)\rightarrow\tfrac{1}{t^{2}}+\tfrac{9}{4}M^{2}.
\end{equation}
We conclude that $c_{\phi^{2}}^{(2)}=-1$ and $B(t)+\tfrac{1}{t^{2}}c_{\phi
^{2}}^{(2)}$ is in fact translationally invariant.

\begin{figure}[htbp]
\begin{center}
\epsfbox{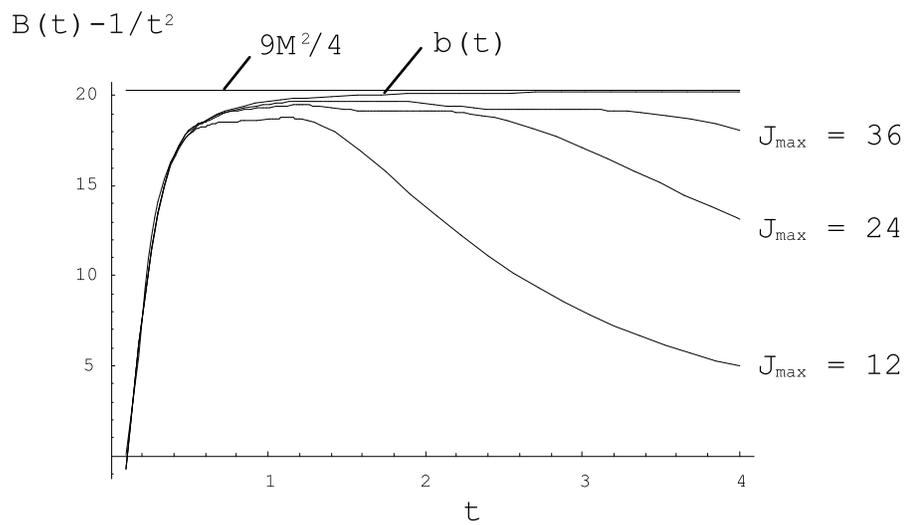}
\caption{Plot of $B(t)-\tfrac{1}{t^{2}}$}
\label{fig:ren4}
\end{center}
\end{figure}

In Figure \ref{fig:ren4}
 we have plotted $B(t)-\tfrac{1}{t^{2}}$, computed numerically for
various values of the high-spin cutoff $J_{\max}$. We have also plotted the
limiting values $b(t)$ and $\tfrac{9}{4}M^{2}$. In our plot $t$ is measured in
units of $m^{-1}$ and $B(t)-\tfrac{1}{t^{2}}$ is in units of $m^{2}$, where
$m$ is an arbitrary mass scale such that $M=3m$. As expected, the errors are
of size $\frac{M^{4}t^{2}}{J_{\max}^{2}}$. There is clearly a deviation from
$\tfrac{9}{4}M^{2}$ for $t\lesssim M^{-1}$ but the integral of the deviation
is negligible as $M\rightarrow\infty$.

We now turn to the four-point function in Figure \ref{fig:ren3}. 
The amplitude can be
written as $t_{1}^{3}t_{2}^{3}C(t_{1},t_{2})$ where%
\begin{equation}
C(t_{1},t_{2})\propto%
{\textstyle\sum_{l,n,m}}
\tfrac{(s_{l}^{M}(t_{1}))^{2}(s_{l}^{M}(t_{2}))^{2}}{(l+1)^{2}}\left[
\tfrac{t_{1}^{l}}{t_{2}^{l+2}}\theta(t_{2}-t_{1})+\tfrac{t_{2}^{l}}%
{t_{1}^{l+2}}\theta(t_{1}-t_{2})\right]  ^{2}. \label{sv}%
\end{equation}
Again constants of proportionally are not important and so we will define
$C(t_{1},t_{2})$ to be equal to the right side of (\ref{sv}). We can write
$C(t_{1},t_{2})$ in terms of the regulated correlator $G_{M^{2}}(\vec{t}%
_{1},\vec{t}_{2}),\footnote{We recall that the regulated correlator goes with
$\phi_{l,n,m}^{\prime}$ rather than $\phi_{l,n,m}$. But this is not important
here since $\phi_{0,0,0}^{\prime}$ $=\phi_{0,0,0}$.}$%
\begin{equation}
C(t_{1},t_{2})\propto\int d^{3}\hat{t}_{1}d^{3}\hat{t}_{2}\left[  G_{M^{2}%
}(\vec{t}_{1},\vec{t}_{2})\right]  ^{2}\propto\int d^{3}\hat{t}_{1}\left[
G_{M^{2}}(\vec{t}_{1},\vec{t}_{2})\right]  ^{2}.
\end{equation}
Since the coupling constant counterterm
\begin{equation}
c_{\phi^{4}}^{(0)}\delta^{4}(\vec{t}_{1}-\vec{t}_{2})
\end{equation}
is translationally invariant, the amplitude by itself should be
translationally invariant. Let us define
\begin{equation}
\int d^{4}\vec{t}_{2}e^{-i\vec{p}\cdot(\vec{t}_{1}-\vec{t}_{2})}\left[
G_{M^{2}}(\vec{t}_{1},\vec{t}_{2})\right]  ^{2}=f(\vec{p}^{2}),
\end{equation}
so that
\begin{equation}
\int d^{4}\vec{t}_{2}e^{i\vec{p}\cdot\vec{t}_{2}}\left[  G_{M^{2}}(\vec{t}%
_{1},\vec{t}_{2})\right]  ^{2}=e^{i\vec{p}\cdot\vec{t}_{1}}f(\vec{p}^{2}).
\end{equation}
Integrating over $\hat{t}_1$, we find%
\begin{equation}
\int dt_{2}\,t_{2}^{2}J_{1}(pt_{2})C(t_{1},t_{2})\propto\tfrac{1}{t_{1}}%
J_{1}(pt_{1})f(\vec{p}^{2}).
\end{equation}
Let us define%
\begin{equation}
C(t)=%
{\textstyle\int}
dt_{2}\,t_{2}^{2}J_{1}(pt_{2})C(t,t_{2}).
\end{equation}
We now check that in fact
\begin{equation}
C(t)\propto\tfrac{1}{t_{1}}J_{1}(pt_{1})\text{.}%
\end{equation}

\begin{figure}[htbp]
\begin{center}
\epsfbox{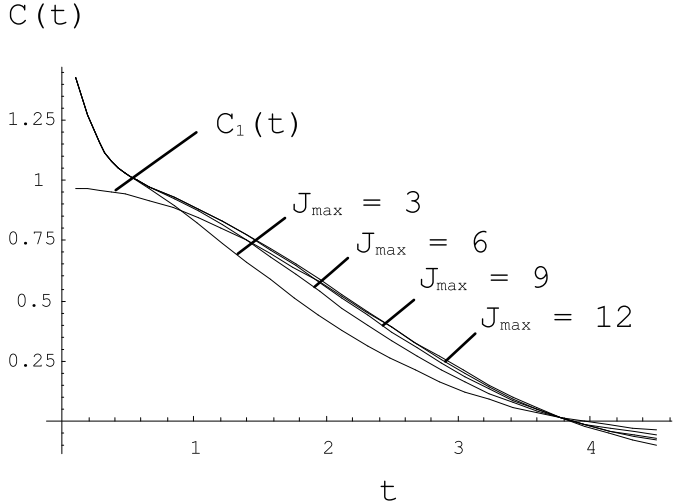}
\caption{Plot of $C(t)$}
\label{fig:ren5}
\end{center}
\end{figure}

In the absence of a high-spin cutoff, we find that $C(t)$ is given
by\footnote{This calculation is somewhat lengthy. Details can be obtained upon
request from the authors.}%
\begin{equation}
C(t)=\tfrac{1}{t_{1}}J_{1}(pt_{1})\left[  \tfrac{1}{2}\log\tfrac{M^{2}}{p^{2}%
}+c\right]  +\cdots,
\end{equation}
where the ellipsis represents terms which vanish as $M\rightarrow\infty$ and
\begin{equation}
c=324\left[  \int_{0}^{1/2}dk\left(  \tfrac{(\sin k-k\cos k)^{4}}{4k^{13}%
}-\tfrac{1}{324k}\right)  +\int_{1/2}^{\infty}dk\tfrac{(\sin k-k\cos k)^{4}%
}{4k^{13}}\right]  .
\end{equation}
In Figure \ref{fig:ren5}
 we plot $C(t)$ for different values of the high-spin cutoff
$J_{\max}$ as well as the large-$M$ limit value
\begin{equation}
C_{1}(t)=\tfrac{1}{t_{1}}J_{1}(pt_{1})\left[  \tfrac{1}{2}\log\tfrac{M^{2}%
}{p^{2}}+c\right]  .
\end{equation}
In our plot $t$ is measured in units of $p^{-1}$ and $M=3p$. From (\ref{est})
the expected error is of size $\frac{M^{8}t^{8}}{J_{\max}^{8}}.$ We see that
the data is consistent with the results expected. Again the deviation for
$t\lesssim M^{-1}$ integrates to a negligible contribution as $M\rightarrow
\infty$.

\section{Summary}

We have examined several important features of non-perturbative
renormalization in the spherical field formalism and answered the three
questions posed in the introduction. Ultraviolet divergences can be cancelled
by a finite number of local counterterms in a manner such that the
renormalized theory is translationally invariant. Using angle-smearing
regularization we find that the counterterms for $\phi^{4}$ theory in four
dimensions can be parameterized by five unknown constants as shown in
(\ref{an}). Aside from our remarks about Ward identity constraints in gauge
theories, the extension to other field theories is straightforward. We hope
that these results will be useful for future studies of general renormalizable
theories by spherical field techniques.\bigskip

\chapter[Massless Thirring model]
{The massless Thirring model in spherical field theory \footnote{N. Salwen, D. Lee, Phys. Lett. B468 (1999) 118.}}

\section{Introduction}

The massless Thirring model \cite{thirring} is an exactly soluble system of a
single self-interacting massless fermion in two dimensions. \ There are a
number of solutions\ in the literature based on properties of the
Euler-Lagrange equations and fermion currents or bosonization techniques
\cite{johnson, sommerfield, hagen, k1, k2, k3, k4, k5, k6, k7, mueller}. \ Given its simplicity and solubility, the
model has become a popular testing ground for new ideas and methods in field theory.

From a computational point of view, however, the massless Thirring model still
presents a significant challenge. In the lattice field formalism, the
difficulties are due to the appearance of fermion doubler states and singular
inversion problems associated with integrating out massless fermions. \ In
this work we use the model to illustrate new non-perturbative methods in the
spherical field formalism \cite{a1, fermion, a3, renorm}. \ The techniques we
present are quite general and can also be applied to other modal expansion
methods such as periodic field theory \cite{periodic}.

As noted in \cite{fermion}, we will not need to worry about fermion doubling.
\ This is true for any modal field theory and follows from the fact that space
is not discretized but retained as a continuous variable. \ Since our model is
not super-renormalizable we will use a procedure called angle-smearing, a
regularization method designed for spherical field theory \cite{renorm}.
\ With angle-smearing regularization and a small set of local counterterms, we
are able to remove all ultraviolet divergences in a manner such that the
renormalized theory is finite and translationally invariant. \ Comparison of
our results with the known Thirring model solution will serve as a consistency
check for our regularization and renormalization procedures.

The organization of this paper is as follows. \ We begin with a short summary
of the massless Thirring model, following the solution of Hagen \cite{hagen}.
Using angle-smearing regularization we obtain the spherical field Hamiltonian
and construct a matrix representation of the Hamiltonian. We reduce the space
of states using a two-parameter auxiliary cutoff procedure. \ In this reduced
space we compute the time evolution of quantum states using a fourth-order
Runge-Kutta-Fehlberg algorithm. \ We calculate the two point correlator for
several values of the coupling and find agreement with the known analytic solution.

\section{The model}

We start with a list of our notational conventions. \ Our analysis will be in
two-dimensional Euclidean space, and we use both cartesian and polar
coordinates,
\begin{equation}
\vec{t}=(t\cos\theta,t\sin\theta)=(x,y).
\end{equation}
The components of the spinors $\psi$ and $\bar{\psi}$ are written as
\begin{equation}
\psi=\left[
\begin{array}
[c]{c}%
\psi^{\uparrow}\\
\psi^{\downarrow}%
\end{array}
\right]  \qquad\bar{\psi}=\left[
\begin{array}
[c]{cc}%
\bar{\psi}^{\uparrow} & \bar{\psi}^{\downarrow}%
\end{array}
\right]  .
\end{equation}
Our representation for the Dirac matrices is%

\begin{equation}
\vec{\gamma}=i\vec{\sigma},
\end{equation}
and so $\vec{\gamma}$ satisfies
\begin{equation}
\left\{  \gamma^{i},\gamma^{j}\right\}  =-2\delta^{ij},\qquad i,j=1,2.
\end{equation}

The massless Thirring model is formally defined by the Lagrange density
\begin{equation}
\mathcal{L}=i\bar{\psi}\vec{\gamma}\cdot\vec{\nabla}\psi-\tfrac{\lambda}%
{2}\,\vec{j}\cdot\vec{j},
\end{equation}
where $\vec{j}$ is the fermion vector current. \ Johnson \cite{johnson}
emphasized the importance of defining the regularized current precisely, and
this was further clarified by the work of Sommerfield \cite{sommerfield} and
Hagen \cite{hagen}. We will use a regularization technique, introduced in
\cite{renorm}, called angle-smearing. We define the regularized current as
\begin{equation}
\vec{j}=\tfrac{1}{2}\left(  \bar{\psi}_{s}\vec{\gamma}\psi_{s}-Tr[\vec{\gamma
}\psi_{s}\bar{\psi}_{s}]\right)  ,
\end{equation}
where
\begin{equation}
\psi_{s}(t,\theta)=\tfrac{Mt}{2}\int_{-\frac{1}{Mt}}^{\frac{1}{Mt}%
}d\varepsilon\psi(t,\theta+\varepsilon). \label{sme}%
\end{equation}
We identify the radial variable $t$ as our time parameter, and our definition
of the current is local with respect to $t$.

Hagen \cite{hagen} described the solution of the Thirring model in the
Hamiltonian formalism with currents defined as products of the canonical
operators at equal times. Though our equal-time surface is curved, the
curvature of the integration segment in (\ref{sme}) scales as $\frac{1}{M}$
while the ultraviolet divergences in this model are only logarithmic in $M$.
In the $M\rightarrow\infty$ limit we therefore recover the standard results.
As discussed in \cite{hagen}, there exists a one parameter class of solutions
to the Thirring model depending on the preferred definition of the regularized
vector and axial vector currents. We will use the conventions used in
\cite{johnson} and \cite{sommerfield}, which in Hagen's notation corresponds
with the parameter values $\xi=\eta=\frac{1}{2}$. \ With this choice the
Hamiltonian density takes the form
\begin{equation}
\mathcal{H}=\mathcal{H}_{free}+\tfrac{\pi c}{1-c}(\hat{t}\cdot\vec{j}%
)^{2}+\tfrac{\pi c}{1+c}(\hat{\theta}\cdot\vec{j})^{2},\label{la}%
\end{equation}
where
\begin{equation}
c=\tfrac{\lambda}{2\pi}\text{.}%
\end{equation}

The only counterterms needed in this model are wavefunction renormalization
counterterms, a result of our careful definition for the regularized
interaction in (\ref{la}). As in \cite{hagen} we calculate correlation
functions using an unrenormalized Hamiltonian. \ The divergent wavefunction
normalizations will appear simply as overall factors in the correlators.\qquad

\section{Spherical field Hamiltonian}

In this section we derive the form of the spherical field Hamiltonian. We
first expand the fermion current in terms of components of the spinors,
\begin{equation}
\hat{t}\cdot\bar{\psi}_{s}\vec{\gamma}\psi_{s}=i\bar{\psi}_{s}\left[
\begin{array}
[c]{cc}%
0 & e^{-i\theta}\\
e^{i\theta} & 0
\end{array}
\right]  \psi_{s}=ie^{-i\theta}\bar{\psi}_{s}^{\uparrow}\psi_{s}^{\downarrow
}+ie^{i\theta}\bar{\psi}_{s}^{\downarrow}\psi_{s}^{\uparrow}%
\end{equation}%
\begin{equation}
\hat{\theta}\cdot\bar{\psi}_{s}\vec{\gamma}\psi_{s}=\bar{\psi}_{s}\left[
\begin{array}
[c]{cc}%
0 & e^{-i\theta}\\
-e^{i\theta} & 0
\end{array}
\right]  \psi_{s}=e^{-i\theta}\bar{\psi}_{s}^{\uparrow}\psi_{s}^{\downarrow
}-e^{i\theta}\bar{\psi}_{s}^{\downarrow}\psi_{s}^{\uparrow}.
\end{equation}
The anti-commutators of the regulated fields are\footnote{Our definition of
the Euclidean fermion fields and anti-commutation relations follows the
conventions of \cite{fubini}.}
\begin{equation}
\left\{  \bar{\psi}_{s}^{\uparrow},\psi_{s}^{\downarrow}\right\}  =\tfrac
{1}{t}\left(  \tfrac{Mt}{2}\right)  ^{2}%
{\textstyle\int_{-\frac{1}{Mt}}^{\frac{1}{Mt}}}
e^{i(\theta+\varepsilon)}d\varepsilon=A(t)e^{i\theta} \label{an1}%
\end{equation}%
\begin{equation}
\left\{  \bar{\psi}_{s}^{\downarrow},\psi_{s}^{\uparrow}\right\}  =\tfrac
{1}{t}\left(  \tfrac{Mt}{2}\right)  ^{2}%
{\textstyle\int_{-\frac{1}{Mt}}^{\frac{1}{Mt}}}
e^{-i(\theta+\varepsilon)}d\varepsilon=A(t)e^{-i\theta}, \label{an2}%
\end{equation}
where
\begin{equation}
A(t)=\tfrac{M^{2}t}{2}\sin(\tfrac{1}{Mt}). \label{a}%
\end{equation}
From the anti-commutation relations, the $\hat{t}$ component of the current
is
\begin{align}
\hat{t}\cdot\vec{j}  &  =\tfrac{1}{2}\left[  ie^{-i\theta}(\bar{\psi}%
_{s}^{\uparrow}\psi_{s}^{\downarrow}-\psi_{s}^{\downarrow}\bar{\psi}%
_{s}^{\uparrow})+ie^{i\theta}(\bar{\psi}_{s}^{\downarrow}\psi_{s}^{\uparrow
}-\psi_{s}^{\uparrow}\bar{\psi}_{s}^{\downarrow})\right] \\
&  =ie^{-i\theta}\bar{\psi}_{s}^{\uparrow}\psi_{s}^{\downarrow}+ie^{i\theta
}\bar{\psi}_{s}^{\downarrow}\psi_{s}^{\uparrow}-iA(t),\nonumber
\end{align}
and so
\begin{equation}
(\hat{t}\cdot\vec{j})^{2}=A(t)\left[  e^{-i\theta}\bar{\psi}_{s}^{\uparrow
}\psi_{s}^{\downarrow}+e^{i\theta}\bar{\psi}_{s}^{\downarrow}\psi
_{s}^{\uparrow}\right]  -2\bar{\psi}_{s}^{\uparrow}\psi_{s}^{\downarrow}%
\bar{\psi}_{s}^{\downarrow}\psi_{s}^{\uparrow}-A^{2}(t).
\end{equation}
Similarly we find
\begin{equation}
(\hat{\theta}\cdot\vec{j})^{2}=A(t)\left[  e^{-i\theta}\bar{\psi}%
_{s}^{\uparrow}\psi_{s}^{\downarrow}+e^{i\theta}\bar{\psi}_{s}^{\downarrow
}\psi_{s}^{\uparrow}\right]  -2\bar{\psi}_{s}^{\uparrow}\psi_{s}^{\downarrow
}\bar{\psi}_{s}^{\downarrow}\psi_{s}^{\uparrow}.
\end{equation}
The Hamiltonian can now be written as
\begin{equation}
H=H_{free}+\int d\theta\,t\left\{  \tfrac{2\pi c}{1-c^{2}}\left[  A(t)\left[
e^{-i\theta}\bar{\psi}_{s}^{\uparrow}\psi_{s}^{\downarrow}+e^{i\theta}%
\bar{\psi}_{s}^{\downarrow}\psi_{s}^{\uparrow}\right]  -2\bar{\psi}%
_{s}^{\uparrow}\psi_{s}^{\downarrow}\bar{\psi}_{s}^{\downarrow}\psi
_{s}^{\uparrow}\right]  \right\}  .
\end{equation}
We have omitted the constant term proportional to $A^{2}(t)$.

Let us define the partial wave modes
\begin{align}
\psi_{n}(t)  &  =\tfrac{1}{\sqrt{2\pi}}\int d\theta\,e^{-in\theta}\psi(\vec
{t}),\qquad\psi_{s,n}(t)=\tfrac{1}{\sqrt{2\pi}}\int d\theta\,e^{-in\theta}%
\psi_{s}(\vec{t}),\\
\bar{\psi}_{n}(t)  &  =\tfrac{1}{\sqrt{2\pi}}\int d\theta\,e^{-in\theta}%
\bar{\psi}(\vec{t}),\qquad\bar{\psi}_{s,n}(t)=\tfrac{1}{\sqrt{2\pi}}\int
d\theta\,e^{-in\theta}\bar{\psi}_{s}(\vec{t}).
\end{align}
It is straightforward to show that for $n\neq0,$
\begin{equation}
\psi_{s,n}(t)=\tfrac{\sin(\frac{n}{Mt})}{\left(  \frac{n}{Mt}\right)  }%
\psi_{n}(t)\qquad\bar{\psi}_{s,n}(t)=\tfrac{\sin(\frac{n}{Mt})}{\left(
\frac{n}{Mt}\right)  }\bar{\psi}_{n}(t). \label{a1}%
\end{equation}
We can extend this result to the case $n=0$ using the convenient shorthand
\begin{equation}
\tfrac{\sin(\frac{0}{Mt})}{\left(  \frac{0}{Mt}\right)  }\equiv1.
\end{equation}
At this point it is convenient to rescale $\bar{\psi}$,
\begin{equation}
\bar{\psi}_{n}^{i\prime}=t\bar{\psi}_{n}^{i}.
\end{equation}
In terms of the partial waves,
\begin{align}
H=  &  \tfrac{1}{t}\sum_{n}\left[  \left(  n+1+\tfrac{b\pi tA(t)\sin(\frac
{n}{Mt})\sin(\frac{n+1}{Mt})}{\left(  \frac{n}{Mt}\right)  \left(  \frac
{n+1}{Mt}\right)  }\right)  \bar{\psi}_{-n}^{\uparrow\prime}\psi
_{n+1}^{\downarrow}\right] \\
&  -\tfrac{1}{t}\sum_{n}\left[  \left(  n-\tfrac{b\pi tA(t)\sin(\frac{n}%
{Mt})\sin(\frac{n+1}{Mt})}{\left(  \frac{n}{Mt}\right)  \left(  \frac{n+1}%
{Mt}\right)  }\right)  \bar{\psi}_{-n-1}^{\downarrow\prime}\psi_{n}^{\uparrow
}\right] \nonumber\\
&  -\sum_{-n_{1}+n_{2}-n_{3}+n_{4}=0}\left[  \tfrac{b}{t}\bar{\psi}_{-n_{1}%
}^{\uparrow\prime}\psi_{n_{2}+1}^{\downarrow}\bar{\psi}_{-n_{3}-1}%
^{\downarrow\prime}\psi_{n_{4}}^{\uparrow}\tfrac{\sin(\frac{n_{1}}{Mt}%
)\sin(\frac{n_{2}+1}{Mt})\sin(\frac{n_{3}+1}{Mt})\sin(\frac{n_{4}}{Mt}%
)}{\left(  \frac{n_{1}}{Mt}\right)  \left(  \frac{n_{2}+1}{Mt}\right)  \left(
\frac{n_{3}+1}{Mt}\right)  \left(  \frac{n_{4}}{Mt}\right)  }\right]
,\nonumber
\end{align}
where
\begin{equation}
b=\tfrac{2c}{1-c^{2}}. \label{bc}%
\end{equation}
Since $b$ is the parameter appearing in the Hamiltonian, it is somewhat more
convenient to express $c$ in terms of $b,$%
\begin{equation}
c=\tfrac{\sqrt{1+b^{2}}-1}{b}. \label{cb}%
\end{equation}

Let us define the ladder operators\footnote{This notation is slightly
different from that used in \cite{fermion}. The translation is as follows:
$a_{n}^{\downarrow},a_{n}^{\downarrow\dagger}=a_{n}^{\downarrow-}%
,a_{n}^{\downarrow+}$; $a_{n}^{\uparrow},a_{n}^{\uparrow\dagger}%
=a_{-n}^{\uparrow-},a_{-n}^{\uparrow+}.$}
\begin{align}
a_{-n}^{\uparrow},a_{-n}^{\uparrow\dagger}  &  \equiv\psi_{n}^{\uparrow}%
,\bar{\psi}_{-n-1}^{\downarrow\prime}\\
a_{n+1}^{\downarrow},a_{n+1}^{\downarrow\dagger}  &  \equiv\psi_{n+1}%
^{\downarrow},\bar{\psi}_{-n}^{\uparrow\prime}.
\end{align}
These operators satisfy the anti-commutation relations%

\begin{equation}
\left\{  a_{n_{1}}^{\downarrow},a_{n_{2}}^{\downarrow\dagger}\right\}
=\left\{  a_{n_{1}}^{\uparrow},a_{n_{2}}^{\uparrow\dagger}\right\}
=\delta_{n_{1}n_{2}},
\end{equation}
with all other anti-commutators equal to zero. We can now recast the
Hamiltonian as
\begin{align}
H  &  =\tfrac{1}{t}\sum_{n}\left[  n+\tfrac{b\pi tA(t)\sin(\frac{n}{Mt}%
)\sin(\frac{n-1}{Mt})}{\left(  \frac{n}{Mt}\right)  \left(  \frac{n-1}%
{Mt}\right)  }\right]  \left(  a_{n}^{\downarrow\dagger}a_{n}^{\downarrow
}+a_{n}^{\uparrow\dagger}a_{n}^{\uparrow}\right) \\
&  -\sum_{-n_{1}+n_{2}+n_{3}-n_{4}=0}\left[  \tfrac{b}{t}a_{n_{1}}%
^{\downarrow\dagger}a_{n_{2}}^{\downarrow}a_{n_{3}}^{\uparrow\dagger}a_{n_{4}%
}^{\uparrow}\tfrac{\sin(\frac{n_{1}-1}{Mt})\sin(\frac{n_{2}}{Mt})\sin
(\frac{n_{3}-1}{Mt})\sin(\frac{n_{4}}{Mt})}{\left(  \frac{n_{1}-1}{Mt}\right)
\left(  \frac{n_{2}}{Mt}\right)  \left(  \frac{n_{3}-1}{Mt}\right)  \left(
\frac{n_{4}}{Mt}\right)  }\right]  .\nonumber
\end{align}

We will implement a high spin cutoff by removing terms in the interaction
containing operators $a_{n}^{\downarrow}$, $a_{n}^{\uparrow},$ $a_{n}%
^{\downarrow\dagger}$, or $a_{n}^{\uparrow\dagger}$ for $\left|  n\right|
>J_{\max}$. This has the effect of removing high spin modes, which correspond
with large tangential momentum states. \ We should emphasize, however, that
$J_{\max}$ is an auxiliary cutoff. It does not play a role in the
regularization scheme since the interactions have already been rendered finite
using angle-smearing. \ The contribution of these high spin modes is
negligible so long as
\begin{equation}
\tfrac{J_{\max}}{t}\gg M,
\end{equation}
where $t$ is the characteristic radius of the process being measured.
\ Returning back to (\ref{an1}) and (\ref{an2}) and removing the contribution
of these partial waves, we find that $A(t)$ is replaced by
\begin{equation}
A_{J_{\max}}(t)=\tfrac{1}{2\pi t}\sum_{n=-J_{\max}}^{J_{\max}}\tfrac
{\sin(\frac{n}{Mt})}{\left(  \frac{n}{Mt}\right)  }\tfrac{\sin(\frac{n-1}%
{Mt})}{\left(  \frac{n-1}{Mt}\right)  }. %
\end{equation}

Let $\left|  0\right\rangle _{free}$ be the ground state of the free massless
fermion Hamiltonian.\footnote{The ground state of the free massless
Hamiltonian is actually degenerate due to s-wave excitations, but this is
remedied by taking the $m\longrightarrow0$ limit of the massive fermion
theory.} For $n>0,$ we find%

\begin{equation}
a_{n}^{\downarrow}\left|  0\right\rangle _{free}=a_{n}^{\uparrow}\left|
0\right\rangle _{free}=0,
\end{equation}
and for $n\leq0,$
\begin{equation}
a_{n}^{\downarrow\dagger}\left|  0\right\rangle _{free}=a_{n}^{\uparrow
\dagger}\left|  0\right\rangle _{free}=0.
\end{equation}
It is convenient to define the normal-ordered products
\begin{equation}
\text{{}}\text{:{}}a_{n}^{\downarrow\dagger}a_{n}^{\downarrow}\text{:}%
=\left\{
\begin{array}
[c]{c}%
a_{n}^{\downarrow\dagger}a_{n}^{\downarrow}\text{ for }n>0\\
-a_{n}^{\downarrow}a_{n}^{\downarrow\dagger}\text{ for }n\leq0
\end{array}
\right.  \qquad\text{:{}}a_{n}^{\uparrow\dagger}a_{n}^{\uparrow}%
\text{:}=\left\{
\begin{array}
[c]{c}%
a_{n}^{\uparrow\dagger}a_{n}^{\uparrow}\text{ for }n>0\\
-a_{n}^{\uparrow}a_{n}^{\uparrow\dagger}\text{ for }n\leq0.
\end{array}
\right.
\end{equation}
The ordering for other operators is immaterial since the anti-commutators are
zero. We can now rewrite $H$ in terms of normal-ordered products,%

\begin{align}
H  &  =\left(  \tfrac{n}{t}+O(J_{\max}^{-2})\right)  \left(  a_{n}%
^{\downarrow\dagger}a_{n}^{\downarrow}+a_{n}^{\uparrow\dagger}a_{n}^{\uparrow
}\right) \\
&  -\sum_{-n_{1}+n_{2}+n_{3}-n_{4}=0}\left[  \tfrac{b}{t}\text{:}a_{n_{1}%
}^{\downarrow\dagger}a_{n_{2}}^{\downarrow}a_{n_{3}}^{\uparrow\dagger}%
a_{n_{4}}^{\uparrow}\text{:}\tfrac{\sin(\frac{n_{1}-1}{Mt})\sin(\frac{n_{2}%
}{Mt})\sin(\frac{n_{3}-1}{Mt})\sin(\frac{n_{4}}{Mt})}{\left(  \frac{n_{1}%
-1}{Mt}\right)  \left(  \frac{n_{2}}{Mt}\right)  \left(  \frac{n_{3}-1}%
{Mt}\right)  \left(  \frac{n_{4}}{Mt}\right)  }\right]  .\nonumber
\end{align}
There is an $O(J_{\max}^{-2})$ term due to a small asymmetry in our cutoff
procedure with respect to the two boundaries $-J_{\max}$ and $J_{\max}%
$.\footnote{If desired we could eliminate this term and the asymmetry by a
slight change in the angle-smearing procedure for $\bar{\psi}$.} We will
neglect this term in the limit $J_{\max}\rightarrow\infty$.

\section{Two-point correlator}

We wish to study the properties of the two-point correlator. The massless
Thirring model is invariant under the discrete transformation%

\begin{equation}
\psi^{\downarrow}(\vec{t}),\bar{\psi}^{\uparrow}(\vec{t})\rightarrow
-\psi^{\downarrow}(\vec{t}),-\bar{\psi}^{\uparrow}(\vec{t}),
\end{equation}
as well as the transformation
\begin{equation}
\psi^{\uparrow}(t,\theta),\bar{\psi}^{\uparrow}(t,\theta)\leftrightarrow
\psi^{\downarrow}(t,-\theta),\bar{\psi}^{\downarrow}(t,-\theta).
\end{equation}
From these we deduce
\begin{equation}
\left\langle 0\right|  T\left[  \bar{\psi}^{\uparrow}(\vec{t})\psi^{\uparrow
}(0)\right]  \left|  0\right\rangle =\left\langle 0\right|  T\left[  \bar
{\psi}^{\downarrow}(\vec{t})\psi^{\downarrow}(0)\right]  \left|
0\right\rangle =0
\end{equation}
and
\begin{equation}
\left\langle 0\right|  T\left[  \bar{\psi}^{\uparrow}(t,\theta)\psi
^{\downarrow}(0)\right]  \left|  0\right\rangle =\left\langle 0\right|
T\left[  \bar{\psi}^{\downarrow}(t,-\theta)\psi^{\uparrow}(0)\right]  \left|
0\right\rangle . \label{nn}%
\end{equation}
It therefore suffices to consider just the correlator on the left side of
(\ref{nn}).

In the limit $M\rightarrow\infty$ the form of the correlator is given by
\begin{equation}
\left\langle 0\right|  T\left[  \bar{\psi}^{\uparrow}(t,\theta)\psi
^{\downarrow}(0)\right]  \left|  0\right\rangle =\tfrac{e^{i\theta}}{2\pi
}(k(c)M)^{\frac{-2c^{2}}{1-c^{2}}}t^{\frac{-1-c^{2}}{1-c^{2}}},\label{exact}%
\end{equation}
where $k(c)$ is a dimensionless parameter. \ Standard analytic methods do not
yield a simple closed form expression for $k(c).$ \ We will therefore extract
$k(c)$ from the computed value of the correlator at a specific renormalization
point $t=t_{0}$.\footnote{In some regularization schemes $k(c)$ can be
calculated analytically \cite{sommerfield}\cite{mueller}, and it may be
worthwhile to use these techniques in future work. \ In this first analysis,
however, we prefer to present a more straightforward and typical example of
the angle-smearing regularization method.}

We define
\begin{equation}
f(t)=\left\langle 0\right|  T\left[  \bar{\psi}_{1}^{\uparrow}(t)\psi
_{0}^{\downarrow}(0)\right]  \left|  0\right\rangle .
\end{equation}
Since%

\begin{equation}
\left\langle 0\right|  T\left[  \bar{\psi}^{\uparrow}(t,\theta)\psi
^{\downarrow}(0)\right]  \left|  0\right\rangle =\tfrac{e^{i\theta}}{2\pi
}\left\langle 0\right|  T\left[  \bar{\psi}_{1}^{\uparrow}(t)\psi
_{0}^{\downarrow}(0)\right]  \left|  0\right\rangle ,
\end{equation}
we conclude that
\begin{equation}
f(t)=(k(c)M)^{\frac{-2c^{2}}{1-c^{2}}}t^{\frac{-1-c^{2}}{1-c^{2}}}.
\label{analyt}%
\end{equation}
\qquad

We now compute $f(t)$ using the spherical field Hamiltonian. \ We first need a
matrix representation for the Grassmann ladder operators. \ We will use tensor
products of the $2\times2$ identity matrix and Pauli matrices: %

\begin{align}
a_{n}^{\downarrow} &  =%
{\textstyle\bigotimes\limits_{i=J_{\max},-J_{\max}}}
\sigma_{z}%
{\textstyle\bigotimes\limits_{i=J_{\max},n+1}}
\sigma_{z}\otimes\left(  \tfrac{1}{2}\sigma_{x}+\tfrac{i}{2}\sigma
_{y}\right)
{\textstyle\bigotimes\limits_{i=n-1,-J_{\max}}}
1,\label{rep}\\
a_{n}^{\uparrow} &  =%
{\textstyle\bigotimes\limits_{i=J_{\max},n+1}}
\sigma_{z}\otimes\left(  \tfrac{1}{2}\sigma_{x}+\tfrac{i}{2}\sigma
_{y}\right)
{\textstyle\bigotimes\limits_{i=n-1,-J_{\max}}}
1%
{\textstyle\bigotimes\limits_{i=J_{\max},-J_{\max}}}
1.\nonumber
\end{align}
The representations for $a_{n}^{\downarrow\dagger}$ and $a_{n}^{\uparrow
\dagger}$ are defined by the conjugate transposes of these matrices. \ We can
now calculate the correlator $f(t)$ using the relation \cite{fermion}
\begin{equation}
f(t)=\lim_{t_{2}\rightarrow\infty}\lim_{t_{1}\rightarrow0}\tfrac{Tr\left[
T\exp\left\{  -\int_{t}^{t_{2}}dt\,H(t)\right\}  \frac{1}{t}a_{0}%
^{\downarrow\dagger}T\exp\left\{  -\int_{t_{1}}^{t}dt\,H(t)\right\}
a_{0}^{\downarrow}\right]  }{Tr\left[  T\exp\left\{  -\int_{t_{1}}^{t_{2}%
}dt\,H(t)\right\}  \right]  }.\label{ex}%
\end{equation}
A straightforward calculation of (\ref{ex}), however, is rather inefficient.
There are several techniques which we will first use to simplify the calculation.

The time evolution of the system at large $t$ is dominated by the contribution
of the ground state or, more precisely, the adiabatic flow of the
$t$-dependent ground state. \ As discussed in \cite{a1, fermion, a3} a similar
phenomenon occurs at small $t$, due to the divergence of energy levels near
$t=0.$ \ It is therefore not necessary to compute the full matrix trace in the
numerator and denominator of (\ref{ex}). \ It is instead sufficient to compute
the corresponding ratio for a single matrix element. \ After making this
reduction, we can then go a step further and eliminate states which do not
contribute to the matrix element.

The high spin parameter $J_{\max}$ was used to remove high spin modes with
$\left|  n\right|  >J_{\max}$. \ This, however, is not a uniform cutoff in the
space of states and most of the remaining states are still high kinetic energy
states. \ Although none of the individual modes are energetic, many of the
modes can be simultaneously excited. \ Let us define $N_{n}^{\downarrow}$ and
$N_{n}^{\uparrow}$ to be bit switches, 1 or 0, depending on whether or not the
corresponding mode is excited. \ Let us also define a cutoff parameter
$K_{\max}$. We will remove all high kinetic energy states such that
\begin{equation}%
{\textstyle\sum_{n}}
\left[  \left|  n\right|  (N_{n}^{\downarrow}+N_{n}^{\uparrow})\right]  \geq
K_{\max}.
\end{equation}
For consistency $K_{\max}$ should be about the same size as $J_{\max}$.

\section{Results}

The CPU time and memory requirement for calculating (\ref{ex}) scales linearly
with the number of transitions in $H$ (i.e., non-zero elements in our matrix
representation). \ In Table 1 we have shown the number of states and
transitions for different values of $J_{\max}$.%

\[
\overset{\text{Table 1}}{%
\begin{tabular}
[c]{|l|l|l|l|l|l|l|}\hline
$J_{\max}$ & 2 & 4 & 6 & 8 & 10 & 12\\\hline
$\text{states}$ & 6 & 40 & 210 & 920 & 3600 & 13000\\\hline
transitions & 38 & 500 & 4200 & 26000 & 1.4E5 & 3.9E5\\\hline
\end{tabular}
}%
\]
We have calculated $f(t)$ for $J_{\max}\leq12$ and several values of the
coupling $b$. \ The total run time was about 100 hours on a 350 MHz PC with
256 MB RAM.

\begin{figure}[htbp]
\begin{center}
\epsfbox{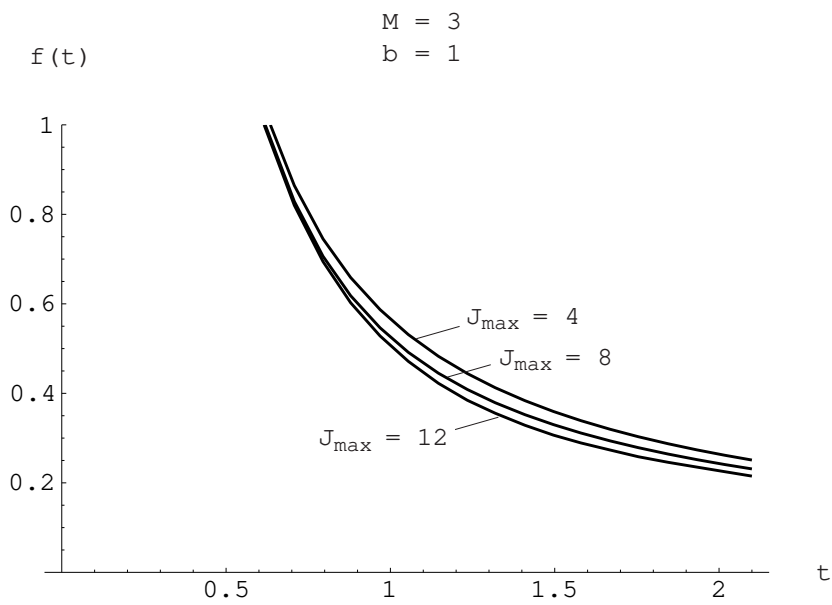}
\caption{Plot of $f_{J_{\max}}(t)$ for $b=1$ and $J_{\max}=4,8,12$}
\label{fig:thi1}
\end{center}
\end{figure}

The matrix time evolution equations in (\ref{ex}) were computed using a
fourth-order Runge-Kutta-Fehlberg algorithm. \ We have set
\begin{equation}
K_{\max}=2J_{\max}+2.
\end{equation}
We will use the notation $f_{J_{\max}}(t)$ to identify the corresponding
result for a given value of $J_{\max}$. \ In Figure \ref{fig:thi1}
 we have plotted
$f_{J_{\max}}(t)$ for $b=1$ and $J_{\max}=4,8,12$. \ We have scaled $t$ and
$f(t)$ in dimensional units chosen such that $M=3$. \ For finite $J_{\max}$ we
expect deviations from the $J_{\max}\rightarrow\infty$ limit to be of size
$O$($\frac{M^{2}t^{2}}{J_{\max}^{2}})$. \ The curves shown in 
Figure \ref{fig:thi1} 
appear
consistent with this rate of convergence.

\begin{figure}[htbp]
  \begin{center}
\epsfbox{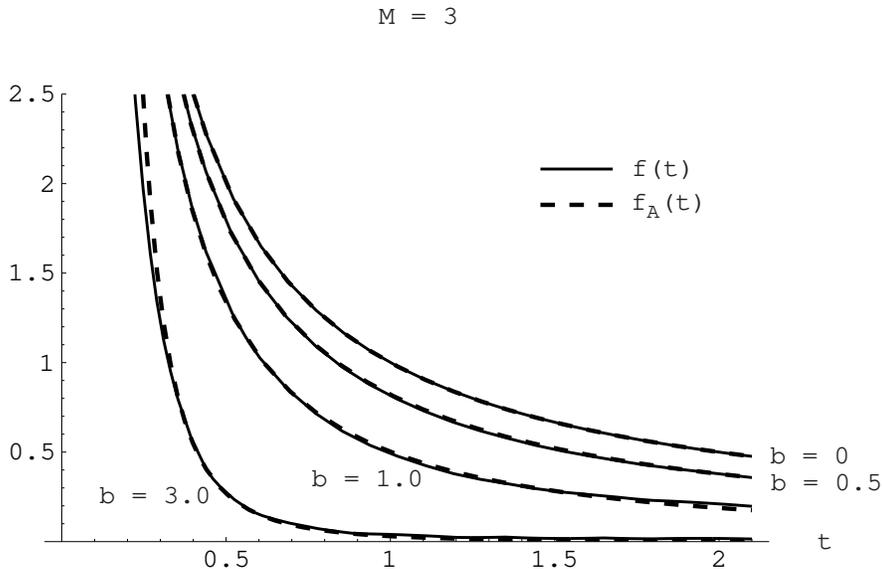}
\caption{Plot of $f_{A}(t)$ and $f(t)$ for $b=0,.5,1,3$ and $M=3$}
\label{fig:thi2}
  \end{center}
\end{figure}

We can extrapolate to the limit $J_{\max}\rightarrow\infty$ using the
asymptotic form
\begin{equation}
f_{J_{\max}}(t)=f_{\infty}(t)+J_{\max}^{-2}f^{(2)}(t)+\cdots.
\end{equation}
For $b=0,0.5,1.0,3.0$ and $M=3$ we have calculated $f(t)$ using this
extrapolation technique for $J_{\max}=10$ and 12.\footnote{Both our results
and the analytic solution are even in $b$, and so we consider only positive
values.} The results are shown in Figure \ref{fig:thi2}.
 For comparison we have plotted the
analytic solution
\begin{equation}
f_{A}(t)=(k(c)M)^{\frac{-2c^{2}}{1-c^{2}}}t^{\frac{-1-c^{2}}{1-c^{2}}%
}.\label{analyt2}%
\end{equation}
The relation between $b$ and $c$ can be found in (\ref{bc}) and (\ref{cb}).
\ The parameter $k(c)$ is fitted to the value of the correlator at the
renormalization point $t=0.6.\footnote{The relative error is expected to be
small in the vicinity of this point.}$ The agreement appears quite good.
\ Some deviations from the analytic solution are due to $O(\frac{1}{M^{2}%
t^{2}})$ residual terms, which were left out of the derivation of
(\ref{analyt2}). \ These effects are significant in the small $t$ region,
$t\lesssim M^{-1}$, especially for larger values of $b$. \ The values we find
for $k(c)$ are shown in Table 2.$\footnote{In some regularization schemes
$k(c)$ can be shown to be independent of the coupling. \ Our regularization
method seems to be rather close to this, with only a slow variation with
respect to the coupling strength.}$%
\[
\overset{\text{Table 2}}{%
\begin{tabular}
[c]{|l|l|l|l|}\hline
$b$ & 0.5 & 1.0 & 3.0\\\hline
$k(c)$ & 1.77 & 1.77 & 1.68\\\hline
\end{tabular}
}%
\]
We can compare our results at small $b$ with a simple perturbative
calculation. Evaluating the corresponding regulated two-loop diagram we
obtain, for small $b$, $k(c)\approx1.75$.  This appears consistent with the
results in Table 2.

\section{Summary}

We derived the angle-smeared spherical Hamiltonian for the massless Thirring
model and constructed an explicit matrix representation. We discarded
negligible high energy states using auxiliary cutoff parameters $J_{\max}$ and
$K_{\max}$. \ In this reduced space we computed the time evolution of quantum
states and calculated the two-point correlator for several values of the
coupling. The results of our computation are in close agreement with the known
analytic solution. \ In addition to demonstrating new computational methods,
our analysis also serves as a consistency check of the regularization and
renormalization methods introduced in \cite{renorm}.

We believe that this work represents a significant new direction in the
non-perturbative computation of fermion dynamics. \ Future work will study the
application of these methods to systems of interacting bosons and fermions. \

\chapter[Modal expansions]{Modal expansions and non-perturbative quantum field theory in Minkowski space\footnote{N. Salwen, D. Lee, Phys.\ Rev. D62 (2000) 025006.}}

\section{Introduction}

Modal expansion methods have recently been used to study non-perturbative
phenomena in quantum field theory \cite{a1, fermion, a3, renorm}. Modal field
theory, the name for the\ general procedure, consists of two main parts. The
first is to approximate field theory as a finite-dimensional quantum
mechanical system. The second is to analyze the properties of the reduced
system using one of several computational techniques. \ The quantum mechanical
approximation is generated by decomposing field configurations into free wave
modes. \ This technique has been explored using both spherical partial waves
(spherical field theory \cite{a1, fermion, a3, renorm, spthirring}) 
and periodic box modes
(periodic field theory \cite{periodic}).

Having reduced field theory to a more tractable quantum mechanical system, we
have several different ways to proceed. \ Boson interactions in Euclidean
space, for example, can be modeled using the method of diffusion Monte Carlo.
\ In many situations, however, Monte Carlo techniques are inadequate. \ These
include unquenched fermion systems, processes in Minkowski space, and the
phenomenology of multi-particle states. \ Difficulties arise when the
functional integral measure cannot be treated as a probability distribution or
when information must be extracted from excited states obscured by dominant
lower lying states. \ Fortunately there are several alternative methods in the
modal field formalism which avoid these problems. \ Matrix Runge-Kutta
techniques were introduced in \cite{spthirring} as a method for calculating
unquenched fermion interactions. \ Here we discuss a different approach, one
which directly calculates the spectrum and eigenstates of the Hamiltonian.
\ For this approach it is essential that the Hamiltonian is time-independent,
and so we will consider periodic rather than spherical field theory. \ As we
demonstrate, these methods naturally accommodate the study of multi-particle
states and Minkowskian dynamics.

We apply the spectral approach to $1+1$ dimensional $\phi^{4}$ theory in a
periodic box and calculate the real and imaginary parts of the $\phi$
propagator. \ Some interesting properties of $\phi_{2}^{4}$ theory such as the
phase transition at large coupling were already discussed within the modal
field formalism using Euclidean Monte Carlo techniques \cite{periodic}. \ The
purpose of this analysis is of a more general and exploratory nature. \ Our
aim is to test the viability of modal diagonalization techniques for quantum
field Hamiltonians. \ We would like to know whether we can clearly see
multi-particle phenomena, the size of the errors and computational limitations
with current computer resources, and how such methods might be extended to
more complicated higher dimensional field theories.

The spectral method presented in the first part of our analysis is similar to
the work of Brooks and Frautschi \cite{brooks1, brooks2},\footnote{We 
thank the referee
of the original manuscript for providing information on this reference.} who
considered a $1+1$ dimensional Yukawa model in a periodic box and deserves
credit for the first application of diagonalization techniques using plane
wave modes in a periodic box. \ Our calculations are also similar in spirit to
diagonalization-based Hamiltonian lattice formulations 
\cite{hollenberg, hollenberg1} and
Tamm-Dancoff light-cone and discrete\ light-cone quantization
\cite{perry, hiller, bender, lightcone}. \ There are, however, some differences which we should
mention. \ As in \cite{brooks1, brooks2} we are using a momentum 
lattice rather than a
spatial lattice. \ We find this convenient to separate out invariant subspaces
according to total momentum quantum numbers. \ Since we are using an equal
time formulation our eigenvectors are not boost invariant as they would be on
the light cone. \ Also we are using a simple momentum cutoff scheme rather
than a regularization scheme which includes Tamm-Dancoff Fock-space
truncation. \ As a result our renormalization procedure is relatively
straightforward, but we will have to confront the problem of diagonalizing
large Fock spaces from the very beginning. \ In the latter part of the paper
we mention current work on quasi-sparse eigenvector methods which can handle
even exceptionally large Fock spaces. \ Despite the differences among the
various diagonalization approaches to field theory, the issues and problems
discussed in our analysis are of a general nature. \ We hope that the ideas
presented here will be of use for the various different approaches.

\section{Spectral method}

The field configuration $\phi$ in $1+1$ dimensions is subject to periodic
boundary conditions $\phi(t,x-L)=\phi(t,x+L).$ \ Expanding in terms of
periodic box modes, we have
\begin{equation}
\phi(t,x)=\sqrt{\tfrac{1}{2L}}%
{\displaystyle\sum\limits_{n=0,\pm1,...}}
\phi_{n}(t)e^{in\pi x/L}.
\end{equation}
The sum over momentum modes is regulated by choosing some large positive
number $N_{\max}$ and throwing out all high-momentum modes $\phi_{n}$ such
that $\left|  n\right|  >N_{\max}$. In this theory renormalization can be
implemented by normal ordering the $\phi^{4}$ interaction term. After a
straightforward calculation (details are given in \cite{periodic}), we find
that the counterterm Hamiltonian has the form
\begin{equation}
\tfrac{6\lambda b}{4!2L}%
{\displaystyle\sum\limits_{n=-N_{\max},N_{\max}}}
\phi_{-n}\phi_{n},
\end{equation}
where
\begin{equation}
b=%
{\displaystyle\sum\limits_{n=-N_{\max},N_{\max}}}
\tfrac{1}{2\omega_{n}},\qquad\omega_{n}=\sqrt{\tfrac{n^{2}\pi^{2}}{L^{2}}%
+\mu^{2}}.
\end{equation}
We represent the canonical conjugate pair $\phi_{n}$ and $\frac{d\phi_{-n}%
}{dt}$ using the Schr\"{o}dinger operators $q_{n}$ and $-i\tfrac{\partial
}{\partial q_{n}}$. \ Then the functional integral for $\phi^{4}$ theory is
equivalent to that for a quantum mechanical system with Hamiltonian
\begin{align}
H  &  =%
{\displaystyle\sum\limits_{n=-N_{\max},N_{\max}}}
\left[  -\tfrac{1}{2}\tfrac{\partial}{\partial q_{-n}}\tfrac{\partial
}{\partial q_{n}}+\tfrac{1}{2}(\omega_{n}^{2}-\tfrac{\lambda b}{4L}%
)q_{-n}q_{n}\right] \\
&  +\tfrac{\lambda}{4!2L}%
{\displaystyle\sum\limits_{\genfrac{}{}{0pt}{1}{n_{i}=-N_{\max},N_{\max
}}{n_{1}+n_{2}+n_{3}+n_{4}=0}}}
q_{n_{1}}q_{n_{2}}q_{n_{3}}q_{n_{4}}.\nonumber
\end{align}

We now consider the Hilbert space of our quantum mechanical system. \ Given
$d$, an array of non-negative integers,%
\begin{equation}
d=\left\{  d_{-N_{\max}},\cdots d_{N_{\max}}\right\}  ,
\end{equation}
we denote $p_{d}(q)$ as the following monomial with total degree $\left|
d\right|  $,%

\begin{equation}
p_{d}(q)=%
{\displaystyle\prod\limits_{n=-N_{\max},N_{\max}}}
q_{n}^{d_{n}},\qquad%
{\displaystyle\sum\limits_{n}}
d_{n}=\left|  d\right|  .
\end{equation}
We define $G_{\zeta}(q)$ to be a Gaussian of the form\footnote{$G_{\zeta}(q)$
has been defined such that $G_{\mu}(q)$ is the ground state of the free
theory.}%
\begin{equation}
G_{\zeta}(q)=%
{\displaystyle\prod\limits_{n=-N_{\max},N_{\max}}}
\exp\left[  -\tfrac{q_{-n}q_{n}\sqrt{\zeta^{2}+n^{2}\pi^{2}/L^{2}}}{2}\right]
.
\end{equation}
$\zeta$ is an adjustable parameter which we will set later. \ Any
square-integrable function $\psi(q)$ can be written as a superposition%

\begin{equation}
\psi(q)=%
{\displaystyle\sum\limits_{d}}
c_{d}\,p_{d}(q)G_{\zeta}(q). \label{sum}%
\end{equation}
In this analysis we consider only the zero-momentum subspace. \ We impose this
constraint by restricting the sum in (\ref{sum}) to monomials satisfying
\begin{equation}%
{\displaystyle\sum\limits_{n}}
nd_{n}=0.
\end{equation}
\ 

We will restrict the space of functions $\psi(q)$ further by removing high
energy states in the following manner. \ Let
\begin{equation}
k(d)=%
{\displaystyle\sum_{n}}
\left|  n\right|  d_{n}.
\end{equation}
$k(d)$ was first introduced in \cite{spthirring} and provides an estimate of the
kinetic energy associated with a given state. Let us define two auxiliary
cutoff parameters, $K_{\max}$ and $D_{\max}$. \ We restrict the sum in
(\ref{sum}) to monomials such that $k(d)<K_{\max}$ and $\left|  d\right|  \leq
D_{\max}$. \ We will refer to the corresponding subspace as $V_{K_{\max
},D_{\max}}$. \ The cutoff $K_{\max}$ removes states with high kinetic energy
and the cutoff $D_{\max}$ eliminates states with a large number of excited
modes.\footnote{In the case of spontaneous symmetry breaking, the broken
symmetry of the vacuum may require retaining a large number of $q_{0}$ modes.
\ This however is remedied by shifting the variable, $q_{0}^{\prime}%
=q_{0}-\left\langle 0\right|  q_{0}\left|  0\right\rangle $.} \ We should
stress that $K_{\max}$ and $D_{\max}$ are only auxiliary cutoffs. \ We
increase these parameters until the physical results appear close to the
asymptotic limit $K_{\max},D_{\max}\rightarrow\infty$. \ In our scheme
ultraviolet regularization is provided only by the momentum cutoff parameter
$N_{\max}$.

Our plan is to analyze the spectrum and eigenstates of $H$ restricted to this
approximate low energy subspace, $V_{K_{\max},D_{\max}}$. For any fixed $L$
and $N_{\max}$, $H$ is the Hamiltonian for a finite-dimensional quantum
mechanical system and the results should converge in the limit $K_{\max
},D_{\max}\rightarrow\infty$. \ We obtain the desired field theory result by
then taking the limit $L,\frac{N_{\max}}{L}\rightarrow\infty$.

\section{Results}

We have calculated the matrix elements of $H$ restricted to $V_{K_{\max
},D_{\max}}$ using a symbolic differentiation-integration
algorithm\footnote{All codes can be obtained by request from the authors.} and
diagonalized $H$, obtaining both eigenvalues and eigenstates. \ Let $\Delta$
be the full propagator,
\begin{equation}
\Delta(p^{2})=i\int d^{2}x\,e^{ip_{\nu}x^{\nu}}\left\langle 0\right|  T\left[
\phi(x^{\mu})\phi(0)\right]  \left|  0\right\rangle .
\end{equation}
We have computed $\Delta$ by inserting our complete set of eigenstates
(complete in $V_{K_{\max},D_{\max}}$)$.$ \ Let $\Delta_{\text{mp}}$ be the
multi-particle contribution to $\Delta$,%
\begin{equation}
\Delta_{\text{mp}}(p^{2})=\Delta(p^{2})-\Delta_{\text{pole}}(p^{2}),
\end{equation}
where $\Delta_{\text{pole}}$ is the single-particle pole contribution. \ We
are primarily interested in $\Delta_{\text{mp}}$, a quantity that cannot be
obtained for $p^{2}>0$ using Monte Carlo methods. \ Since the imaginary part
of $\Delta_{\text{pole}}$ is a delta function, it is easy to distinguish the
single-particle and multi-particle contributions in a plot of the imaginary
part of $\Delta$. \ The real part of $\Delta$, however, is dominated by the
one-particle pole. \ For this reason we have chosen to plot the real part of
$\Delta_{\text{mp}}$ rather than that of $\Delta$.

Although we have referred to multi-particle states, it should be noted that in
our finite periodic box there are no true continuum multi-particle states.
\ Instead we find densely packed discrete spectra with level separation of
size $\sim L^{-1}$ which become continuum states in the limit $L\rightarrow
\infty$. \ We can approximate the contribution of these $L\rightarrow\infty$
continuum states by a simple smoothing process. We included a small width
$\Gamma\sim L^{-1}$ to each of the would-be continuum states and averaged over
a range of values for $L$. \ For the results we report here we have averaged
over values $L=2.0\pi,2.1\pi,\cdots2.8\pi.$ \ For convenience all units have
been scaled such that $\mu=1$.

The parameter $\zeta$ was adjusted to reduce the errors due to the finite
cutoff values $K_{\max}$ and $D_{\max}$. \ Since the spectrum of $H$ is
bounded below, errors due to finite $K_{\max}$ and $D_{\max}$ generally drive
the estimated eigenvalues higher. \ One strategy, therefore, is to optimize
$\zeta$ by minimizing the trace of $H$ restricted to the subspace $V_{K_{\max
},D_{\max}}$. \ The approach used here is a slight variation of this --- we
have minimized the trace of $H$ restricted to a smaller subspace $V_{K_{\max
}^{\prime},D_{\max}^{\prime}}\subset V_{K_{\max},D_{\max}}$. \ The aim is to
accelerate the convergence of the lowest energy states rather than the entire
space $V_{K_{\max},D_{\max}}$. \ Throughout our analysis we used $K_{\max
}^{\prime},D_{\max}^{\prime}=8,3.$

\begin{figure}[htbp]
\begin{center}
\epsfbox{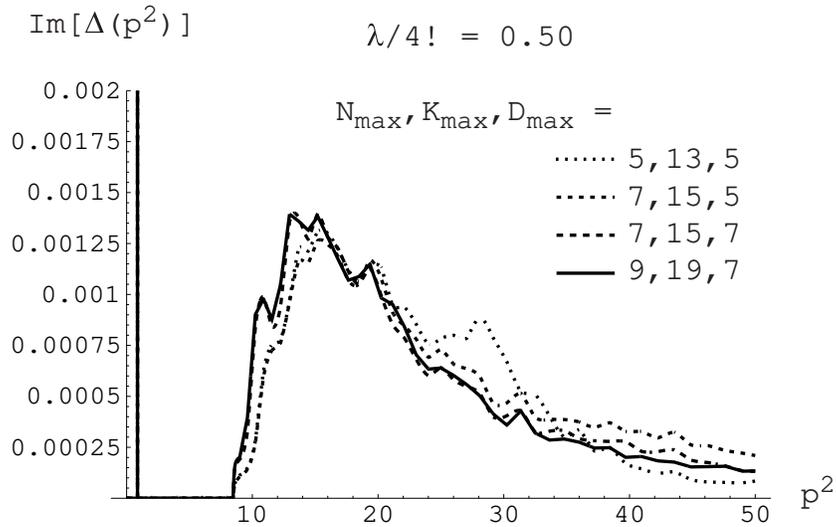}
\caption{Imaginary part of $\Delta(p^{2})$ for $\frac{\lambda}%
{4!}=0.50$ and several values for $N_{\max},K_{\max},D_{\max}$}
\label{fig:spe1}
\end{center}
\end{figure}

\begin{figure}[htbp]
\begin{center}
\epsfbox{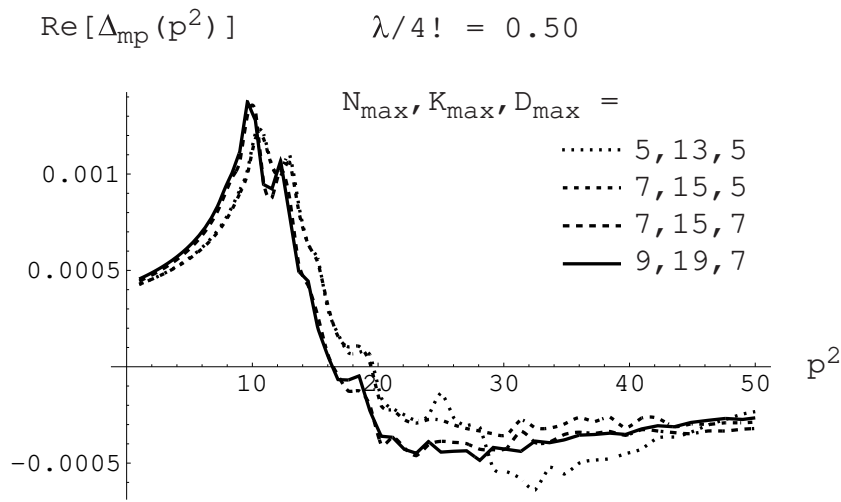}
\caption{Real part of $\Delta_{\text{mp}}(p^{2})$ for $\frac{\lambda
}{4!}=0.50$ and several values for $N_{\max},K_{\max},D_{\max}$}
\label{fig:spe2}
\end{center}
\end{figure}

For $\frac{\lambda}{4!}=0.50$ we have plotted the imaginary part of $\Delta$
in Figure \ref{fig:spe1}
 and the real part of $\Delta_{\text{mp}}$ in Figure \ref{fig:spe2}.
\ The
value $\frac{\lambda}{4!}=0.50$ is above the threshold for reliable
perturbative approximation\footnote{For momenta $\left|  p^{2}\right|
\gtrsim1$.} ($\frac{\lambda}{4!}\lesssim0.25$) but below the critical value at
which $\phi\rightarrow-\phi$ symmetry breaks spontaneously ($\frac{\lambda
}{4!}\approx2.5).$ \ The contribution of the one-particle state appears near
$p^{2}=(0.93)^{2}$ and the three-particle threshold is at approximately
$p^{2}=(2.9)^{2}$. \ We have chosen several different values for $N_{\max
},K_{\max},D_{\max}$ to show the convergence as these parameters become large.
\ The plot for $N_{\max},K_{\max},D_{\max}=9,19,7$ appears relatively close to
the asymptotic limit. \ The somewhat bumpy texture of the curves is due to the
finite size of our periodic box and diminishes with increasing $L$. \ From
dimensional power counting, we expect errors for finite $N_{\max}$ to scale as
$N_{\max}^{-2}$. \ Assuming that $K_{\max}$ and $D_{\max}$ also function as
uniform energy cutoffs, we expect a similar error dependence -- and it appears
plausible from the results in Figures \ref{fig:spe1} and
\ref{fig:spe2}. 
\ A more systematic analysis of
the errors and extrapolation methods for finite $N_{\max},K_{\max},D_{\max}$,
and $L,$ will be discussed in future work.

\begin{figure}[htbp]
\begin{center}
\epsfbox{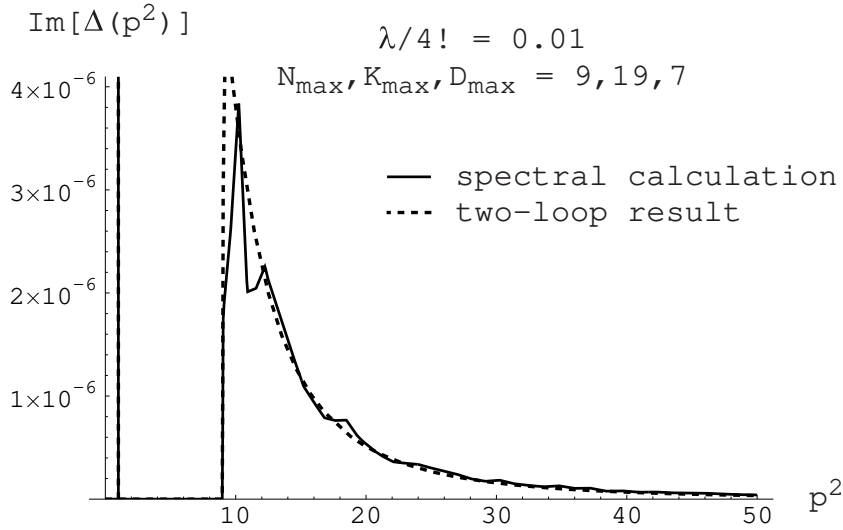}
\caption{Imaginary part of $\Delta(p^{2})$ for $\frac{\lambda}%
{4!}=0.01$ and comparison with the two-loop result}
\label{fig:spe3}
\end{center}
\end{figure}

\begin{figure}[htbp]
\begin{center}
\epsfbox{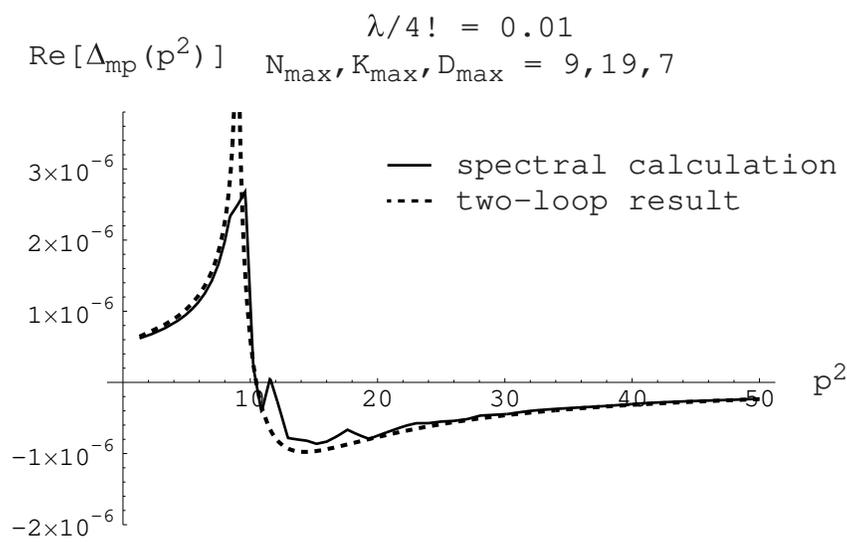}
\caption{Real part of $\Delta_{\text{mp}}(p^{2})$ for $\frac{\lambda
}{4!}=0.01$ and comparison with the two-loop result}
\label{fig:spe4}
\end{center}
\end{figure}

In Figures \ref{fig:spe3} and \ref{fig:spe4}
 we have compared our spectral calculations with the
two-loop perturbative result for $\frac{\lambda}{4!}=0.01$. \ We have used
$N_{\max},K_{\max},D_{\max}=9,19,7$, and the agreement appears good. \ For
small $\frac{\lambda}{4!}$ the propagator has a very prominent logarithmic
cusp at the three-particle threshold, which can be seen clearly in
Figures \ref{fig:spe3} 
and \ref{fig:spe4}.

\begin{figure}[htbp]
\begin{center}
\epsfbox{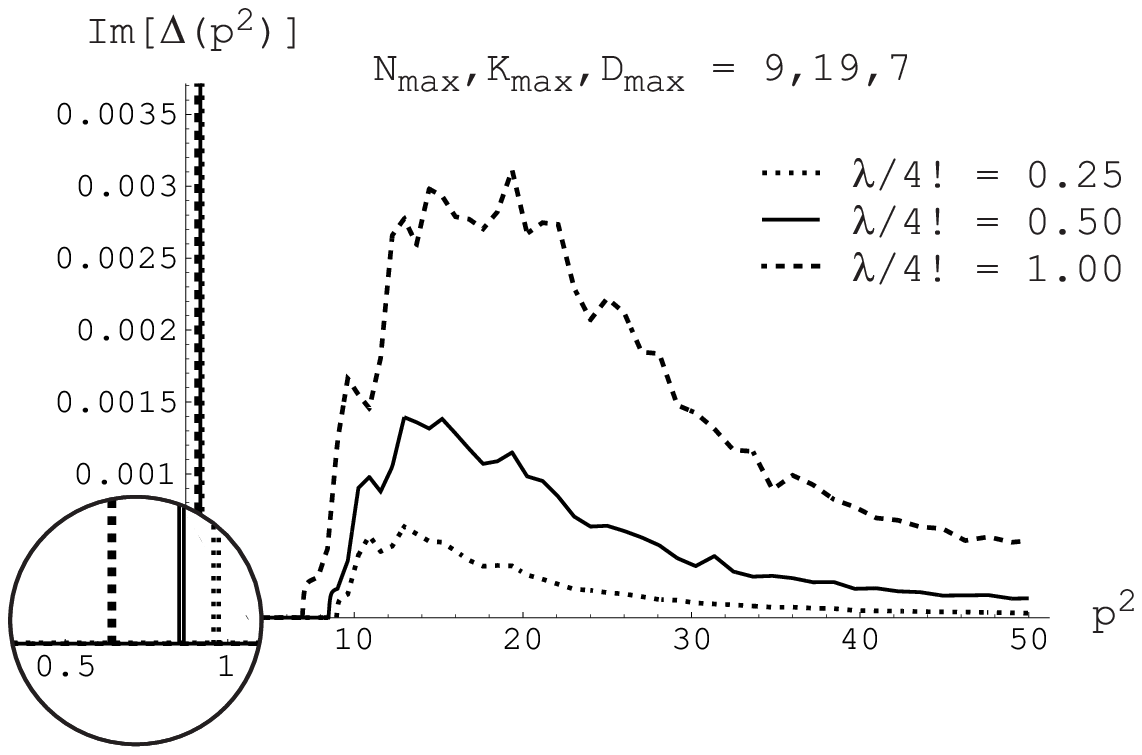}
\caption{Imaginary part of $\Delta(p^{2})$ for $\frac{\lambda}%
{4!}=0.25,0.50,1.00$}
\label{fig:spe5}
\end{center}
\end{figure}

\begin{figure}[htbp]
\begin{center}
\epsfbox{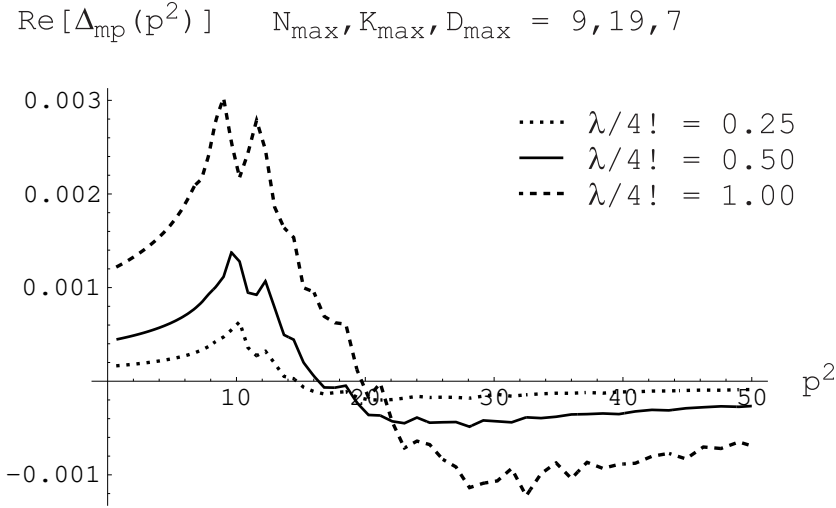}
\caption{Real part of $\Delta_{\text{mp}}(p^{2})$ for $\frac{\lambda
}{4!}=0.25,0.50,1.00$}
\label{fig:spe6}
\end{center}
\end{figure}

In Figures \ref{fig:spe5} and \ref{fig:spe6}
 we have compared results for $\frac{\lambda}{4!}=0.25$,
$0.50$, $1.00$. \ We have again used $N_{\max},K_{\max},D_{\max}=9,19,7$. \ In
contrast with the quadratic scaling in the perturbative regime, the results
here scale approximately linearly with $\frac{\lambda}{4!}$. \ An interesting
and perhaps related observation is that the magnitude of the multi-particle
contribution to $\Delta$ remains small $(\lesssim0.003$) even for the rather
large coupling value $\frac{\lambda}{4!}=1.00$.

\section{Limitations and new ideas}

We now address the computational limits of the diagonalization techniques
presented in this work. \ These techniques are rather straightforward and can
in principle be generalized to any field theory. \ In practise however the
Fock space $V_{K_{\max},D_{\max}}$ becomes prohibitively large, especially for
higher dimensional theories. \ The data in Figures \ref{fig:spe1} and
\ref{fig:spe2}  and crosschecks
with Euclidean Monte Carlo results\footnote{See \cite{periodic} for a
discussion of these methods.} suggest that for $N_{\max}=9\ $and
$L=2.0\pi,\cdots2.8\pi$ our spectral results with $K_{\max},D_{\max}=19,7$ and
$\frac{\lambda}{4!}<1$ are within 20\% of the $K_{\max},D_{\max}%
\rightarrow\infty$ limit. $\ $In this case $V_{K_{\max},D_{\max}}$ is a $2036$
dimensional space and requires about 100 MB of RAM using general (dense)
matrix methods.

Sparse matrix techniques such as the Lanczos or Arnoldi schemes allow us to
push the dimension of the Fock space to about 10$^{5}$ states. \ This may be
sufficient to do accurate calculations near the critical point $\frac{\lambda
}{4!}\approx2.5$ for larger values of $L$ and $N_{\max}.$ \ It is, however,
near the upper limit of what is possible using current computer technology and
existing algorithms. \ Unfortunately field theories in $2+1$ and $3+1$
dimensions will require much larger Fock spaces, probably at least 10$^{12}$
and 10$^{18}$ states respectively. \ In order to tackle these larger Fock
spaces it is necessary to venture beyond standard diagonalization approaches.
\ The problem of large Fock spaces ($\gg$10$^{6}$ states) is beyond the
intended scope of this analysis. \ But since it is of central importance to
the diagonalization approach to field theory we would like to briefly comment
on current work being done which may resolve many of the difficulties. \ The
new approach takes advantage of the sparsity of the Fock-space Hamiltonian and
the approximate (quasi-)sparsity of the eigenvectors. \ A detailed description
will be provided in a future publication \cite{qse}.

We start with some observations about the eigenvectors of the $\phi_{1+1}^{4}$
Hamiltonian for $N_{\max}=9,$ $L=2.5\pi$ and $K_{\max},D_{\max}=19,7.$ \ To
make certain that we are probing non-perturbative physics we will set
$\frac{\lambda}{4!}=2.5,$ the approximate critical point value. \ We label the
normalized eigenvectors as $\left|  v_{0}\right\rangle $, $\left|
v_{1}\right\rangle $,$\cdots,$ ascending in order with respect to energy. \ We
also define $\left|  b_{0}\right\rangle $, $\left|  b_{1}\right\rangle
$,$\cdots$ as the normalized eigenvectors of the free, non-interacting theory.
\ For any $v_{i}$ we know
\begin{equation}
\sum_{j}\left|  \left\langle b_{j}|v_{i}\right\rangle \right|  ^{2}=1.
\end{equation}
Let us define $\left\|  \left|  v_{i}\right\rangle \right\|  _{(n)}$ as the
partial sum%
\begin{equation}
\left\|  \left|  v_{i}\right\rangle \right\|  _{(n)}=\sum_{k=1,\cdots
n}\left|  \left\langle b_{j_{k}}|v_{i}\right\rangle \right|  ^{2},
\end{equation}
where the inner products have been sorted from largest to smallest
\begin{equation}
\left|  \left\langle b_{j_{1}}|v_{i}\right\rangle \right|  \geq\left|
\left\langle b_{j_{2}}|v_{i}\right\rangle \right|  \geq\cdots.
\end{equation}
Table 1 shows $\left\|  \left|  v_{i}\right\rangle \right\|  _{(n)}$ for
several eigenvectors and different values of $n$.
\[
\overset{\text{Table 1}}{%
\begin{array}
[c]{c}%
\\%
\begin{tabular}
[c]{l|l|l|l|l|}%
$\left\|  \left|  v_{i}\right\rangle \right\|  _{(n)}$ & $n=10$ & $n=20$ &
$n=40$ & $n=80$\\\hline
$\left|  v_{0}\right\rangle $ & 0.75 & 0.84 & 0.90 & 0.94\\\hline
$\left|  v_{1}\right\rangle $ & 0.89 & 0.92 & 0.95 & 0.97\\\hline
$\left|  v_{5}\right\rangle $ & 0.87 & 0.91 & 0.94 & 0.96\\\hline
$\left|  v_{10}\right\rangle $ & 0.77 & 0.86 & 0.90 & 0.94\\\hline
\end{tabular}
\end{array}
}%
\]
\ Despite the non-perturbative coupling and complex phenomena associated with
the phase transition, we see from the table that each of the eigenvectors can
be approximated by just a small number of its largest Fock-space components.
\ We recall that the Fock space for this system has 2036 dimensions. \ The
eigenvectors are therefore quasi-sparse in this space, a consequence of the
sparsity of the Hamiltonian. \ If we write the Hamiltonian as a matrix in the
free Fock-space basis, a typical row or column contains only about 200
non-zero entries, a number we refer to as $N_{\text{transition}}.$ \ In
\cite{qse} we show that a typical eigenvector will be dominated by the largest
$\sqrt{N_{\text{transition}}}$ elements.\footnote{There are some special
exceptions to this rule and they are discussed in \cite{qse}. \ But these are
typically not relevant for the lower energy eigenstates of a quantum field
Hamiltonian.} \ The key point is that $\sqrt{N_{\text{transition}}}$ is quite
manageable --- on the order of $10^{3}$ and 10$^{5}$ for $2+1$ and $3+1$
dimensional field theories respectively. \ Although the size of the Fock space
for these systems are enormous, the extreme sparsity of the Hamiltonian
suggests that the eigenvectors can be approximated using current computational resources.

With this simplification, the task is to find the important basis states for a
given eigenvector. \ Since the important basis states for one eigenvector are
generally different from that of another, each eigenvector is determined
independently. \ This provides a starting point for parallelization, and many
eigenvectors can be determined at the same time using massively parallel
computers. \ In \cite{qse} we present a simple stochastic algorithm where the
exact eigenvectors act as stable fixed points of the update process.

\section{Summary}

We have introduced a spectral approach to periodic field theory and used it to
calculate the propagator in $1+1$ dimensional $\phi^{4}$ theory. \ We find
that the straightforward application of these methods with existing computer
technology can be useful for describing the multi-particle properties of the
theory, information difficult to obtain using Euclidean Monte Carlo methods.
\ However the extension to higher dimensional theories is made difficult by
the large size of the corresponding Fock space. \ As a possible solution to
this problem, we note that each eigenvector of the $\phi_{1+1}^{4}$
Hamiltonian can be well-approximated using relatively few components and
discuss some current work on quasi-sparse eigenvector
methods.{\normalsize \bigskip}

\chapter[Quasi sparse methods]{Quasi sparse methods in the
diagonalization of quantum field Hamiltonians\footnote{Dean Lee,
  N. Salwen, Dan Lee, Phys. Lett. B 503(2001) 223-235}}

\section{Introduction}

Most computational work in non-perturbative quantum field theory and many body
phenomena rely on one of two general techniques, Monte Carlo or
diagonalization. \ These methods are nearly opposite in their strengths and
weaknesses. \ Monte Carlo requires relatively little storage, can be performed
using parallel processors, and in some cases the computational effort scales
reasonably with system size. \ But it has great difficulty for systems with
sign or phase oscillations and provides only indirect information on
wavefunctions and excited states. \ In contrast diagonalization methods do not
suffer from fermion sign problems, can handle complex-valued actions, and can
extract details of the spectrum and eigenstate wavefunctions. \ However the
main problem with diagonalization is that the required memory and CPU time
scales exponentially with the size of the system.

In view of the complementary nature of the two methods, we consider the
combination of both diagonalization and Monte Carlo within a computational
scheme. \ We propose a new approach which takes advantage of the strengths of
the two computational methods in their respective domains. \ The first half of
the method involves finding and diagonalizing the Hamiltonian restricted to an
optimal subspace. \ This subspace is designed to include the most important
basis vectors of the lowest energy eigenstates. \ Once the most important
basis vectors are found and their interactions treated exactly, Monte Carlo is
used to sample the contribution of the remaining basis vectors. \ By this
two-step procedure much of the sign problem is negated by treating the
interactions of the most important basis states exactly, while storage and CPU
problems are resolved by stochastically sampling the collective effect of the
remaining states.

In our approach diagonalization is used as the starting point of the Monte
Carlo calculation. \ Therefore the two methods should not only be efficient
but work well together. \ On the diagonalization side there are several
existing methods using Tamm-Dancoff truncation \cite{perry}, similarity
transformations \cite{glazek1, glazek2}, density matrix renormalization group
\cite{white1, white2}, or variational algorithms such as 
stochastic diagonalization
\cite{husslein, deraedt}. 
\ However we find that each of these methods is either not
sufficiently general, not able to search an infinite or large dimensional
Hilbert space, not efficient at finding important basis vectors, or not
compatible with the subsequent Monte Carlo part of the calculation. \ The
Monte Carlo part of our diagonalization/Monte Carlo scheme is discussed
separately in a companion paper \cite{sec}. \ In this paper we consider the
diagonalization part of the scheme. \ We introduce a new diagonalization
method called quasi-sparse eigenvector (QSE) diagonalization. \ It is a
general algorithm which can operate using any basis, either orthogonal or
non-orthogonal, and any sparse Hamiltonian, either real, complex, Hermitian,
non-Hermitian, finite-dimensional, or infinite-dimensional. \ It is able to
find the most important basis states of several low energy eigenvectors
simultaneously, including those with identical quantum numbers, from a random
start with no prior knowledge about the form of the eigenvectors.

Our discussion is organized as follows. \ We first define the notion of
quasi-sparsity in eigenvectors and introduce the quasi-sparse eigenvector
method. \ We discuss when the low energy eigenvectors are likely to be
quasi-sparse and make an analogy with Anderson localization. \ We then
consider three examples which test the performance of the algorithm. \ In the
first example we find the lowest energy eigenstates for a random sparse real
symmetric matrix. \ In the second example we find the lowest eigenstates
sorted according to the real part of the eigenvalue for a random sparse
complex non-Hermitian matrix. \ In the last example we consider the case of an
infinite-dimensional Hamiltonian defined by $1+1$ dimensional $\phi^{4}$
theory in a periodic box. \ We conclude with a summary and some comments on
the role of quasi-sparse eigenvector diagonalization within the context of the
new diagonalization/Monte Carlo approach.

\section{Quasi-sparse eigenvector method}

Let $\left|  e_{i}\right\rangle $ denote a complete set of basis vectors.
\ For a given energy eigenstate
\begin{equation}
|v\rangle=\sum_{i}c_{i}\left|  e_{i}\right\rangle ,
\end{equation}
we define the important basis states of $|v\rangle$ to be those $\left|
e_{i}\right\rangle $ such that for fixed normalizations of $|v\rangle$ and the
basis states, $\left|  c_{i}\right|  $ exceeds a prescribed threshold value.
\ If $|v\rangle$ can be well-approximated by the contribution from only its
important basis states we refer to the eigenvector $|v\rangle$ as
\textit{quasi-sparse} with respect to $\left|  e_{i}\right\rangle $.

Standard sparse matrix algorithms such as the Lanczos or Arnoldi methods allow
one to find the extreme eigenvalues and eigenvectors of a sparse matrix
efficiently, without having to store or manipulate large non-sparse matrices.
\ However in quantum field theory or many body theory one considers very large
or infinite dimensional spaces where even storing the components of a general
vector is impossible. \ For these more difficult problems the strategy is to
approximate the low energy eigenvectors of the large space by diagonalizing
smaller subspaces. \ If one has sufficient intuition about the low energy
eigenstates it may be possible to find a useful truncation of the full vector
space to an appropriate smaller subspace. \ In most cases, however, not enough
is known \textit{a priori }about the low energy eigenvectors. \ The dilemma is
that to find the low energy eigenstates one must truncate the vector space,
but in order to truncate the space something must be known about the low
energy states.

Our solution to this puzzle is to find the low energy eigenstates and the
appropriate subspace truncation at the same time by a recursive process. \ We
call the method quasi-sparse eigenvector (QSE) diagonalization, and we
describe the steps of the algorithm as follows. \ The starting point is any
complete basis for which the Hamiltonian matrix $H_{ij}$ is sparse. \ The
basis vectors may be non-orthogonal and/or the Hamiltonian matrix may be
non-Hermitian. \ The following steps are now iterated:

\begin{enumerate}
\item  Select a subset of basis vectors $\left\{  e_{i_{1}},\cdots,e_{i_{n}%
}\right\}  $ and call the corresponding subspace $S$.

\item  Diagonalize $H$ restricted to $S$ and find one eigenvector $v$.

\item  Sort the basis components of $v$ according to their magnitude and
remove the least important basis vectors.

\item  Replace the discarded basis vectors by new basis vectors. \ These are
selected at random according to some weighting function from a pool of
candidate basis vectors which are connected to the old basis vectors through
non-vanishing matrix elements of $H$.

\item  Redefine $S$ as the subspace spanned by the updated set of basis
vectors and repeat steps 2 through 5.
\end{enumerate}

If the subset of basis vectors is sufficiently large, the exact low energy
eigenvectors will be stable fixed points of the QSE update process. \ We can
show this as follows. \ Let $\left|  i\right\rangle $ be the eigenvectors of
the submatrix of $H$ restricted to the subspace $S$, where $S$ is the span of
the subset of basis vectors after step 3 of the QSE algorithm. \ Let $\left|
A_{j}\right\rangle $ be the remaining basis vectors in the full space not
contained in $S$. \ We can represent $H$ as
\begin{equation}
\left[
\begin{array}
[c]{cccccc}%
\lambda_{1} & 0 & \cdots & \left\langle 1\right|  H\left|  A_{1}\right\rangle
& \left\langle 1\right|  H\left|  A_{2}\right\rangle  & \cdots\\
0 & \lambda_{2} & \cdots & \left\langle 2\right|  H\left|  A_{1}\right\rangle
& \left\langle 2\right|  H\left|  A_{2}\right\rangle  & \cdots\\
\vdots & \vdots & \ddots & \vdots & \vdots & \cdots\\
\left\langle A_{1}\right|  H\left|  1\right\rangle  & \left\langle
A_{1}\right|  H\left|  2\right\rangle  & \cdots &  E\cdot\lambda_{A_{1}} &
\left\langle A_{1}\right|  H\left|  A_{2}\right\rangle  & \cdots\\
\left\langle A_{2}\right|  H\left|  1\right\rangle  & \left\langle
A_{2}\right|  H\left|  2\right\rangle  & \cdots & \left\langle A_{2}\right|
H\left|  A_{1}\right\rangle  & E\cdot\lambda_{A_{2}} & \cdots\\
\vdots & \vdots & \vdots & \vdots & \vdots & \ddots
\end{array}
\right]  . \label{matrix}%
\end{equation}
We have used Dirac's bra-ket notation to represent the terms of the
Hamiltonian matrix. \ In cases where the basis is non-orthogonal and/or the
Hamiltonian is non-Hermitian, the meaning of this notation may not be clear.
\ When writing $\left\langle A_{1}\right|  H\left|  1\right\rangle $, for
example, we mean the result of the dual vector to $\left|  A_{1}\right\rangle
$ acting upon the vector $H\left|  1\right\rangle $. \ In (\ref{matrix}) we
have written the diagonal terms for the basis vectors $\left|  A_{j}%
\right\rangle $ with an explicit factor $E$. \ We let $\left|  1\right\rangle
$ be the approximate eigenvector of interest and have shifted the diagonal
entries so that $\lambda_{1}=0.$ \ Our starting hypothesis is that $\left|
1\right\rangle $ is close to some exact eigenvector of $H$ which we denote as
$\left|  1_{\text{full}}\right\rangle $. \ More precisely we assume that the
components of $\left|  1_{\text{full}}\right\rangle $ outside $S$ are small
enough so that we can expand in inverse powers of the introduced parameter $E.$

We now expand the eigenvector as
\begin{equation}
\left|  1_{\text{full}}\right\rangle =\left[
\begin{array}
[c]{c}%
1\\
c_{2}^{\prime}E^{-1}+\cdots\\
\vdots\\
c_{A_{1}}^{\prime}E^{-1}+\cdots\\
c_{A_{2}}^{\prime}E^{-1}+\cdots\\
\vdots
\end{array}
\right]  \label{eigvec}%
\end{equation}
and the corresponding eigenvalue as
\begin{equation}
\lambda_{\text{full}}=\lambda_{1}^{\prime}E^{-1}+\cdots.
\end{equation}
In (\ref{eigvec}) we have chosen the normalization of $\left|  1_{\text{full}%
}\right\rangle $ such that $\left\langle 1\right.  \left|  1_{\text{full}%
}\right\rangle =1$. \ From the eigenvalue equation
\begin{equation}
H\left|  1_{\text{full}}\right\rangle =\lambda_{\text{full}}\left|
1_{\text{full}}\right\rangle
\end{equation}
we find at lowest order
\begin{equation}
c_{A_{j}}^{\prime}=-\tfrac{\left\langle A_{j}\right|  H\left|  1\right\rangle
}{\lambda_{A_{j}}}.
\end{equation}
We see that at lowest order the component of $\left|  1_{\text{full}%
}\right\rangle $ in the $\left|  A_{j}\right\rangle $ direction is independent
of the other vectors $\left|  A_{j^{\prime}}\right\rangle $. \ If $\left|
1\right\rangle $ is sufficiently close to $\left|  1_{\text{full}%
}\right\rangle $ then the limitation that only a fixed number of new basis
vectors is added in step 4 of the QSE algorithm is not relevant. \ At lowest
order in $E^{-1}$ the comparison of basis components in step 3 (in the next
iteration) is the same as if we had included all remaining vectors $\left|
A_{j}\right\rangle $ at once. \ Therefore at each update only the truly
largest components are kept and the algorithm converges to some optimal
approximation of $\left|  1_{\text{full}}\right\rangle $. \ This is consistent
with the actual performance of the algorithm as we will see in some examples
later. \ In those examples we also demonstrate that the QSE algorithm is able
to find several low energy eigenvectors simultaneously. \ The only change is
that when diagonalizing the subspace $S$ we find more than one eigenvector and
apply steps 3 and 4 of the algorithm to each of the eigenvectors.

\section{Quasi-sparsity and Anderson localization}

As the name indicates the accuracy of the quasi-sparse eigenvector method
depends on the quasi-sparsity of the low energy eigenstates in the chosen
basis. \ If the eigenvectors are quasi-sparse then the QSE method provides an
efficient way to find the important basis vectors. \ In the context of our
diagonalization/Monte Carlo approach, this means that diagonalization does
most of the work and only a small amount of correction is needed. \ This
correction is found by Monte Carlo sampling the remaining basis vectors, a
technique called stochastic error correction \cite{sec}. \ If however the
eigenvectors are not quasi-sparse then one must rely more heavily on the Monte
Carlo portion of the calculation.

The fastest and most reliable way we know to determine whether the low energy
eigenstates of a Hamiltonian are quasi-sparse with respect to a chosen basis
is to use the QSE algorithm and look at the results of the successive
iterations. \ But it is also useful to consider the question more intuitively,
and so we consider the following example.

\begin{figure}[htbp]
\begin{center}
\epsfxsize=20pc \epsfbox{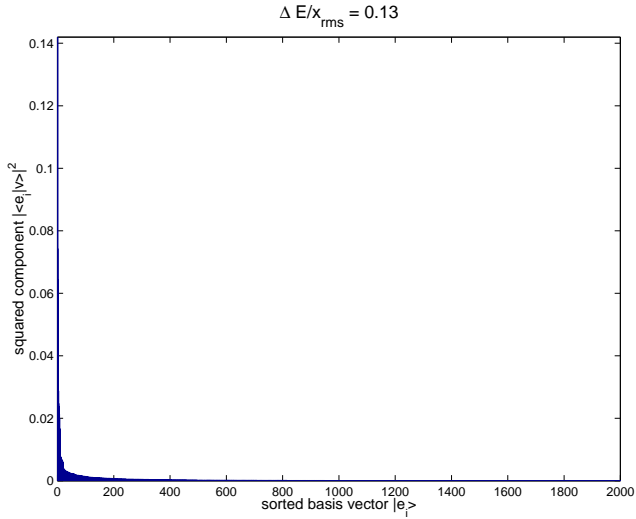}
%
%
%
%
%
%
%
\caption{Distribution of basis components for an eigenvector where the spacing
between consecutive levels is $\Delta E=0.13x_{\text{rms}}$. }%
\label{fig:1}
\end{center}
\end{figure}

\begin{figure}[htbp]
\begin{center}
\epsfxsize=20pc \epsfbox{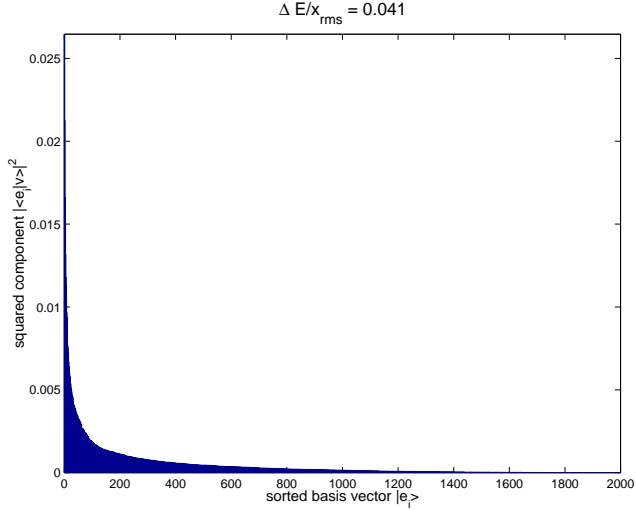}
\caption{Distribution of basis components for an eigenvector where $\Delta
E=0.041x_{\text{rms}}$. }%
\label{fig:2}
\end{center}
\end{figure}

\begin{figure}[htbp]
\begin{center}
\epsfxsize=20pc \epsfbox{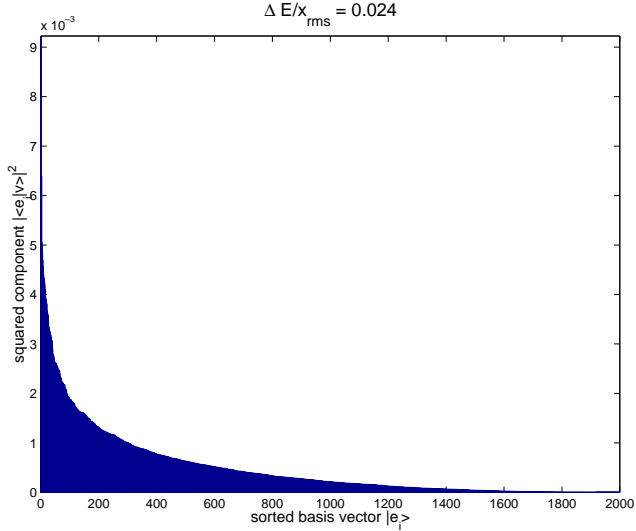}
\caption{Distribution of basis components for an eigenvector where $\Delta
E=0.024x_{\text{rms}}$. }%
\label{fig:3}
\end{center}
\end{figure}

Let $H$ be a sparse Hermitian $2000\times2000$ matrix defined by
\begin{equation}
H_{jk}=\log(j)\cdot\delta_{jk}+x_{jk}\cdot M_{jk}, \label{sym}%
\end{equation}
where $j$ and $k$ run from $1$ to $2000$, $x_{jk}$ is a Gaussian random real
variable centered at zero with standard deviation $x_{\text{rms}}=0.25$,
and\ $M_{jk}$ is a sparse symmetric matrix consisting of random $0$'s and
$1$'s such that the density of $1$'s is $5\%$. \ The reason for introducing
the $\log(j)$ term in the diagonal is to produce a large variation in the
density of states. With this choice the density of states increases
exponentially with energy. \ Our test matrix is small enough that all
eigenvectors can be found without difficulty. \ We will consider the
distribution of basis components for the eigenvectors of $H$. \ In
Figure \ref{fig:1}
 we
show the square of the basis components for a given low energy eigenvector
$\left|  v\right\rangle .$ \ The basis components are sorted in order of
descending importance. \ The ratio of $\Delta E$, the average spacing between
neighboring energy levels, to $x_{\text{rms}}$ is $0.13$. \ We see that the
eigenvector is dominated by a few of its most important basis components. \ In
Figure \ref{fig:2} 
we show the same plot for another eigenstate but one where the
spacing between levels is three times smaller, $\Delta E/x_{\text{rms}%
}=0.041.$ \ This eigenvector is not nearly as quasi-sparse. \ The effect is
even stronger in Figure \ref{fig:3},
 where we show an eigenvector such that the spacing
between levels is $\Delta E/x_{\text{rms}}=0.024$.

Our observations show a strong effect of the density of states on the
quasi-sparsity of the eigenvectors. \ States with a smaller spacing between
neighboring levels tend to have basis components that extend throughout the
entire space, while states with a larger spacing tend to be quasi-sparse.
\ The relationship between extended versus localized eigenstates and the
density of states has been studied in the context of Anderson localization and
metal-insulator transitions \cite{biswas, souk, ziman, inui}.\ 
The simplest example is the
tight-binding model for a single electron on a one-dimensional lattice with
$Z$ sites,%

\begin{equation}
H=\sum_{j}d_{j}\left|  j\right\rangle \left\langle j\right|  +\sum
_{\left\langle jj^{\prime}\right\rangle }t_{jj^{\prime}}\left|  j\right\rangle
\left\langle j^{\prime}\right|  .
\end{equation}
$\left|  j\right\rangle $ denotes the atomic orbital state at site $j,$
$d_{j}$ is the on-site potential, and $t_{jj^{\prime}}$ is the hopping term
between nearest neighbor sites $j$ and $j^{\prime}$. \ If both terms are
uniform ($d_{j}=d,$ $t_{jj^{\prime}}=t$) then the eigenvalues and eigenvectors
of $H$ are
\begin{align}
Hv_{n}  &  =(d+2t\cos\tfrac{2\pi n}{Z})v_{n},\\
v_{n}  &  =\tfrac{1}{\sqrt{Z}}\sum_{j}e^{i\tfrac{2\pi nj}{Z}}\left|
j\right\rangle ,
\end{align}
where $n=1,\cdots,Z$ labels the eigenvectors. \ In the absence of diagonal and
off-diagonal disorder, the eigenstates of $H$ extend throughout the entire
lattice. \ The eigenvalues are also approximately degenerate, all lying within
an interval of size 4$t$. \ However, if diagonal and/or off-diagonal disorder
is introduced, the eigenvalue spectrum becomes less degenerate. \ If the
disorder is sufficiently large, the eigenstates become localized to only a few
neighboring lattice sites giving rise to a transition of the material from
metal to insulator.

We can regard a sparse quantum Hamiltonian as a similar type of system, one
with both diagonal and general off-diagonal disorder. \ If the disorder is
sufficient such that the eigenvalues become non-degenerate, then the
eigenvectors will be quasi-sparse. \ We reiterate that the most reliable way
to determine if the low energy states are quasi-sparse is to use the QSE
algorithm. \ Intuitively, though, we expect the eigenstates to be quasi-sparse
with respect to a chosen basis if the spacing between energy levels is not too
small compared with the size of the off-diagonal entries of the Hamiltonian matrix.

\section{Finite matrix examples}

\begin{figure}[htbp]
\begin{center}
\epsfxsize=20pc \epsfbox{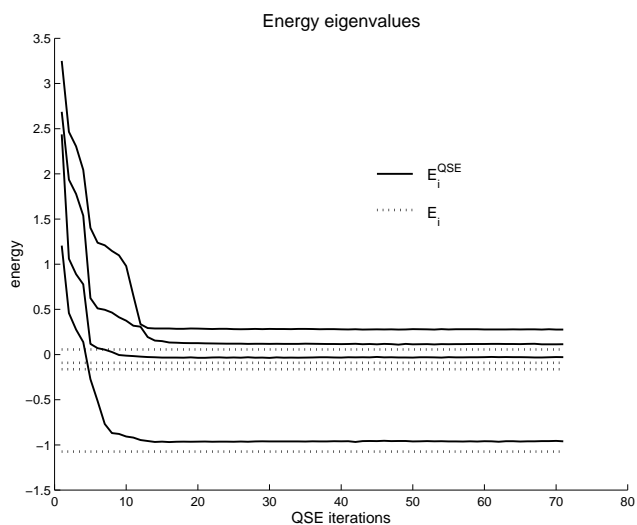}
\caption{Comparison of the four lowest exact energies $E_{i}$ and QSE results
$E_{i}^{\text{QSE}}$ as functions of iteration number. }%
\label{fig:4}
\end{center}
\end{figure}

\begin{figure}[htbp]
\begin{center}
\epsfxsize=20pc \epsfbox{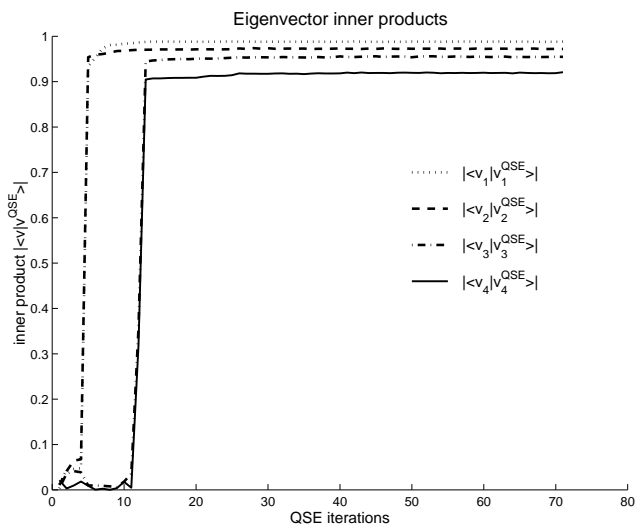}
\caption{Inner products between the normalized exact eigenvectors $\left|
v_{i}\right\rangle $ and the QSE results $\left|  v_{i}^{\text{QSE}%
}\right\rangle $ as functions of iteration number. }%
\label{fig:5}
\end{center}
\end{figure}

As a first test of the QSE method, we will find the lowest four energy states
of the random symmetric matrix $H$ defined in (\ref{sym}). \ So that there is
no misunderstanding, we should repeat that diagonalizing a $2000\times2000$
matrix is not difficult. \ The purpose of this test is to analyze the
performance of the method in a controlled environment. \ One interesting twist
is that the algorithm uses only small pieces of the matrix and operates under
the assumption that the space may be infinite dimensional. \ A sample MATLAB
program similar to the one used here has been printed out as a tutorial
example in \cite{tutorial}.

\ The program starts from a random configuration, 70 basis states for each of
the four eigenvectors. \ With each iteration we select $10$ replacement basis
states for each of the eigenvectors. \ In Figure \ref{fig:4}
 we show the exact energies
and the results of the QSE\ method as functions of iteration number. \ In
Figure \ref{fig:5}
 we show the inner products of the normalized QSE eigenvectors with
the normalized exact eigenvectors. \ We note that all of the eigenvectors were
found after about 15 iterations and remained stable throughout successive
iterations. \ Errors are at the $5$ to $10\%$ level, which is about the
theoretical limit one can achieve using this number of basis states. \ The QSE
method has little difficulty finding several low lying eigenvectors
simultaneously because it uses the distribution of basis components for each
of the eigenvectors to determine the update process. \ This provides a
performance advantage over variational-based techniques such as stochastic
diagonalization in finding eigenstates other than the ground state. \ 

\begin{figure}[htbp]
\begin{center}
\epsfxsize=20pc \epsfbox{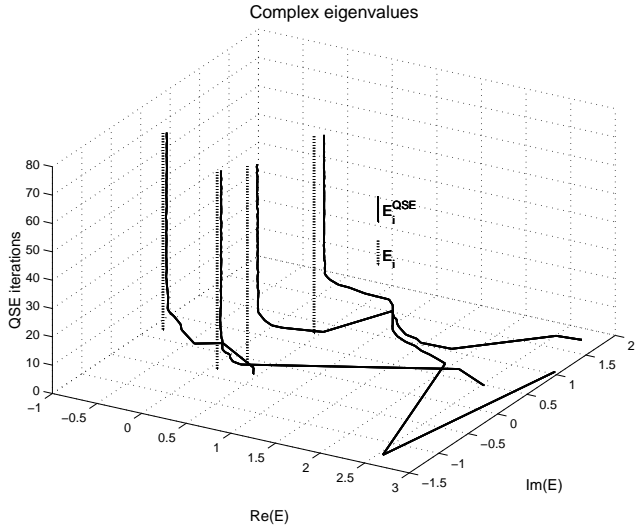}
\caption{Comparison of the four lowest exact eigenvalues $E_{i}$ (sorted by
real part) and QSE results $E_{i}^{\text{QSE}}$ in the complex plane as
functions of iteration number. }%
\label{fig:6}
\end{center}
\end{figure}

\begin{figure}[htbp]
\begin{center}
\epsfxsize=20pc \epsfbox{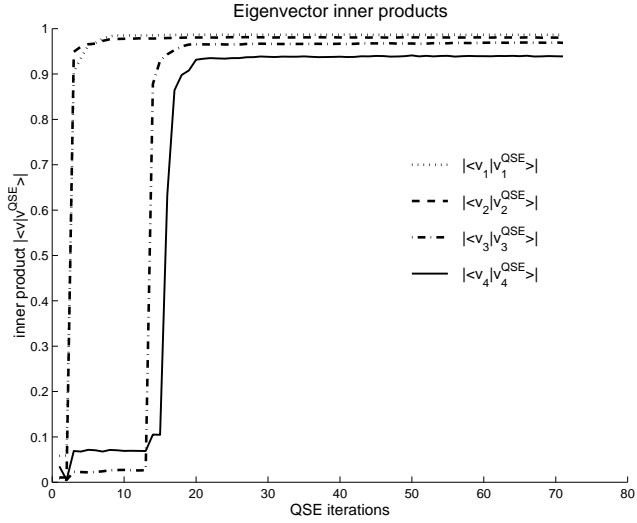}
\caption{Inner products between the normalized exact eigenvectors $\left|
v_{i}\right\rangle $ and the QSE results $\left|  v_{i}^{\text{QSE}%
}\right\rangle $ as functions of iteration number. }%
\label{fig:7}
\end{center}
\end{figure}

As a second test we consider a sparse non-Hermitian matrix with complex
eigenvalues. \ This type of matrix is not amenable to variational-based
methods. \ We will find the four eigenstates corresponding with eigenvalues
with the lowest real part for the random complex non-Hermitian matrix%

\begin{equation}
H_{jk}^{\prime}=(1+i\cdot c_{jk})H_{jk}.
\end{equation}
$H_{jk}$ is the same matrix used previously and$\ c_{jk}$ is a uniform random
variable distributed between $-1$ and 1. \ As before the program is started
from a random configuration, 70 basis states for each of the four
eigenvectors. \ For each iteration $10$ replacement basis vectors are selected
for each of the eigenvectors. \ In Figure \ref{fig:6}
 the exact eigenvalues and the
results of the QSE run are shown in the complex plane as functions of
iteration number. \ In Figure \ref{fig:7}
 we show the inner products of the QSE
eigenvectors with the exact eigenvectors. \ All of the eigenvectors were found
after about 20 iterations and remained stable throughout successive
iterations. \ Errors were again at about the $5$ to $10\%$ level.

\section{$\phi^{4}$ theory in $1+1$ dimensions}

We now apply the QSE method to an infinite dimensional quantum Hamiltonian.
\ We consider $\phi^{4}$ theory in $1+1$ dimensions, a system that is familiar
to us from previous studies using Monte Carlo \cite{periodic} and explicit
diagonalization \cite{spectral}. \ The Hamiltonian density for $\phi^{4}$
theory in $1+1$ dimensions has the form
\[
\mathcal{H}=\tfrac{1}{2}\left(  \tfrac{\partial\phi}{\partial t}\right)
^{2}+\tfrac{1}{2}\left(  \tfrac{\partial\phi}{\partial x}\right)  ^{2}%
+\tfrac{\mu^{2}}{2}\phi^{2}+\tfrac{\lambda}{4!}\text{:}\phi^{4}\text{:},
\]
where the normal ordering is with respect to the mass $\mu$. \ We consider the
system in a periodic box of length $2L$. \ We then expand in momentum modes
and reinterpret\ the problem as an equivalent Schr\"{o}dinger equation
\cite{periodic}. \ The resulting Hamiltonian is
\begin{align}
H  &  =-\tfrac{1}{2}%
{\displaystyle\sum_{n}}
\tfrac{\partial}{\partial q_{-n}}\tfrac{\partial}{\partial q_{n}}+\tfrac{1}{2}%
{\displaystyle\sum_{n}}
\left(  \omega_{n}^{2}(\mu)-\tfrac{\lambda b(\mu)}{8L}\right)  \,q_{-n}q_{n}\\
&  +\tfrac{\lambda}{4!2L}%
{\displaystyle\sum_{n_{1}+n_{2}+n_{3}+n_{4}=0}}
q_{n_{1}}q_{n_{2}}q_{n_{3}}q_{n_{4}}\nonumber
\end{align}
where
\begin{equation}
\omega_{n}(\mu)=\sqrt{\tfrac{n^{2}\pi^{2}}{L^{2}}+\mu^{2}}%
\end{equation}
and $b(\mu)$ is the coefficient for the mass counterterm
\begin{equation}
b(\mu)=%
{\displaystyle\sum_{n}}
\tfrac{1}{2\omega_{n}(\mu)}.
\end{equation}

It is convenient to split the Hamiltonian into free and interacting parts with
respect to an arbitrary mass $\mu^{\prime}$:%

\begin{equation}
H_{free}=-\tfrac{1}{2}%
{\displaystyle\sum_{n}}
\tfrac{\partial}{\partial q_{-n}}\tfrac{\partial}{\partial q_{n}}+\tfrac{1}{2}%
{\displaystyle\sum_{n}}
\omega_{n}^{2}(\mu^{\prime})\,q_{-n}q_{n},
\end{equation}%
\begin{align}
H  &  =H_{free}+\tfrac{1}{2}%
{\displaystyle\sum_{n}}
\left(  \mu^{2}-\mu^{\prime2}-\tfrac{\lambda b(\mu)}{8L}\right)  q_{-n}q_{n}\\
&  +\tfrac{\lambda}{4!2L}%
{\displaystyle\sum_{n_{1}+n_{2}+n_{3}+n_{4}=0}}
q_{n_{1}}q_{n_{2}}q_{n_{3}}q_{n_{4}}.\nonumber
\end{align}
$\mu^{\prime}$ is used to define the basis states of our Fock space. \ Since
$H$ is independent of $\mu^{\prime}$, we perform calculations for different
$\mu^{\prime}$ to obtain a reasonable estimate of the error. \ It is also
useful to find the range of values for $\mu^{\prime}$ which maximizes the
quasi-sparsity of the eigenvectors and therefore improves the accuracy of the
calculation. \ For the calculations presented here, we set the length of the
box to size $L=5\pi\mu^{-1}$. \ We restrict our attention to momentum modes
$q_{n}$ such that $\left|  n\right|  \leq N_{\max}$, where $N_{\max}=20$.
\ This corresponds with a momentum cutoff scale of $\Lambda=4\mu.$

To implement the QSE algorithm on this infinite dimensional Hilbert space, we
first define ladder operators with respect to $\mu^{\prime}$,
\begin{align}
a_{n}(\mu^{\prime})  &  =\tfrac{1}{\sqrt{2\omega_{n}(\mu^{\prime})}}\left[
q_{n}\omega_{n}(\mu^{\prime})+\tfrac{\partial}{\partial q_{-n}}\right] \\
a_{n}^{\dagger}(\mu^{\prime})  &  =\tfrac{1}{\sqrt{2\omega_{n}(\mu^{\prime})}%
}\left[  q_{-n}\omega_{n}(\mu^{\prime})-\tfrac{\partial}{\partial q_{n}%
}\right]  .
\end{align}
The Hamiltonian can now be rewritten as%

\begin{align}
H  &  =%
{\displaystyle\sum_{n}}
\omega_{n}(\mu^{\prime})a_{n}^{\dagger}a_{n}+\tfrac{1}{4}(\mu^{2}-\mu
^{\prime2}-\tfrac{\lambda b}{8L})%
{\displaystyle\sum_{n}}
\tfrac{\left(  a_{-n}+a_{n}^{\dagger}\right)  \left(  a_{n}+a_{-n}^{\dagger
}\right)  }{\omega_{n}(\mu^{\prime})}\label{ha}\\
&  +\tfrac{\lambda}{192L}%
{\displaystyle\sum_{n_{1}+n_{2}+n_{3}+n_{4}=0}}
\left[  \tfrac{\left(  a_{n_{1}}+a_{-n_{1}}^{\dagger}\right)  }{\sqrt
{\omega_{n_{1}}(\mu^{\prime})}}\tfrac{\left(  a_{n_{2}}+a_{-n_{2}}^{\dagger
}\right)  }{\sqrt{\omega_{n_{2}}(\mu^{\prime})}}\tfrac{\left(  a_{n_{3}%
}+a_{-n_{3}}^{\dagger}\right)  }{\sqrt{\omega_{n_{3}}(\mu^{\prime})}}%
\tfrac{\left(  a_{n_{4}}+a_{-n_{4}}^{\dagger}\right)  }{\sqrt{\omega_{n_{4}%
}(\mu^{\prime})}}\right]  .\nonumber
\end{align}
In (\ref{ha}) we have omitted constants contributing only to the vacuum
energy. \ We represent any momentum-space Fock state as a string of occupation
numbers, $\left|  o_{-N_{\max}},\cdots,o_{N_{\max}}\right\rangle $, where
\begin{equation}
a_{n}^{\dagger}a_{n}\left|  o_{-N_{\max}},\cdots,o_{N_{\max}}\right\rangle
=o_{n}\left|  o_{-N_{\max}},\cdots,o_{N_{\max}}\right\rangle .
\end{equation}
From the usual ladder operator relations, it is straightforward to calculate
the matrix element of $H$ between two arbitrary Fock states.

Aside from calculating matrix elements, the only other fundamental operation
needed for the QSE algorithm is the generation of new basis vectors. \ The new
states should be connected to some old basis vector through non-vanishing
matrix elements of $H$. \ Let us refer to the old basis vector as $\left|
e\right\rangle $. \ For this example there are two types of terms in our
interaction Hamiltonian, a quartic interaction
\begin{equation}%
{\displaystyle\sum_{n_{1},n_{2},n_{3}}}
\left(  a_{n_{1}}+a_{-n_{1}}^{\dagger}\right)  \left(  a_{n_{2}}+a_{-n_{2}%
}^{\dagger}\right)  \left(  a_{n_{3}}+a_{-n_{3}}^{\dagger}\right)  \left(
a_{-n_{1}-n_{2}-n_{3}}+a_{n_{1}+n_{2}+n_{3}}^{\dagger}\right)  ,
\end{equation}
and a quadratic interaction
\begin{equation}%
{\displaystyle\sum_{n}}
\left(  a_{-n}+a_{n}^{\dagger}\right)  \left(  a_{n}+a_{-n}^{\dagger}\right)
.
\end{equation}
To produce a new vector from $\left|  e\right\rangle $ we simply choose one of
the possible operator monomials
\begin{align}
&  a_{n_{1}}a_{n_{2}}a_{n_{3}}a_{-n_{1}-n_{2}-n_{3}},\,a_{-n_{1}}^{\dagger
}a_{n_{2}}a_{n_{3}}a_{-n_{1}-n_{2}-n_{3}},\cdots,\\
&  a_{-n}a_{n},\,a_{n}^{\dagger}a_{-n}^{\dagger},\cdots\nonumber
\end{align}
and act on $\left|  e\right\rangle $. \ Our experience is that the
interactions involving the small momentum modes are generally more important
than those for the large momentum modes, a signal that the ultraviolet
divergences have been properly renormalized. \ For this reason it is best to
arrange the selection probabilities such that the smaller values of $\left|
n_{1}\right|  $, $\left|  n_{2}\right|  $, $\left|  n_{3}\right|  $ and
$\left|  n\right|  $ are chosen more often.

\begin{figure}[htbp]
\begin{center}
\epsfxsize=20pc \epsfbox{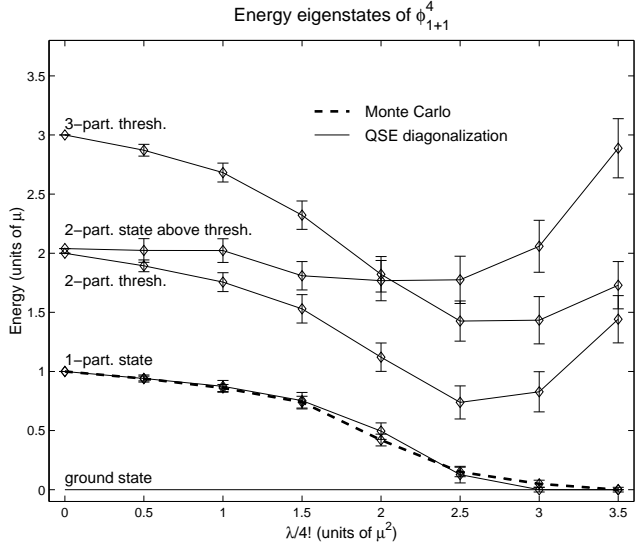}
\caption{Energy eigenvalues of $\phi_{1+1}^{4}$ as functions of the coupling
constant. }%
\label{fig:8}
\end{center}
\end{figure}

For each QSE iteration, $50$ new basis vectors were selected for each
eigenstate and $250$ basis vectors were retained. \ The results for the lowest
energy eigenvalues are shown in Figure \ref{fig:8}.
 \ The error bars were estimated by
repeating the calculation for different values of the auxiliary mass parameter
$\mu^{\prime}$.

From prior Monte Carlo calculations we know that the theory has a phase
transition at $\frac{\lambda}{4!}\approx2.5\mu^{2}$ corresponding with
spontaneous breaking of the $\phi\rightarrow-\phi$ reflection symmetry. \ In
the broken phase there are two degenerate ground states and we refer to these
as the even and odd vacuum states. \ In Figure \ref{fig:8} we see signs of a second
order phase transition near $\frac{\lambda}{4!}\approx2.5\mu^{2}$. \ Since we
are working in a finite volume the spectrum is discrete, and we can track the
energy eigenvalues as functions of the coupling. \ Crossing the phase
boundary, we see that the vacuum in the symmetric phase becomes the even
vacuum in the broken phase while the one-particle state in the symmetric phase
becomes the odd vacuum. \ The energy difference between the states is also in
agreement with a Monte Carlo calculation of the same quantities. \ The state
marking the two-particle threshold in the symmetric phase becomes the
one-particle state above the odd vacuum, while the state at the three-particle
threshold becomes the one-particle state above the even vacuum. \ These
one-particle states should be degenerate in the infinite volume limit. \ One
rather unusual feature is the behavior of the first two-particle state above
threshold in the symmetric phase. \ In the symmetric phase this state lies
close to the two-particle threshold. \ But as we cross the phase boundary the
state which was the two-particle threshold is changed into a one-particle
state. \ Thus our two-particle state is pushed up even further to become a
two-particle state above the even vacuum and we see a pronounced level crossing.

We note that while the one-particle mass vanishes near the critical point, the
energies of the two-particle and three-particle thresholds reach a minimum but
do not come as close to zero energy. \ It is known that this model is
repulsive in the two-particle scattering channel. \ In a large but finite
volume the ground state and one-particle states do not feel significant finite
volume effects. \ The two-particle state at threshold, however, requires that
the two asymptotic particles be widely separated. \ In our periodic box of
length 2$L$ the maximal separation distance is $L$ and we expect an increase
in energy with respect to twice the one-particle mass of size $\sim V(L)$,
where $V$ is the potential energy between particles. \ Likewise a
three-particle state will increase in energy an amount $\sim3V(2L/3)$. \ Our
results indicate that finite volume effects for the excited states are
significant for this value of $L$.

\section{Summary}

We have proposed a new approach which combines both diagonalization and Monte
Carlo within a computational scheme. \ The motivation for our approach is to
take advantage of the strengths of the two computational methods in their
respective domains. We remedy sign and phase oscillation problems by handling
the interactions of the most important basis states exactly using
diagonalization, and we deal with storage and CPU problems by stochastically
sampling the contribution of the remaining states. \ We discussed the
diagonalization part of the method in this paper. \ The goal of
diagonalization within our scheme is to find the most important basis vectors
of the low energy eigenstates and treat the interactions among them exactly.
\ We have introduced a new diagonalization method called quasi-sparse
eigenvector diagonalization which achieves this goal efficiently and can
operate using any basis, either orthogonal or non-orthogonal, and any sparse
Hamiltonian, either real, complex, Hermitian, non-Hermitian,
finite-dimensional, or infinite-dimensional. \ Quasi-sparse eigenvector
diagonalization is the only method we know which can address all of these problems.

We considered three examples which tested the performance of the algorithm.
\ We found the lowest energy eigenstates for a random sparse real symmetric
matrix, the lowest eigenstates (sorted according to the real part of the
eigenvalue) for a random sparse complex non-Hermitian matrix, and the lowest
energy eigenstates for an infinite-dimensional Hamiltonian defined by $1+1$
dimensional $\phi^{4}$ theory in a periodic box.

We regard QSE diagonalization as only a starting point for the Monte Carlo
part of the calculation. \ Once the most important basis vectors are found and
their interactions treated exactly, a technique called stochastic error
correction is used to sample the contribution of the remaining basis vectors.
\ This method is introduced in \cite{sec}.

\paragraph*{Acknowledgments}

We thank P. van Baal, H. S. Seung, H. Sompolinsky, and M. Windoloski for
useful discussions.

\chapter[Stochastic error correction]
{Introduction to stochastic error correction methods\footnote{D. Lee,
  N. Salwen, M. Windoloski, Phys. Lett. B 502(2001) 329-337}}

\section{Introduction}

In \cite{qse} a new approach was proposed for finding the low-energy
eigenstates of very large or infinite-dimensional quantum Hamiltonians. \ This
proposal combines both diagonalization and Monte Carlo methods, each being
used to solve a portion of the problem for which the technique is most
efficient. \ The first part of the proposal is to diagonalize the Hamiltonian
restricted to a subspace containing the most important basis vectors for each
low energy eigenstate. \ This may be accomplished either through variational
techniques or an \textit{ab initio} method such as quasi-sparse eigenvector
(QSE) diagonalization.\ \ The second step is to include the contribution of
the remaining basis vectors\ by Monte Carlo sampling. \ The use of
diagonalization allows one to consider systems with fermion sign oscillations
and extract information about wavefunctions and excited states. \ The use of
Monte Carlo provides tools to handle the exponential increase in the number of
basis states for large volume systems.

The first half of this proposal was discussed in \cite{qse}. \ An adaptive
diagonalization algorithm known as the quasi-sparse eigenvector method was
introduced to find the most important basis vectors for each low energy
eigenstate. \ This technique is especially valuable when little is known about
the low energy states. \ It is also the only method available which can handle
non-orthogonal bases, non-Hermitian Hamiltonians, infinite dimensional
systems, and which can find several low energy states with like quantum
numbers simultaneously. \ In this paper we discuss the second half of the
diagonalization/Monte Carlo scheme. \ We introduce several new Monte Carlo
techniques which we call stochastic error correction (SEC). \ There are two
general varieties of stochastic error correction, methods based on a series
expansion and those which are not. \ The series method starts with an
eigenvector of the Hamiltonian restricted to some starting subspace and then
includes the contribution of the remaining basis states as terms in an ordered
expansion. \ The idea is to form a perturbative expansion centered around a
good non-perturbative starting point.

As an example of a non-series method we discuss a technique called the
stochastic Lanczos method. \ This method again starts with eigenvectors of a
Hamiltonian submatrix. \ Using these as starting vectors, we define Krylov
vectors, $\left|  j\right\rangle ,H\left|  j\right\rangle ,H^{2}\left|
j\right\rangle \cdots$, similar to standard Lanczos diagonalization. \ The new
ingredient is that matrix elements between Krylov vectors, $\left\langle
j^{\prime}\right|  H^{n}\left|  j\right\rangle ,$ are computed using matrix
diffusion Monte Carlo. \ Since the method does not rely on a series expansion,
it has the advantage that the starting vectors need not be close to the exact eigenvectors.

One can generate a large class of stochastic error correction methods based on
other non-series algorithms, various ways of resumming the series expansion,
or combinations of the two techniques. \ In this introductory paper we
concentrate on describing the basic principles and implementation of the
series and non-series approaches. \ We also present three test problems which
demonstrate the potential of the new approach for a range of different
problems. \ In the first example we determine the low energy spectrum of
$\phi_{2+1}^{4}$ theory using QSE diagonalization and first order corrections
using the series method. \ In the second example we find the low energy
spectrum of compact $U(1)$ in $2+1$ dimensions using the stochastic Lanczos
method. \ In the last example we find the ground state of the $2+1$
dimensional Hubbard model using QSE diagonalization and first order series
stochastic error correction. \ In each case we compare with published results
in the literature. \ We conclude with a summary and some general comments on
the new computational scheme.

\section{Series method}

Let $\left|  i\right\rangle $ be the eigenvectors of a Hamiltonian $H$
restricted to some subspace $S$. \ Let $\left|  A_{j}\right\rangle $ be the
remaining basis vectors in the full space not contained in $S$. \ We can
represent $H$ as
\begin{equation}
\left[
\begin{array}
[c]{cccccc}%
\lambda_{1} & 0 & \cdots & \left\langle 1\right|  H\left|  A_{1}\right\rangle
& \left\langle 1\right|  H\left|  A_{2}\right\rangle  & \cdots\\
0 & \lambda_{2} & \cdots & \left\langle 2\right|  H\left|  A_{1}\right\rangle
& \left\langle 2\right|  H\left|  A_{2}\right\rangle  & \cdots\\
\vdots & \vdots & \ddots & \vdots & \vdots & \cdots\\
\left\langle A_{1}\right|  H\left|  1\right\rangle  & \left\langle
A_{1}\right|  H\left|  2\right\rangle  & \cdots &  E\cdot\lambda_{A_{1}} &
\left\langle A_{1}\right|  H\left|  A_{2}\right\rangle  & \cdots\\
\left\langle A_{2}\right|  H\left|  1\right\rangle  & \left\langle
A_{2}\right|  H\left|  2\right\rangle  & \cdots & \left\langle A_{2}\right|
H\left|  A_{1}\right\rangle  & E\cdot\lambda_{A_{2}} & \cdots\\
\vdots & \vdots & \vdots & \vdots & \vdots & \ddots
\end{array}
\right]  .
\end{equation}
We have used Dirac's bra-ket notation to represent the terms of the matrix.
\ In cases where the basis is non-orthogonal or the Hamiltonian is
non-Hermitian, the precise meaning of terms such as $\left\langle
A_{1}\right|  H\left|  1\right\rangle $ is the action of the dual vector to
$\left|  A_{1}\right\rangle $ upon the vector $H\left|  1\right\rangle $. \ We
have written the diagonal terms for the basis vectors $\left|  A_{j}%
\right\rangle $ with an explicit factor $E$ for reasons to be explained shortly.

Let us assume that $\left|  1\right\rangle $ is close to some exact
eigenvector of $H$ which we denote as $\left|  1_{\text{full}}\right\rangle $.
\ More concretely we assume that the components of $\left|  1_{\text{full}%
}\right\rangle $ outside $S$ are small enough so that we can expand in inverse
powers of the introduced parameter $E.$ \ In order to simplify the expansion we
choose to shift the diagonal entries so that $\lambda_{1}=0.$

The series method of stochastic error correction is based on the $E^{-1}$
expansion,
\begin{equation}
\left|  1_{\text{full}}\right\rangle \propto\left[
\begin{array}
[c]{c}%
1\\
c_{2}^{\prime}E^{-1}+c_{2}^{\prime\prime}E^{-2}+\cdots\\
\vdots\\
c_{A_{1}}^{\prime}E^{-1}+c_{A_{1}}^{\prime\prime}E^{-2}+\cdots\\
c_{A_{2}}^{\prime}E^{-1}+c_{A_{2}}^{\prime\prime}E^{-2}+\cdots\\
\vdots
\end{array}
\right]  , \label{unnorm}%
\end{equation}%
\begin{equation}
\lambda_{\text{full}}=\lambda_{1}^{\prime}E^{-1}+\lambda_{1}^{\prime\prime
}E^{-2}\cdots.
\end{equation}
It is convenient to choose the normalization of the eigenvector such that the
$\left|  1\right\rangle $ component remains 1. \ The convergence of the
expansion is controlled by the proximity of $\left|  1\right\rangle $ to
$\left|  1_{\text{full}}\right\rangle $. \ If $\left|  1\right\rangle $ is not
at all close to $\left|  1_{\text{full}}\right\rangle $ then it will be
necessary to use a non-series method such as the stochastic Lanczos method
discussed in the next section.

At first order in $E^{-1}$ we find
\begin{equation}
c_{A_{j}}^{\prime}=-\tfrac{\left\langle A_{j}\right|  H\left|  1\right\rangle
}{\lambda_{A_{j}}}%
\end{equation}%
\begin{equation}
\lambda_{1}^{\prime}=-\sum_{j}\tfrac{\left\langle 1\right|  H\left|
A_{j}\right\rangle \left\langle A_{j}\right|  H\left|  1\right\rangle
}{\lambda_{A_{j}}} \label{firs}%
\end{equation}%
\begin{equation}
c_{j}^{\prime}=\tfrac{1}{\lambda_{j}}\sum_{k}\tfrac{\left\langle j\right|
H\left|  A_{k}\right\rangle \left\langle A_{k}\right|  H\left|  1\right\rangle
}{\lambda_{A_{k}}}.
\end{equation}
At second order the contributions are
\begin{equation}
c_{A_{j}}^{\prime\prime}=\sum_{k\neq j}\tfrac{\left\langle A_{j}\right|
H\left|  A_{k}\right\rangle \left\langle A_{k}\right|  H\left|  1\right\rangle
}{\lambda_{A_{j}}\lambda_{A_{k}}}-\sum_{l\neq1}\sum_{k}\tfrac{\left\langle
A_{j}\right|  H\left|  l\right\rangle \left\langle l\right|  H\left|
A_{k}\right\rangle \left\langle A_{k}\right|  H\left|  1\right\rangle
}{\lambda_{A_{j}}\lambda_{l}\lambda_{A_{k}}}%
\end{equation}
\qquad%
\begin{equation}
\lambda_{1}^{\prime\prime}=\sum_{j}\sum_{k\neq j}\tfrac{\left\langle 1\right|
H\left|  A_{j}\right\rangle \left\langle A_{j}\right|  H\left|  A_{k}%
\right\rangle \left\langle A_{k}\right|  H\left|  1\right\rangle }%
{\lambda_{A_{j}}\lambda_{A_{k}}}-\sum_{j}\sum_{l\neq1}\sum_{k}\tfrac
{\left\langle 1\right|  H\left|  A_{j}\right\rangle \left\langle A_{j}\right|
H\left|  l\right\rangle \left\langle l\right|  H\left|  A_{k}\right\rangle
\left\langle A_{k}\right|  H\left|  1\right\rangle }{\lambda_{A_{j}}%
\lambda_{l}\lambda_{A_{k}}}%
\end{equation}%

\begin{align}
c_{m}^{\prime\prime}  &  =-\sum_{j}\sum_{k\neq j}\tfrac{\left\langle m\right|
H\left|  A_{j}\right\rangle \left\langle A_{j}\right|  H\left|  A_{k}%
\right\rangle \left\langle A_{k}\right|  H\left|  1\right\rangle }{\lambda
_{m}\lambda_{A_{j}}\lambda_{A_{k}}}+\sum_{j}\sum_{l\neq1}\sum_{k}%
\tfrac{\left\langle m\right|  H\left|  A_{j}\right\rangle \left\langle
A_{j}\right|  H\left|  l\right\rangle \left\langle l\right|  H\left|
A_{k}\right\rangle \left\langle A_{k}\right|  H\left|  1\right\rangle
}{\lambda_{m}\lambda_{A_{j}}\lambda_{l}\lambda_{A_{k}}}\\
&  -\tfrac{1}{\lambda_{m}}\left[  \sum_{j}\tfrac{\left\langle 1\right|
H\left|  A_{j}\right\rangle \left\langle A_{j}\right|  H\left|  1\right\rangle
}{\lambda_{A_{j}}}\right]  \tfrac{1}{\lambda_{m}}\left[  \sum_{k}%
\tfrac{\left\langle m\right|  H\left|  A_{k}\right\rangle \left\langle
A_{k}\right|  H\left|  1\right\rangle }{\lambda_{A_{k}}}\right]  .\nonumber
\end{align}
These contributions can be calculated by straightforward Monte Carlo sampling.
\ All that is required is an efficient way of generating random basis vectors
$\left|  A_{k}\right\rangle $ with known probability rates. \ Let
$P(A_{\text{trial}})$ denote the probability of selecting $\left|
A_{\text{trial}}\right\rangle $ on a given trial. \ If for example we are
calculating the first order correction to the eigenvalue, then we have%
\begin{align}
\lambda_{1}^{\prime}  &  =-\sum_{j}\tfrac{\left\langle 1\right|  H\left|
A_{j}\right\rangle \left\langle A_{j}\right|  H\left|  1\right\rangle
}{\lambda_{A_{j}}}\\
&  =-\lim_{N\rightarrow\infty}\tfrac{1}{N}\sum_{i=1,\cdots,N}\tfrac
{\left\langle 1\right|  H\left|  A_{\text{trial}(i)}\right\rangle \left\langle
A_{\text{trial}(i)}\right|  H\left|  1\right\rangle }{\lambda_{A_{\text{trial}%
(i)}}P(A_{\text{trial}(i)})}.\nonumber
\end{align}

\section{Stochastic Lanczos}

We now consider a method called stochastic Lanczos which does not require the
starting vectors to be close to exact eigenvectors of $H.$ \ This is essential
if the eigenvectors of $H$ are not quasi-sparse and require extremely large
numbers of basis states to represent accurately.

Let $V$ be the full Hilbert space for our system. \ As in the previous section
let $S$ be the subspace over which we have diagonalized $H$ exactly. \ Let
$P_{S}$ be the projection operator for $S$ and let $\lambda_{j}$ and $\left|
j\right\rangle $ be the eigenvalues and eigenvectors of $H$ restricted to $S$
so that
\begin{equation}
P_{S}HP_{S}\left|  j\right\rangle =\lambda_{j}\left|  j\right\rangle .
\end{equation}
\ Let $Z$ be an auxiliary subspace, one which contains $S$ but excludes very
high-energy states. Let $P_{Z}$ be the projection operator for $Z$. \ We will
choose $Z$ such that $P_{Z}HP_{Z}$ is bounded above. \ Let$\ a$ be a real
constant which is greater than the midpoint of the minimum and maximum
eigenvalues of $P_{Z}HP_{Z}$. \ As $n\rightarrow\infty$ the operator $\left[
P_{Z}(H-a)P_{Z}\right]  ^{n}$ maps any given state in $Z$ to the corresponding
lowest-energy eigenvector of $P_{Z}HP_{Z}$\ with non-zero overlap.

The stochastic Lanczos method uses the operators $\left[  P_{Z}(H-a)P_{Z}%
\right]  ^{n}$ to approximate the low-energy eigenvalues and eigenvectors of
$P_{Z}HP_{Z}$. \ The goal is to diagonalize $H$ in a subspace spanned by vectors%

\begin{equation}
\left|  d,j\right\rangle =\left[  P_{Z}(H-a)P_{Z}\right]  ^{d}\left|
j\right\rangle , \label{slbasis}%
\end{equation}
for several values of $d$ and $j$. \ This requires calculating $\left\langle
d^{\prime},j^{\prime}\right|  \left.  d,j\right\rangle $ and $\left\langle
d^{\prime},j^{\prime}\right|  H\left|  d,j\right\rangle $. \ If our
Hamiltonian matrix is Hermitian, both of these terms can be written in the
general form
\begin{equation}
\left\langle j^{\prime}\right|  \left[  P_{Z}(H-a)P_{Z}\right]  ^{n}\left|
j\right\rangle .
\end{equation}
Therefore it suffices to determine the matrix
\begin{equation}
A_{n}\equiv P_{S}\left[  P_{Z}(H-a)P_{Z}\right]  ^{n}P_{S}.
\end{equation}
For non-orthogonal bases and non-Hermitian Hamiltonians, the only change is
that we use vectors
\begin{equation}
\left[  P_{Z}(H-a)P_{Z}\right]  ^{d}\left|  j\right\rangle
\end{equation}
to generate approximate right eigenvectors of $H$ and vectors in the dual
space
\begin{equation}
\left\langle j\right|  \left[  P_{Z}(H-a)P_{Z}\right]  ^{d}%
\end{equation}
to produce approximate left eigenvectors. \ Adding and subtracting
$P_{S}(H-a)P_{S}$, we can rewrite
\begin{equation}
A_{n}=P_{S}\left[  P_{Z}(H-a)P_{Z}-P_{S}(H-a)P_{S}+P_{S}(H-a)P_{S}\right]
^{n}P_{S}.
\end{equation}
$A_{n}$ can now be evaluated recursively as
\begin{equation}
A_{n+1}=B_{n+1}+\sum_{m=0,\cdots,n}B_{m}(H-a)A_{n-m},
\end{equation}
where
\begin{equation}
B_{n}=P_{S}\left[  P_{Z}(H-a)P_{Z}-P_{S}(H-a)P_{S}\right]  ^{n}P_{S}.
\end{equation}

The components of $B_{n}$ are computed by matrix diffusion Monte Carlo. \ One
could also directly evaluate the components of $A_{n}$. \ However the
calculation for $B_{n}$ eliminates the need to sample the matrix
$P_{S}(H-a)P_{S}$, which is already known. \ Any general matrix product
$M^{(1)}M^{(2)}\cdots M^{(n)}$ is a sum of degree $n$ monomials,
\begin{equation}
\left[  M^{(1)}M^{(2)}\cdots M^{(n)}\right]  _{jk}=\sum_{i_{1},\cdots i_{n-1}%
}M_{ji_{1}}^{(1)}M_{i_{1}i_{2}}^{(2)}\cdots M_{i_{n-1}k}^{(n)}. \label{sums}%
\end{equation}
We can interpret (\ref{sums}) as a sum over paths through the set of basis
vectors of $Z,$%
\begin{equation}
\left|  j\right\rangle \rightarrow\left|  i_{1}\right\rangle \rightarrow
\cdots\rightarrow\left|  i_{n-1}\right\rangle \rightarrow\left|
k\right\rangle \text{,}%
\end{equation}
with an associated weight $M_{ji_{1}}^{(1)}M_{i_{1}i_{2}}^{(2)}\cdots
M_{i_{n-1}k}^{(n)}$. \ The components of $B_{n}$ are sampled using ensembles
of random walkers. \ We refer the interested reader to \cite{kosztin} for a
review of methods in diffusion Monte Carlo.

We end the section with a discussion of the fermion sign problem. \ The sign
problem is a general issue for any Monte Carlo calculation. \ For a system
with sign oscillations the evaluation of a Euclidean-time Green's function
involves sums
\begin{equation}
\sum_{i}x_{i}%
\end{equation}
with the property that
\begin{equation}
\frac{\sum_{i}x_{i}}{\sum_{i}\left|  x_{i}\right|  }\sim\exp(-c\cdot V\cdot
T),
\end{equation}
where $V$ is the volume, $T$ is the Euclidean time, and $c$ is a positive
constant. \ We will refer to this term as the cancellation ratio. \ The
exponential dependence on $V$ and $T$ makes computations difficult even for
small systems.

The sign problem will affect the calculation of $B_{n}$ in the stochastic
Lanczos algorithm and terms in the series method discussed in the previous
section. \ The effect however is different from the sign problem in typical
Monte Carlo Green's function calculations. \ Stochastic error correction is a
calculation of eigenvalues and eigenvectors rather than a sampling of the
partition function or the time evolution of a given initial state. \ Therefore
the quantity of interest is not $\exp(-HT)$ but the action of $H$ or $H^{n}$
on approximate eigenvectors of $H$. \ Due to homogeneity in $H$ the explicit
volume dependence does not appear in the cancellation ratio. \ Instead we
find
\begin{equation}
\frac{\sum_{i}x_{i}}{\sum_{i}\left|  x_{i}\right|  }\sim\exp(-k\cdot n),
\end{equation}
where $k$ is a positive constant. \ The sign problem will return if the
starting point of the SEC calculation is very poor and it becomes necessary to
use $n$ such that $k\cdot n$ is large. \ However in many cases $k\cdot n$ can
be kept small even for large $n$ since the most important part of the
Hamiltonian, $P_{S}HP_{S},$ is diagonalized exactly. \ In short the sign
problem is less severe because stochastic error correction uses the result of
subspace diagonalization as its starting point.

\section{$\phi^{4}$ theory in $2+1$ dimensions}

The first example we consider is $\phi^{4}$ theory in $2+1$ dimensions near
the $\phi\rightarrow-\phi$ symmetry restoration phase transition. \ We will
use QSE diagonalization and the series version of stochastic error correction
to probe the low energy spectrum of the theory on both sides of the phase transition.

In \cite{mag} Magruder demonstrated the existence of a phase transition in
$\phi_{2+1}^{4}$ by extending Chang's duality argument for $\phi_{1+1}^{4}$.
\ The statement of the main result is as follows. \ Consider the two Lagrange
densities%
\begin{align}
\mathcal{L}_{+}  &  =\tfrac{1}{2}\partial_{\nu}\phi\partial^{\nu}\phi
-\tfrac{1}{2}\mu_{+}^{2}\phi^{2}-\tfrac{g}{4!}\phi^{4}+\tfrac{1}{2}\delta
_{\mu_{+}}^{2}\phi^{2}\\
\mathcal{L}_{-}  &  =\tfrac{1}{2}\partial_{\nu}\phi\partial^{\nu}\phi
+\tfrac{1}{4}\mu_{-}^{2}\phi^{2}-\tfrac{g}{4!}\phi^{4}+\tfrac{1}{2}\delta
_{\mu_{-}}^{2}\phi^{2}.
\end{align}
The counterterm $\delta_{\mu_{+}}^{2}$ is defined so that in the
$\mathcal{L}_{+}$ system the $\phi$ self-energy graphs vanish at zero-momentum
up to two-loop order. \ By shifting the field%
\begin{equation}
\phi=\phi^{\prime}+\sqrt{\tfrac{3\mu_{-}^{2}}{g}}%
\end{equation}
we note that the same counterterm $\delta_{\mu_{-}}^{2}$ (same mass dependence
but $\mu_{+}$ replaced by $\mu_{-}$) is also sufficient to renormalize
$\mathcal{L}_{-}$. \ By equating $\mathcal{L}_{+}$ and $\mathcal{L}_{-}$ we
obtain a duality constraint between the two theories. \ One feature of this
constraint is that the $g\rightarrow\infty$ limit of $\mathcal{L}_{-}$ is
mapped to the $g\rightarrow0$ limit of $\mathcal{L}_{+}$. \ Therefore
$\mathcal{L}_{-}$, whose reflection symmetry $\phi\rightarrow-\phi$ is broken
at small $g$, must eventually reach the symmetric phase for sufficiently large
coupling.\footnote{The $g\rightarrow\infty$ limit of $\mathcal{L}_{+}$ is
mapped to the $g\rightarrow\infty$ limit of $\mathcal{L}_{-}$ and so there is
no analogous argument for a phase transition in $\mathcal{L}_{+}$. \ Numerical
calculations indicate that there is no phase transition for $\mathcal{L}_{+}$
\cite{markw}.}

The $\mathcal{L}_{-}$ phase transition was studied using quasi-sparse
eigenvector diagonalization with stochastic error correction. \ Quantities
such as the critical coupling, critical exponents, and the low lying energy
spectrum were studied and, where possible, compared with Monte Carlo results.
\ A full discussion methods and results are presented in \cite{markw}. \ We
will very briefly summarize some of the results below.

The two spatial dimensions of our system are taken to be a periodic box of
size $2L$ by $2L$. We will use the modal field formalism to describe the
Hamiltonian for the theory.\footnote{We refer the reader to\
\cite{periodic} for
a short introduction.} \ In the following we let the vectors $\vec{n}$
represent ordered integer pairs $(n_{x},n_{y})$ such that $\left|
n_{x}\right|  ,\left|  n_{y}\right|  \leq N_{\max}$. \ The parameter $N_{\max
}$ corresponds with a momentum cutoff scale of $\Lambda=N_{\max}\pi/L$. \ The
modal field Hamiltonian has the form\footnote{Counterterms were calculated
using finite volume perturbation theory.}%

\begin{align}
H  &  =\sum_{\vec{n}}\left[  -\tfrac{1}{2}\tfrac{\partial}{\partial
q_{-\vec{n}}}\tfrac{\partial}{\partial q_{\vec{n}}}+\tfrac{1}{2}\left(
\tfrac{\vec{n}^{2}\pi^{2}}{L^{2}}-\tfrac{\mu^{2}}{2}\right)  -\tfrac
{6b}{(2L)^{2}}\tfrac{g}{4!}+\tfrac{48}{(2L)^{4}}\left(  \tfrac{g}{4!}\right)
^{2}\alpha_{\vec{n}}\right]  q_{-\vec{n}}q_{\vec{n}}\\
&  +\tfrac{1}{(2L)^{2}}\tfrac{g}{4!}\sum_{\vec{n}_{1}+\vec{n}_{2}+\vec{n}%
_{3}+\vec{n}_{4}=0}q_{\vec{n}_{1}}q_{\vec{n}_{2}}q_{\vec{n}_{3}}q_{\vec{n}%
_{4}}\nonumber
\end{align}
where
\begin{equation}
b=\sum_{\vec{n}}\tfrac{1}{2\omega_{\vec{n}}},\qquad\omega_{\vec{n}}%
=\sqrt{\tfrac{\vec{n}^{2}\pi^{2}}{L^{2}}+\mu^{2}},
\end{equation}
and%
\begin{equation}
\alpha_{\vec{n}}=\sum_{\vec{n}_{1},\vec{n}_{2}}\tfrac{1}{4\omega_{\vec{n}_{1}%
}\omega_{\vec{n}_{2}}\omega_{\vec{n}-\vec{n}_{1}-\vec{n}_{2}}(\omega_{\vec
{n}_{1}}+\omega_{\vec{n}_{2}}+\omega_{\vec{n}-\vec{n}_{1}-\vec{n}_{2}})}.
\end{equation}

\begin{figure}[htbp]
\begin{center}
\epsfxsize=25pc \epsfbox{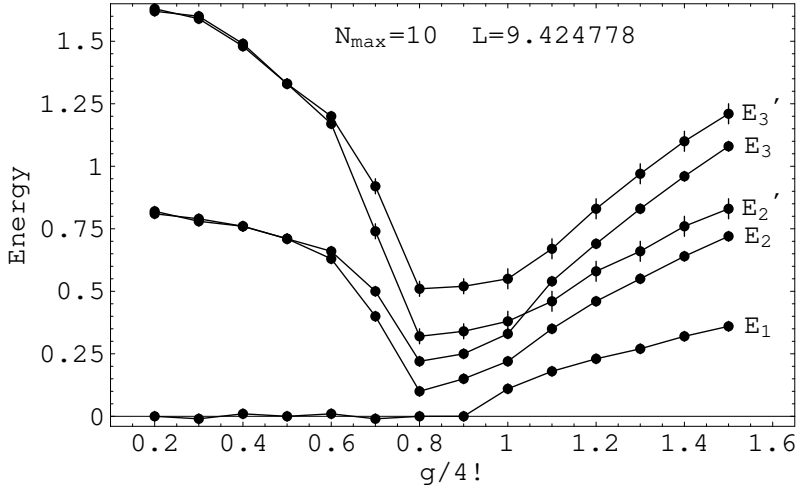}
\caption{Energy eigenvalues as functions of $\frac{g}{4!}$ as calculated by
QSE diagonalization with first-order error corrections.}%
\label{fig:new1}
\end{center}
\end{figure}

In Figure \ref{fig:new1} we have plotted the lowest energy eigenstates
in the rest frame 
for $N_{\max}=10$ and $L=3\pi$ (in units where $\mu=1$). \ This choice of
parameters corresponds with $441$ different momentum modes and a momentum
cutoff scale of $\Lambda=3.33\mu$. \ The states shown in Figure \ref{fig:new1} 
are the
three lowest eigenstates in the even and odd $\phi\rightarrow-\phi$ symmetry
sectors, and the energies are measured relative to the ground state energy.
\ In our calculation QSE diagonalization was used keeping $500$ Fock states,
and the stochastic error correction was computed to first-order using the
series method. \ Error bars shown include statistical error and an estimate of
the contribution from higher order corrections. \ We see clear evidence of a
second-order phase transition near $\frac{g}{4!}=0.9$.\footnote{A more
complete discussion of the critical coupling as well as critical exponents can
be found in \cite{markw}.} \ We have labelled the energies of the states
according to their physical interpretation in the symmetric phase. \ $E_{1}$
is the energy for the one-particle state, $E_{2}(E_{3})$ is for the
two(three)-particle threshold, and $E_{2}^{\prime}(E_{3}^{\prime})$ is for the
first state above the two(three)-particle threshold. \ At finite volume these
energies are continuous functions of the coupling $g$. \ One feature which was
also observed in $\phi_{1+1}^{4}$ \cite{qse} is the crossing of energy levels
due to the double degeneracy of states in the broken symmetry phase. $\ E_{3}$
is connected to a one-particle state in the broken phase while $E_{2}^{\prime
}$ is connected to a two-particle state. \ The levels $E_{2}^{\prime}$ and
$E_{3}$ therefore cross near the critical point.

Another interesting phenomenon is the appearance of a bound state in the
broken symmetry phase. \ In both the odd and even symmetry sectors we can
measure the ratio of the two-particle to one-particle energies:%
\[
\overset{\text{Table 1}}{%
\begin{tabular}
[c]{|l|l|l|}\hline
$\frac{g}{4!}$ & $E_{2}^{\prime}/E_{2}$ & $E_{3}^{\prime}/E_{3}$\\\hline
$0.2$ & $2.01(4)$ & $1.98(4)$\\\hline
$0.3$ & $2.01(4)$ & $2.05(4)$\\\hline
$0.4$ & $1.95(4)$ & $1.96(4)$\\\hline
$0.5$ & $1.87(4)$ & $1.87(4)$\\\hline
$0.6$ & $1.86(4)$ & $1.82(4)$\\\hline
\end{tabular}
}%
\]
These results are consistent with the binding energies reported in
\cite{CHP, CHP1}
and \cite{CHPZ}, which indicate a ratio of $1.83(3)$ near the critical point.

\section{Compact U(1) in 2+1 dimensions}

Compact U(1) lattice gauge theory in $2+1$ dimensions is a simple but
phenomenologically interesting gauge model. \ It is asymptotically free and in
the usual continuum limit describes massless non-interacting photons. \ On the
other hand if the continuum limit is reached by rescaling the mass gap to
remain constant, one instead finds a confining theory of massive bosons
\cite{gopfert}. \ The Hamiltonian has the form%

\begin{equation}
H=-\sum_{l}\tfrac{\partial^{2}}{\partial A_{l}^{2}}-2x\sum_{P}\cos\theta_{P}.
\label{hamil}%
\end{equation}
In (\ref{hamil}) $A_{l}$ are link gauge fields, $\theta_{P}$ is the sum of the
links circuiting a plaquette,%
\begin{equation}
\theta_{P}=A_{l_{1}}+A_{l_{2}}-A_{l_{3}}-A_{l_{4}},
\end{equation}
and%
\begin{equation}
x=e^{-4}a^{-2}%
\end{equation}
is the strong coupling parameter, which tends to infinity as the lattice
spacing $a$ goes to $0$. \ We follow the notation of \cite{hamer} in which an
overall constant factor of $\frac{e^{2}}{2}$ multiplying the right-hand side
of (\ref{hamil}) is suppressed. \ The energy levels we measure are therefore
in units of $\frac{e^{2}}{2}$.

The diagonalization of lattice gauge Hamiltonians is constrained by the
requirements of gauge invariance. \ To preserve gauge invariance it is most
convenient to use a basis which diagonalizes the electric field part of the
Hamiltonian%
\begin{equation}
\left[  -\sum_{l}\tfrac{\partial^{2}}{\partial A_{l}^{2}}\right]
\bigotimes_{l^{\prime}}\left|  n_{l^{\prime}}\right\rangle \cdots=\left[
\sum_{l}n_{l}^{2}\right]  \bigotimes_{l^{\prime}}\left|  n_{l^{\prime}%
}\right\rangle .
\end{equation}
As our next example of stochastic error correction we will address the
$4\times4$ lattice system at $x=1$ using this electric field basis. \ In
\cite{irving} it was noted that this poses a challenge to standard
diagonalization techniques. \ Even on the small $4\times4$ lattice a
surprisingly large number of states, about $10^{7}\sim10^{8}$, are needed to
accurately describe the low energy spectrum at $x=1$. \ This problem can be
circumvented by modifying the basis states to incorporate more of the physics
of the ground state. \ For example one can introduce a disordered background
of magnetic flux as suggested in \cite{heys}, and that approach is followed in
an ongoing project \cite{gauge}. \ However we would like to directly address
the problem described in \cite{irving} and show how the stochastic Lanczos
method handles the proliferation of large numbers of basis states in the
original electric field basis.

We will choose our starting subspace $S$ to include all basis states
\begin{equation}
\bigotimes_{l^{\prime}}\left|  n_{l^{\prime}}\right\rangle
\end{equation}
which satisfy%
\begin{equation}
\sum_{l}n_{l}^{2}\leq8,
\end{equation}
and which can be reached from the strong coupling vacuum by at most two
transitions via the plaquette operators $\exp(\pm i\theta_{P})$. \ We take the
auxiliary space $Z$ to be the subspace spanned by basis vectors%
\begin{equation}
\sum_{l}n_{l}^{2}\leq L_{\max}^{2}.
\end{equation}
Using matrix diffusion Monte Carlo, we diagonalize the subspace formed by the
states%
\begin{equation}
\left|  d,j\right\rangle =\left[  P_{Z}(H-a)P_{Z}\right]  ^{d}\left|
j\right\rangle .
\end{equation}
$\left|  j\right\rangle $ are the eigenvectors of the Hamiltonian restricted
to the original subspace $S$. \ In our calculations we use $a=L_{\max}^{2}$
and $d=0,1,\cdots12$ for cutoff values,
\begin{equation}
L_{\max}^{2}=24,28,32.
\end{equation}
In Table 2 we show the results for the ground state energy $E_{0}$ for the
different cutoff values $L_{\max}^{2}$ and the extrapolated value at $L_{\max
}^{2}=\infty$. \ The errors shown are estimated statistical errors. \ For
comparison we show the results of \cite{hamer} obtained using Green's function
Monte Carlo (GFMC).%

\[
\overset{\text{Table 2}}{%
\begin{tabular}
[c]{|l|l|l|l|l|l|}\hline
& $L_{\max}^{2}=24$ & $L_{\max}^{2}=28$ & $L_{\max}^{2}=32$ & $L_{\max}%
^{2}=\infty$ & GFMC\\\hline
$E_{0}$ & $-7.394(3)$ & $-7.430(3)$ & $-7.438(3)$ & $-7.442(4)$ &
$-7.4432(5)$\\\hline
\end{tabular}
}%
\]
In Table 3 we show the masses for the lightest six particles in the system
extrapolated to the limit $L_{\max}^{2}=\infty$. \ We have labelled the
particles according to their spin $J$ and sign under conjugation
$C:A\rightarrow-A.$ We also include results from \cite{hamer} for the lowest
antisymmetric and symmetric glueballs.%
\[
\overset{\text{Table 3}}{%
\begin{tabular}
[c]{|l|l|l|}\hline
$J^{C}$ & Mass & GFMC\\\hline
$\left|  0^{-}\right\rangle $ & $3.03(2)$ & $3.01(6)$\\\hline
$\left|  0^{+}\right\rangle $ & $4.03(3)$ & $4.05(8)$\\\hline
$\left|  2^{-}\right\rangle $ & $6.8(1)$ & \\\hline
$\left|  2^{+}\right\rangle $ & $6.8(1)$ & \\\hline
$\left|  0^{+}\right\rangle $ & $7.0(2)$ & \\\hline
$\left|  0^{-}\right\rangle $ & $7.1(2)$ & \\\hline
\end{tabular}
}%
\]
The results we find appear in agreement with \cite{hamer}. \ Unlike most Monte
Carlo algorithms, the SEC method is able to find excited states with the same
quantum numbers as lower lying states. \ This was also evident in the
$\phi_{2+1}^{4}$ example where we could track many different states crossing
the phase transition. \ The reason for this advantage goes back to the design
of stochastic error correction as a Monte Carlo improvement of a
diagonalization scheme. \ For the $U(1)$ example one can reliably find the
eigenvalues and eigenvectors for the first twenty or so states in the low
energy spectrum.

\section{Hubbard Model}

The last example we consider is the two-dimensional Hubbard model defined by
the Hamiltonian
\begin{equation}
H=-t\sum_{<i,j>;\;\sigma=\uparrow,\downarrow}(c_{i\sigma}^{\dagger}c_{j\sigma
}+c_{j\sigma}^{\dagger}c_{i\sigma})+U\sum_{i}(c_{i\uparrow}^{\dagger
}c_{i\uparrow}c_{i\downarrow}^{\dagger}c_{i\downarrow}).
\end{equation}
The summation $<i,j>$ is over nearest neighbor pairs. $\ c_{i\sigma}^{\dagger
}$($c_{i\sigma}$) is the creation(annihilation) operator for a spin $\sigma$
electron at site $i$. $\ t$ is the hopping parameter, and $U$ controls the
on-site Coulomb repulsion. The model has attracted considerable attention in
recent years due to its possible connection to $d$-wave pairing and stripe
correlations in high-$T_{c}$ cuprate superconductors. \ In spite of its simple
form, the computational difficulties associated with finding the ground state
of the model are substantial even for small systems. \ Fermion sign problems
render Monte Carlo simulations ineffective for $U$ positive and away from
half-filling, and the collective effect of very large numbers of basis Fock
states make most diagonalization approaches very difficult. A brief overview
of the history and literature pertaining to numerical aspects of the Hubbard
model can be found in \cite{overview}.

In terms of momentum space variables, the Hubbard Hamiltonian on an $N\times
N$ periodic lattice has the form%
\begin{align}
H  &  =-2t\sum_{p_{x},p_{y}=0,\cdots,N-1}(\cos\tfrac{2\pi p_{x}}{N}+\cos
\tfrac{2\pi p_{y}}{N})\left[  c_{p_{x},p_{y}}^{\uparrow\dagger}c_{p_{x},p_{y}%
}^{\uparrow}+c_{p_{x},p_{y},\sigma}^{\downarrow\dagger}c_{p_{x},p_{y}%
}^{\downarrow}\right] \\
&  +\tfrac{U}{N^{2}}\sum_{\substack{p_{x}-q_{x}+r_{x}-s_{x}%
=0\operatorname{mod}N\\p_{y}-q_{y}+r_{y}-s_{y}=0\operatorname{mod}N}%
}c_{p_{x},p_{y}}^{\uparrow\dagger}c_{q_{x,}q_{y}}^{\uparrow}c_{r_{x},r_{y}%
}^{\downarrow\dagger}c_{s_{x},s_{y}}^{\downarrow}.\nonumber
\end{align}
As a test of our methods, we use QSE diagonalization with stochastic error
correction to find the ground state energy of the $4\times4$ Hubbard model
with $5$ electrons per spin. \ The corresponding Hilbert space has about
$2\cdot10^{7}$ dimensions. \ For the QSE diagonalization we use momentum Fock
states which diagonalize the quadratic part of the Hamiltonian. \ The
Hamiltonian is invariant under the symmetry group generated by reflections
about the $x$ and $y$ axes, interchanges between $x$ and $y,$ and interchanges
between $\downarrow$ and $\uparrow$. \ We find it convenient to work with
symmetrized Fock states. \ We will compute stochastic error corrections to
first order using the series method.

In Table 4 we present results for the ground state energy. \ We encountered no
trouble with the sign problem, and in fact one can easily see that each term
in the first order series expression (\ref{firs}) is negative definite. \ The
energies are measured relative to the energy of the Fermi sea at $U=0$. \ The
errors reported are statistical errors associated with the first order SEC
calculation. \ Where available, we compare with the results presented in
\cite{husslein}, which we label as Exact, Projector Quantum Monte-Carlo (PQMC),
and Stochastic Diagonalization (SD). \ Stochastic diagonalization is a
subspace diagonalization technique similar to QSE but one which uses a
different method for selecting the subspace and is based on a variational
principle \cite{deraedt}. \ Although the precise number of basis states used
in the SD calculations is not listed, we infer from numbers reported for a
modified $4\times4$ Hubbard system that roughly $10^{5}$ states were
used.\footnote{The discrete symmetries of the system were not utilized in
their calculations.}%

\[
\overset{\text{Table 4}}{%
\begin{tabular}
[c]{|l|l|l|l|l|l|l|}\hline
Coupling & States & QSE & QSE+SEC & Exact & SD & PQMC\\\hline
$U=2t$ & $%
\begin{array}
[c]{c}%
100\\
500\\
1000
\end{array}
$ & $%
\begin{array}
[c]{c}%
-.4797\\
-.4945\\
-.5006
\end{array}
$ & $%
\begin{array}
[c]{c}%
-.50147(5)\\
-.50181(3)\\
-.50198(1)
\end{array}
$ & $-.50194$ & $-.5010$ & $-.44(5)$\\\hline
$U=4t$ & $%
\begin{array}
[c]{c}%
100\\
500\\
1000
\end{array}
$ & $%
\begin{array}
[c]{c}%
-1.620\\
-1.748\\
-1.800
\end{array}
$ & $%
\begin{array}
[c]{c}%
-1.8113(4)\\
-1.8242(3)\\
-1.8302(1)
\end{array}
$ & $-1.8309$ & $-1.829$ & $-1.8(2)$\\\hline
$U=5t$ & $%
\begin{array}
[c]{c}%
500\\
1000\\
2000
\end{array}
$ & $%
\begin{array}
[c]{c}%
-2.558\\
-2.651\\
-2.685
\end{array}
$ & $%
\begin{array}
[c]{c}%
-2.7073(4)\\
-2.7208(2)\\
-2.7231(1)
\end{array}
$ & $-2.7245$ & $-2.723$ & $-2.9(3)$\\\hline
\end{tabular}
}%
\]
Apparently QSE diagonalization with SEC handles the $4\times4$ system quite
well with relatively few states. \ Much larger systems are being studied using
both higher series corrections and stochastic Lanczos techniques \cite{salwen}.

\section{Summary and comments}

In this paper we presented two versions of stochastic error correction, the
series method and the stochastic Lanczos method. \ The series method starts
with eigenvectors of the Hamiltonian restricted to some optimized subspace and
includes the contribution of the remaining basis states as an ordered
expansion. \ The stochastic Lanczos method starts with eigenvectors of a
Hamiltonian submatrix and constructs matrix elements of Krylov vectors using
matrix diffusion Monte Carlo. \ This method has the advantage that the
starting vectors need not be close to exact eigenvectors.

We presented three different examples which demonstrate the potential of the
new approach for strongly coupled scalar, gauge, and fermionic theories. \ In
the first example we calculated the low energy spectrum of $\phi_{2+1}^{4}$
using the series method, and in the second example we found the spectrum of
compact $U(1)$ in $2+1$ dimensions using the stochastic Lanczos method. \ In
both examples we found agreement with results from the literature. \ We also
found that unlike typical Monte Carlo results, the SEC method is able to find
the eigenvalues and eigenvectors for excited states with the same quantum
numbers as lower lying states. \ This advantage is due to its design as a
Monte Carlo improvement of a diagonalization scheme. \ In the last example we
found the ground state of the $2+1$ dimensional Hubbard model using QSE
diagonalization and first order series stochastic error correction. \ In this
calculation we encountered no fermion sign problem and found that our methods
yielded very accurate results with far less effort than existing techniques.
We believe that the methods we have presented hold considerable potential for
studying a wide range of non-perturbative quantum systems and answering
questions difficult to address using other methods.


\def\vsp#1{}
\chapter[Hubbard Model]{EQSE Diagonalization of the Hubbard
  Model\footnote{Nucl. Phys. B (Proc. Suppl.) 90 (2000) 202-204}}

\section{Introduction}

The enhanced quasi-sparse eigenvector (EQSE) method of solving quantum
field theory Hamiltonians is the combination of quasi-sparse
eigenvector (QSE) method \cite{qse} with a stochastic calculation for
the contribution of the remaining basis states \cite{sec}.  The
Hubbard model was chosen as a laboratory for testing this method for
several reasons. First, we believed the approach could yield results.
The basis vectors can be specified in a few words of data and the
Hamiltonian is sparse is momentum space. On the other hand, the ground
state of the Hubbard model is known for its inclusion of an
extraordinary number of Fock states \cite{and} so the model presents a
non-trivial challenge to the quasi-sparse approach. Finally, the
Hubbard model is thought to be a physically relevant model for
superconductivity \cite{and}. While the description of the model is
simple, solutions have been difficult and any promising new approach
is worthwhile.

We work on a 2-dimensional spatial lattice with the Hubbard Hamiltonian
\[
H=-t\sum _{\footnotesize \begin{array}{c}
\sigma =\uparrow ,\downarrow \\
<i,j>\text {nearest}
\end{array}}(c_{i\sigma }^{\dagger }c_{j\sigma }+c_{j\sigma }^{\dagger }c_{i\sigma })+U\sum (c_{i\uparrow }^{\dagger }c_{i\uparrow }c_{i\downarrow }^{\dagger }c_{i\downarrow })\]

There are 2 species of electron so there are 4 possible states for each lattice
site. Thus the 8x8 Hubbard model has \( 4^{64} \) dimensions. Even after using
particle conservation to partition the space there are more than \( 10^{32} \)
basis vectors. In this large space the Hamiltonian is clearly sparse But the
equality of the off-diagonal elements contributes to an extraordinary number
of Fock states in the ground.

It is thought that the D-wave correlator \( C_{d_{x^{2}-y^{2}}}(r) \) is an
important indicator of superconductivity \cite{husslein}.

\section{Hamiltonian Momentum Lattice Formulation}

After making space periodic we use the Fock states as our basis set. The Hamiltonian
conserves momentum which helps limit the number of relevant basis states. 
\begin{align*}
H_{\text {kin}}=
-2t\sum (\cos \frac{\pi n}{L}+\cos \frac{\pi  p}{L})
a_{np,\sigma }^{\dagger }a_{np,\sigma } \\
V=\frac{U}{4L^{2}}
\sum _{
  \footnotesize
  \begin{array}{c}
    k-l+m-n=0\\
    p-q+r-s=0
  \end{array}
  }
a_{kp\uparrow }^{\dagger }a^{}_{lq\uparrow }a_{mr\downarrow }^{\dagger }a^{\downarrow }_{ns\downarrow}  
\end{align*}
There is a 16 member symmetry group generated by reflections in the \( x \)
and \( y \) planes, \( x\leftrightarrow y \) and\( \downarrow \leftrightarrow \uparrow  \).
For the purpose of finding the ground state, we use only symmetrized basis states.

The first step in the calculation consists of picking a basis set of size \( N \)
and diagonalizing its submatrix of the Hamiltonian using the Lanczos
method \cite{arpack}. 
The \( \frac{N}{5} \)basis vectors which least contribute to the ground are
then discarded, replacements are chosen and the Hamiltonian is again diagonalized.
When the ground energy \( E_{0} \) obtained in this manner converges
the QSE step is complete.

The EQSE is a first order correction to this result.  We calculate it
stochastically. Let $C$ be a set of basis vectors for the complement
of the $N$-dimensional subspace.  Then, choosing representative basis
vectors \( v \in C \) with probability \( P(v) \) .

\begin{align*}
E=E_{0}+\text {Average}\frac{FO(v)}{P(v)}, \qquad \qquad
FO(v)=-\frac{\langle 0|H|v\rangle \langle v|H|0\rangle }{\lambda _{v}}
\end{align*}
where \( |0\rangle  \) is the ground state of the \( N \) dimensional subspace. 
The expectation is thus
\[
E=E_{0}+\sum_{v\in C} P(v) \frac{FO(v)}{P(v)}.\]

\section{Results}
Results were obtained for the ground state energy, wavefunction and
$d$-wave correlator.  The computing time was about 2 days on a 350Mhz
Pentium II. Where available, we compare with Husslein et al
\cite{husslein} results labeled Exact, Projector Quantum Monte-Carlo
(PQMC), and Stochastic Diagonalization (SD) (which uses a different method
of choosing the subspace than QSE).

{\centering \begin{tabular}{|c|c|c|c|c|c|c|}
\multicolumn{7}c{Ground State Energy 4x4 Hubbard Model($\frac{5}{16}$ occupied) }\\
\hline 
Coupling &
States&
QSE&
EQSE&
Exact&
SD&
PQMC\\
\hline 
\hline 
&50 & -.47471 & -.50127(5)& && \\
U=2&100 & -.47967 & -.50147(5)&
-.50194&
-.501&
-.49\\
&500 & -.49454 & -.50181(3) & & &\\
&1000 & -.50062 & -.50198 & &&\\
\hline
& 50 & -1.5707 & -1.80635(5) & && \\
U=4&100 & -1.6203 & -1.8113(4) & -1.8309 & -1.829 & -1.8(2) \\
&500 & -1.7476 & -1.8242(3) & & & \\
&1000 & -1.8003 & -1.8302 & && \\
\hline
&50 & -2.2450 & -2.6663(8) & && \\
U=5&100 & -2.3322 & -2.6724(7)&
-2.7245&
-2.723&
-2.9(3) \\
&500 & -2.5578 & -2.7073(4) & && \\
&1000 & -2.6512 & -2.7208(2)  & &&\\
&2000 & -2.685 & -2.7231  & &&\\
\hline 
&50 & -2.963 & -3.615 & && \\
U=6&100 & -3.103 & -3.635 & && \\
&1000 & -3.452 & -3.697 & && \\
&2000 & -3.595 & -3.723 & && \\
\hline 
\end{tabular}\par}
\vsp{0.3cm}

\vsp{0.3cm}
{\centering \begin{tabular}{|c|c|c|c|}
\multicolumn{4}c{Ground Energy 8x8 Hubbard Model($\frac{25}{64}$ occupied)}\\
\hline 
Coupling &
States&
QSE&
EQSE\\
\hline 
\hline 
&50 & -.281 & -2.58(2)\\
&100 & -.443 & -2.49(2)\\
U=2&500 & -1.221 & -2.40(2)\\
&1000 & -1.751 & -2.406(8)\\
&2000 & -1.956 & -2.423(9)\\
&4000 & -1.958 & -2.427 \\
\hline
& 50 & -.811 & -9.18(6)  \\
&100 & -1.386 & -8.48(6)\\
U=4&500 & -3.449 & -7.58(5) \\
&1000 & -4.798 & -7.59(3) \\
&2000 & -5.374 & -7.620(4) \\
&4000 & -5.387 & -7.621(5) \\
\hline
&50 & -1.222 & -13.33(8) \\
&100 & -1.88 & -12.33(8) \\
U=5&500 & -4.65 & -10.65(6) \\
&1000 & -6.44 & -10.65(3)\\
&2000 & -7.225 & -10.671(5)\\
&4000 & -7.236 & -10.662(8)\\
\hline 
&50 & -1.6 & -18.2 \\
&100 & -2.4 & -16.6 \\
U=6&500 & -5.9 & -13.93(8) \\
&1000 & -8.15 & -13.78(4) \\
&2000 & -9.110 & -13.888(7) \\
&4000 & -9.113 & -13.880(8) \\
\hline 
\end{tabular}\par}
\vsp{0.3cm}

The $d_{x^2-y^2}$ correlator was also obtained using the QSE
algorithm.  The 4x4 result again matched that of Husslein et al
\cite{husslein}. The EQSE calculation has not been completed and we
therefore omit the data.

\section{Conclusion}

As we can see, the ground state for the 4x4 Hubbard model can be well
described with about 1000 symmetrized states and the 50 state results
yield remarkable accuracy when the first order correction is included.
In the 8x8 case it is clear that even 4000 states are not sufficient
to describe the ground state.  The precision of the first-order
values will be determined with the completion of higher order
calculations. 

Further advances will come in the refinement of the enhancement
technique.  Better importance sampling will speed convergence of
second and higher orders contributions.  Other extensions will be
calculation of excited states of the Hamiltonian, correlation
functions, binding energies and other quantities of interest.


\def\ket#1{|#1\rangle}
\def\bra#1{\langle#1|}
\def\br#1{\langle#1}
\def\pket{\ket{\psi}}
\def\pbra{\bra{\psi}}

\chapter[Higher order Hubbard]{Higher order calculations in the Hubbard model}

\section{Introduction}

The last chapter contained an application of the quasi-sparse
eigenvector approach to determining the energy of the Hubbard model.
The work was done with a Fock state basis.  The 4x4 Hubbard model has
few enough states that it is possible to directly diagonalize the
Hamiltonian and we were able to compare our results with the exact
answer.  

The accuracy of the results obviously depended on the dimensionality
of the exactly solved subspace, $Q$.  We summarize here the results for 
a $U/t$ ratio of $4$,
which may be in the physically relevant range for High-Tc
superconductivity.  A subspace spanned by 50 symmetrized 
states yielded the ground state energy to within 15\%.  This slowly
declined as we increased the dimensionality.  A dimensionality of 1000
was sufficient to within 2\%.  On the other hand, a first order
perturbative calculation (EQSE), when applied to
the 50 state space was also sufficient to achieve 1\% accuracy and
when applied to the 1000 state space the answer was accurate to 4
significant figures.

For the 8x8 Hubbard model the situation was much less clear.  While
quasi-sparse and first-order corrected results were obtained, their
accuracy could not assessed.  The EQSE results appear to converge for
large initial subspaces.  However, a close look shows that the
uncorrected quasi-sparse result also appears to converge for $Q$ over
about 1200 states.  The change from 2000 to 4000 states only lowers
the ground energy by about .01 units.  But the first order result is
more than 2 units lower still.  This is evidence of the extreme number
of Fock states in the Hubbard ground.  Even if there is no further
falloff in contribution, one could estimate from this that a
quasi-sparse calculation with 400,000 symmetrized basis states in $Q$
would be required to reach the accuracy of the the first order result.

The purpose of this paper is to further explore the ground state
energy of the Hubbard model using several approaches including
variations of the ordered expansion and the stochastic Lanczos method
presented in chapter 6.

\section{Calculations}

On a 4x4 lattice with 5 occupied sites for both spins, the
perturbative ground is given by
\[
\begin{array}{ccccc}
2 & . & . & . & .\\
1 & . & \otimes  & . & .\\
0 & \otimes  & \otimes  & \otimes  & .\\
-1 & . & \otimes  & . & .\\
 & -1 & 0 & 1 & 2\\
 \multicolumn{5}{c}{\uparrow \text{modes}}
\end{array}\qquad \begin{array}{ccccc}
2 & . & . & . & .\\
1 & . & \otimes  & . & .\\
0 & \otimes  & \otimes  & \otimes  & .\\
-1 & . & \otimes  & . & .\\
 & -1 & 0 & 1 & 2\\
 \multicolumn{5}{c}{\downarrow \text{modes}}
\end{array}=|0\rangle \]

A typical excited state is 
\[
\begin{array}{ccccc}
2 & . & . & . & .\\
1 & \otimes  & \otimes  & . & .\\
0 & \otimes  & \otimes  & . & .\\
-1 & . & \otimes  & . & .\\
 & -1 & 0 & 1 & 2\\
 \multicolumn{5}{c}{\uparrow \text{modes}}
\end{array}\qquad \begin{array}{ccccc}
2 & . & . & . & . \\
1 & . & \otimes  & . & .\\
0 & \otimes  & \otimes  & . & .\\
-1 & \otimes & \otimes  & . & .\\
 & -1 & 0 & 1 & 2\\
 \multicolumn{5}{c}{\downarrow \text{modes}}
\end{array}=a_{-11}^{\uparrow \dagger }a_{10}^{\uparrow }a_{-1-1}^{\downarrow \dagger }a_{10}^{\downarrow }|0\rangle \]

We can define a distance function of a state as the number of
differences from the perturbative ground.  The example above has 2
differences among the $\uparrow$ modes and 2 differences among the 
$\downarrow$ modes.  We say its distance is $(2,2)$.  In cases where we
are only interested in the greater of the up and down distances we
would say the distance is $2$.

When the ground state energy computed in chapter 7 was considered as
a function of the number of states, we found that the slope would suddenly
drop off at around 1200 states for the 8x8 Hubbard model.  For the 4x4
case, the drop in slope occurred at around 30 states.  The fillings we
have used, $5/16$ in the 4x4 case and $25/64$ in the 8x8 case both
have a non-degenerate $U=0$ ground state.  On close examination, it
became clear that the first states found in the quasi-sparse
diagonalization were always distance $(2,2)$.  When these states, for
which $H$ has non-zero matrix elements with the free ground state,
were exhausted, the slope of the energy curve would flatten out.

The calculations presented in this chapter all used the natural cutoff
of distance $\leq$ 2 to define $Q$, the base subspace.  This simple
definition saved time in determining if new states were contained in the
base subspace.  Further, given this definition of $Q$, a
clear relationship between powers of $H$ and distance of a state
became clear.  $H$ has non-zero matrix elements only between states
whose relative distance is less than or equal to $2$.  As an example,
in a calculation of $H^4$ between the base space and itself the first
and third intermediate states would be states with maximum
distance $4$ and the second intermediate state would have maximum
distance $6$.

The first extension of the previous results was inclusion of higher
order contributions in the ordered expansion.  This was done in two
ways.  Second order Monte Carlo calculations were attempted using the
ordered expansion described in chapter 6.  The results were on the
same order as the first order results and are not presented here.
Higher order calculations in this expansion were difficult because of
the many different terms that contribute.  Instead, the
Brillouin-Wigner degenerate perturbation method was used to calculate
higher terms.  This approach yielded some interesting results which
will be discussed.

The other method of extracting data was to use Monte Carlo to
calculate the matrix elements of powers of the Hamiltonian.  Once
these are known there are several ways they can be used.  The
stochastic Lanczos method uses these matrix elements to
generate a matrix for $H$ in a Krylov subspace generated from the
ground of the base subspace.  In theory one could diagonalize the
Krylov subspace to obtain the eigenvectors and eigenvalues of $H$.
The usefulness of this method was limited by the very small
determinants of the matrix thus created and the uncertainties of the
matrix elements calculated using Monte Carlo.  

A more careful choice of subspace, using states generated by higher
eigenvectors of $Q$ may solve this problem.  Although some
work was done in this direction, it is not presented here.  Instead,
we used the power method to extract the eigenvalue using the same
matrix elements for $H$.

\section{Brillouin-Wigner perturbation theory}

The Brillouin-Wigner perturbation method is described in Appendix A.
The general idea is to solve a self-consistent equation for the
eigenvalue.  For the desired eigenvector $\pket$ in $Q$, we find the
eigenvalue and components as follows.

\begin{eqnarray*}
  E \br{i}\pket
  &=
  (H_{ij} + A^{(1)}_{ij} + A^{(2)}_{ij} + A^{(3)}_{ij}  + \ldots
  )\br{j}\pket\\
  H_{ij} &= \bra{i}H\ket{j} \\
  A^{(1)}_{ij} &= \sum_{\ket{k} \notin Q} 
  \frac{\bra{i}H\ket{k} \bra{k}H\ket{i}}{E-\bra{k}H\ket{k}} \\
  A^{(2)}_{ij} &= \sum_{
  \scriptsize
  \begin{array}{c}
    \ket{k},\ket{l} \notin Q \\ \ket{k}\ne \ket{l}
  \end{array}}
  \frac{\bra{i}H\ket{k} \bra{k}H\ket{l} \bra{l}H\ket{i}}
  {(E-\bra{k}H\ket{k})(E-\bra{l}H\ket{l})} \\
  A^{(3)}_{ij} &= \sum_{
  \scriptsize
  \begin{array}{c}
    \ket{k},\ket{l},\ket{m} \notin Q \\ \ket{k}\ne \ket{l}, \ket{l}\ne \ket{m}
  \end{array}}
  \frac{\bra{i}H\ket{k} \bra{k}H\ket{l} \bra{l}H\ket{m}\bra{m}H\ket{i}}
  {(E-\bra{k}H\ket{k})(E-\bra{l}H\ket{l})(E-\bra{m}H\ket{m})} \\
  \text{etc.}
\end{eqnarray*}

This could be solved by feeding the $n$-th order result for $E$ into
the $(n+1)$-th order calculation's denominators.  Alternatively we
could keep a running estimate for $E$ as the Monte Carlo proceeds and,
at each step, use the current estimate.  Instead, we chose to
calculate the left hand side $E_l$ as a function of the right hand side
$E_r$.  We then take the intersection of the graph of this function
and the line $E_l = E_r$ as the eigenvalue.

\begin{figure}[htbp]
  \begin{center}
   \epsfxsize=30pc \epsfbox{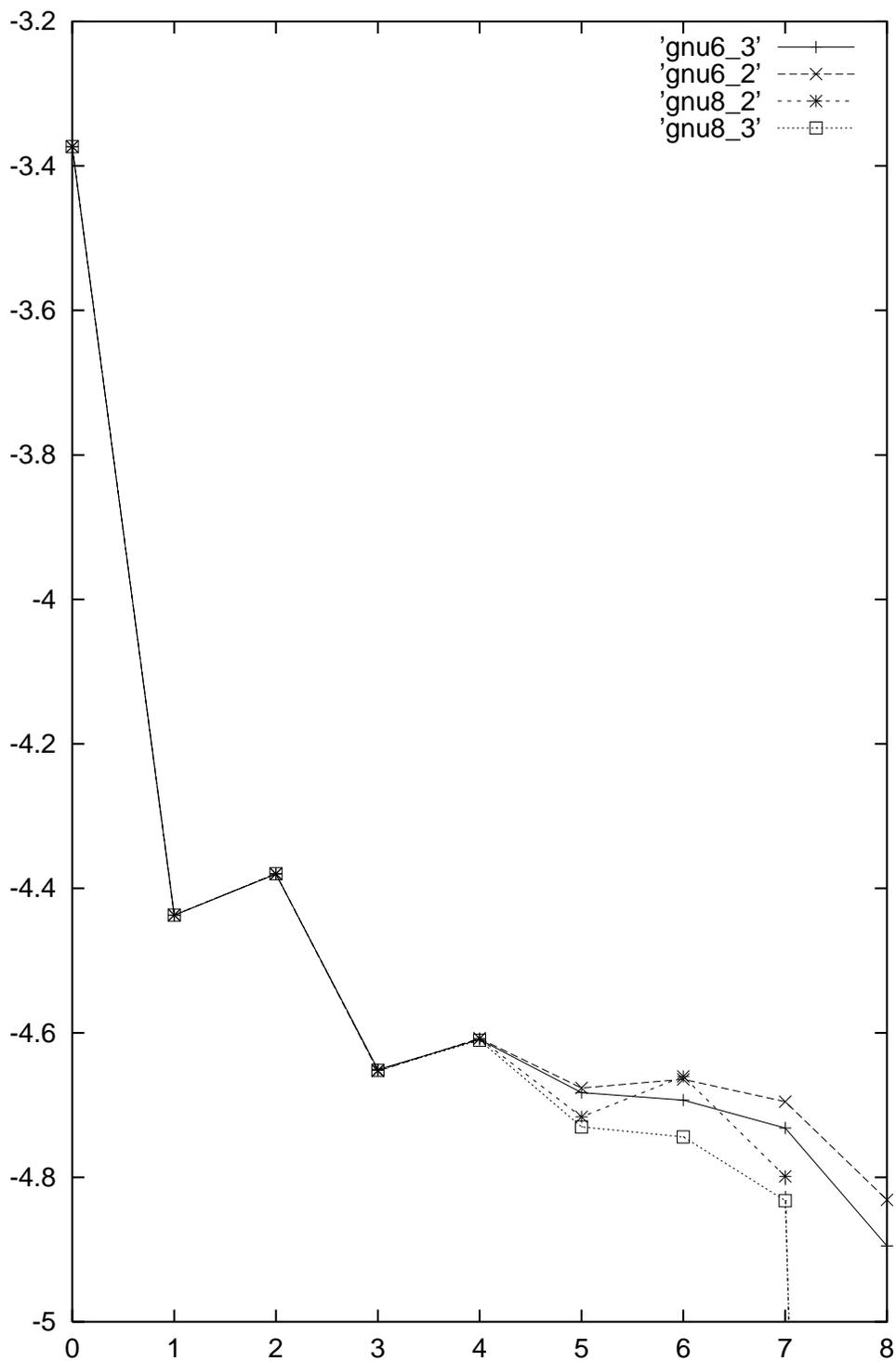}
    \caption{Brillouin-Wigner calculation for energy as a function of
  order.  gnu6\_2 and gnu6\_3 have max distance 6.  The others have
  max distance 8.}
    \label{fig:B-W}
  \end{center}
\end{figure}

Each Monte Carlo trial consists of a trajectory beginning and ending
in $Q$.  At each step the weight is adjusted by the appropriate
probabilities and, outside of $Q$ the appropriate denominator is
divided.  Figure \ref{fig:B-W} 
presents data for the 6x6 Hubbard model
with $E_r=7$.  We would have to run this again with $E_r=5$ to draw a
curve and find the actual eigenvalue.  Two of the runs used a distance
cutoff of $6$ and two used a distance cutoff of $8$.

A guiding scheme was used to increase the likelihood that paths return
to $Q$ and to prevent them from going beyond the cutoff distance.
Were it not for the denominators the guiding would be perfect in the
sense that all paths would have an equal contribution to the matrix
elements.  A better guiding scheme would take account of the diagonal
energy and better sample those states with smaller denominators.

The convergence is excellent up to order 4 and is good up to order 6
or 7.  After that we consistently find values for the minimum
eigenvalue that are much too low.  One can account for this by the
non-linear process of diagonalizing that is used to find the
eigenvalue from the $A_{ij}$ matrix.  The distance cutoff $6$ and
distance cutoff $8$ results remain close at order 5 and 6.  After
order 7 the noise in the cutoff=$8$ results becomes dominant.

\section{Power method}

The power method is very simple in theory.  We start with the ground
state of $H$ restricted to $Q$, run trajectories with $H-a$ and tabulate
the results.  If the energy shift, $a$, must be chosen such that $|E_0-a| >
|E_j-a|$ for all other eigenvalues $E_j$, then the ratio of successive
powers of $H$ will approach the value $(E_0 - a)$.

It is not surprising that the calculation suffers from a sign
problem.  The guiding, described in the previous section, results in
the contributions of different paths at each order being close to each
other in magnitude but of varying sign.  Functionally, after around 7
or 8 steps this problem significantly adds to the convergence time.

Several modifications were made to minimize the impact of long paths.
The first was that only paths outside of $Q$ were tabulated.  By
restricting to such ``connected'' paths we decrease the contribution
of long paths.
Let $\hat Q$ be the projector onto the space $Q$ and $\hat P = 1- \hat
Q$.  And let $C_{n,ij}$ be the $n$-step matrix element between
$i$ and $j$ that we get by restricting to paths outside of $Q$.
\begin{eqnarray*}
  C_1 =& \hat Q H \hat Q \\
  C_n =& (\hat Q H \hat P)(\hat P H \hat P)^{n-2}(\hat P H \hat Q) \\
  H^n =& C_n + C_{n-1}H^1 + \ldots C_1 H^{n-1}
\end{eqnarray*}
The disadvantage of this approach is that it increases the memory needs.

If an intermediate state $\pket \notin Q$ is chosen with known
probability we can calculate the contribution of paths that pass
through it at the $m$-th step.  We multiply the Monte Carlo results
for $\bra{i} H^m \pket$ and $\pbra H^n \ket{j}$ together to find the
contribution to $H^{m+n}_{ij}$.  Effectively, this allows the
calculation of twice as many steps before the sign problem kicks in.
The concern with this approach is that more time is spent on each
state at the $m$-th step so the space of possible $m$-th states is not
as well sampled.  If the contribution to $C_{m+n}$ is a relatively
smooth function of the $m$-th state this is not a problem.  In
practice, this method was only worthwhile for higher order terms.

\begin{figure}[htbp]
\begin{center}
\epsfxsize=30pc \epsfbox{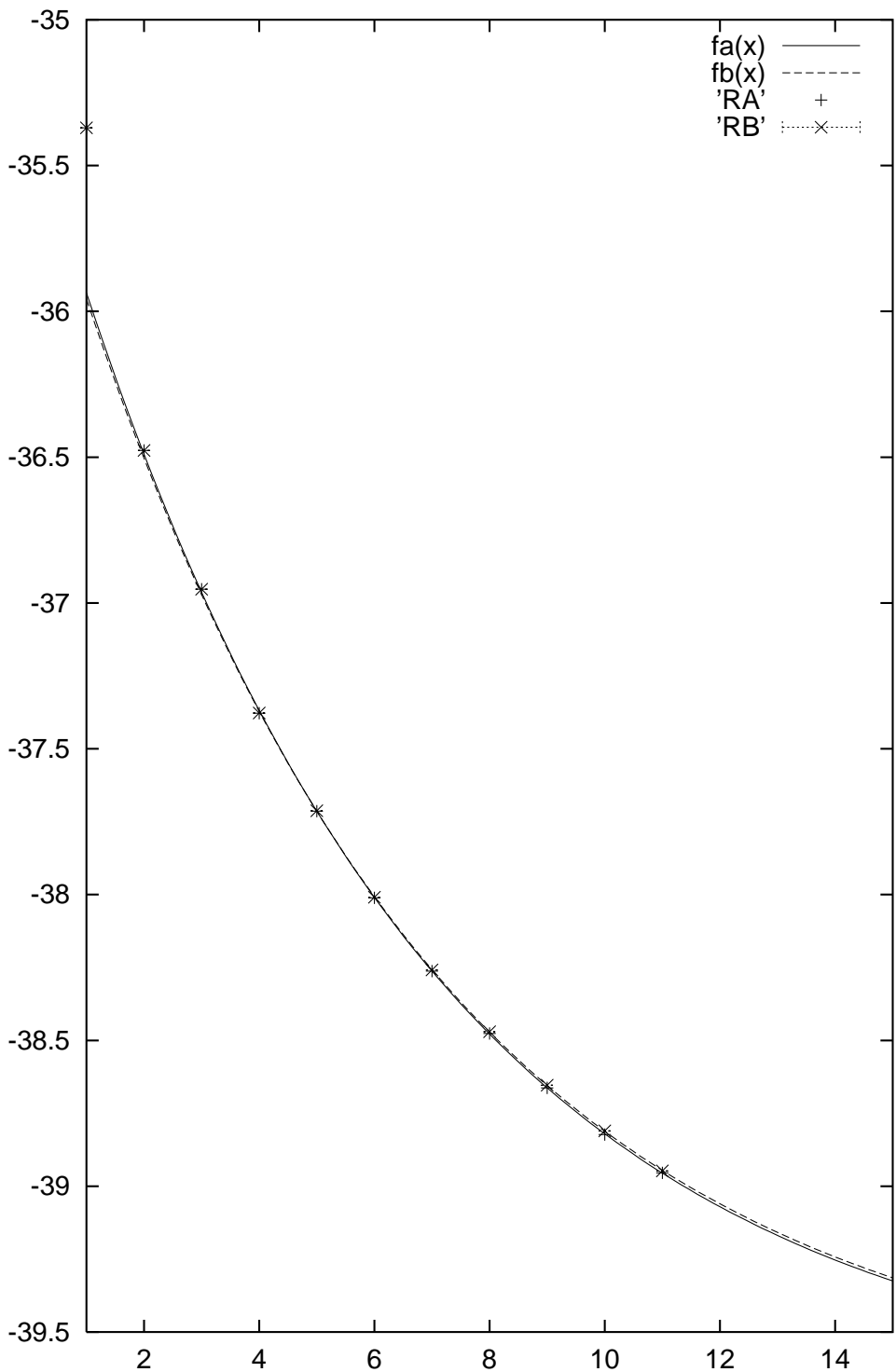}
\caption{8x8 Hubbard model: Ratio of 
  $\pbra H^n \pket$ over $\pbra H^{n-1} \pket$ as a
  function of $n$.  Two trials are shown as well
  as their best fit curves.}
    \label{fig:8x8}
\end{center}
\end{figure}

We present the results for the 8x8 Hubbard model in figure
  \ref{fig:8x8}.
The
distance function was limited to a maximum of $8$.  We plot the ratio
of $H^n/H^{n-1}$ for $n=1$ to $11$.  The energy shift is set to $30$.
Two trials were run.  $C_1$ through $C_7$ were calculated with paths
starting in $Q$ and $C_8$ through $C_{11}$ used the intermediate state
method.  As expected, the paths approach a decaying exponential.
Using gnuplot to fit points $5$ through $11$ yields an asymptotic
value of $-39.74$ for one trial and $-39.75$ for the other.  Other
choices of points to fit result in small changes in these results.
Overall, for a distance cutoff of $8$, the 8x8 Hubbard model yields a
ground state energy of $-9.74 \pm .05$.

\begin{figure}[t]
\begin{center}
\epsfxsize=30pc \epsfbox{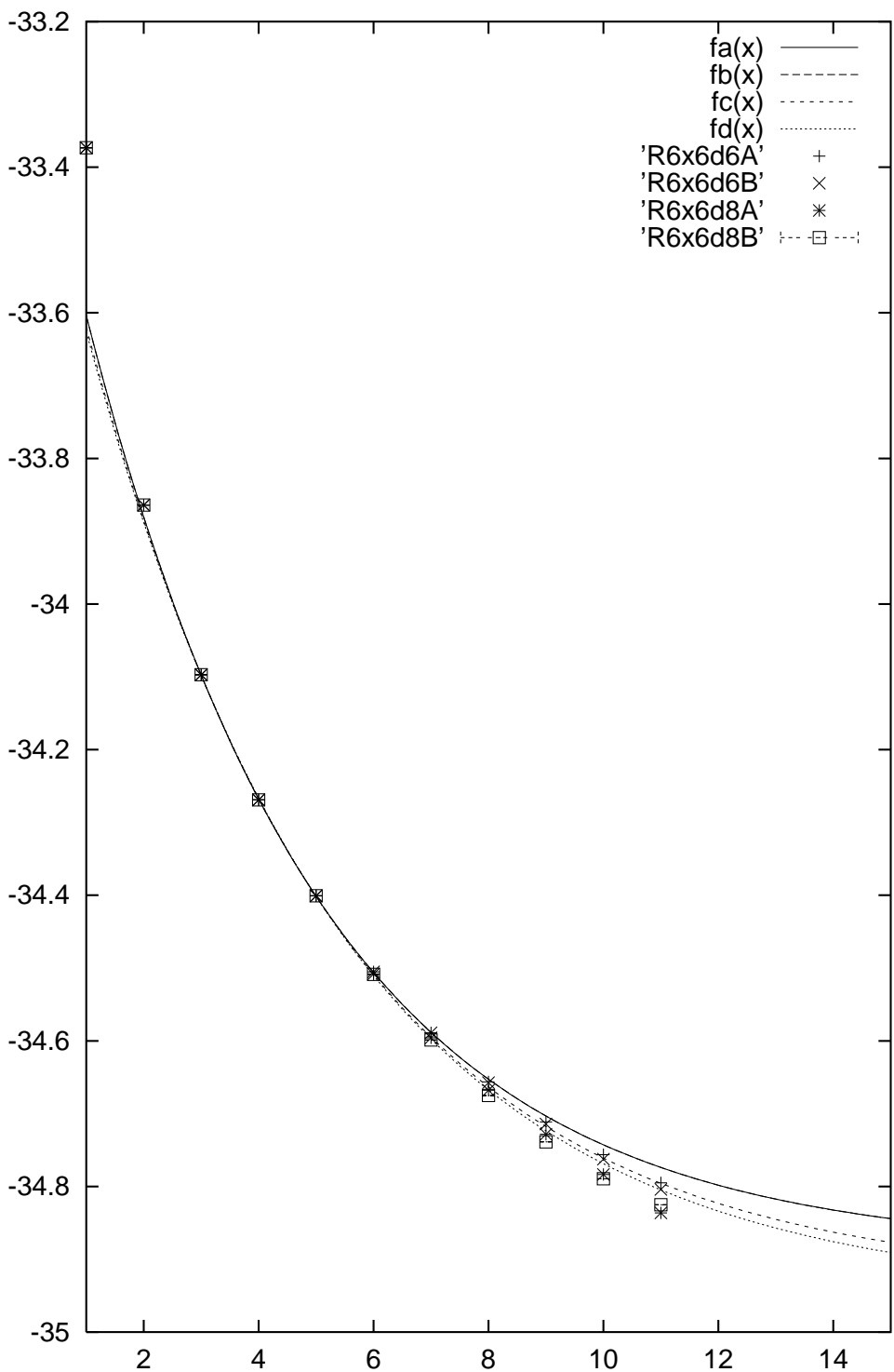}
\caption{6x6 Hubbard model: Ratio of 
  $\pbra H^n \pket$ over $\pbra H^{n-1} \pket$ as a
  function of $n$.  Two trials with max distance 6 and two trials with
  max distance 8 are shown as well
  as their best fit curves.}
\label{fig:6x6}
\end{center}
\end{figure}

We
present results for the 6x6 Hubbard model in figure \ref{fig:6x6}.
  With a
distance cutoff of 6, both curves are on top of each other and yield
an energy of -4.887.  With a distance cutoff of 8 the answer is -4.94.
In these calculations the intermediate state method was not used so
the results at $n>7$ show larger errors and are not included in the
fit. 

\section{Conclusions}
For the computing power we had available, the 8x8 Hubbard model
poses a serious challenge.  We have been able to extract a good
estimate for the ground state energy in the $\frac{25}{64}$ filled
sector.  The same techniques may also work for calculating correlators
in the ground state.  The most powerful approach we have used is the
power method but the matrices $C_n$ extracted for this method are also
available for other methods of analysis such as the modified
stochastic Lanczos method alluded to above.  While the
Brillouin-Wigner method may hold promise for the future, it was less
precise for our purposes than the power method.  The non-linearity of
the inversion step makes it particularly sensitive to noisy data.
Improvements in guiding or sampling strategies and filters to allow
linear treatment of noisy data may help both the
Brillouin-Wigner and power method approaches.

\[
\overset{\text{8x8 Hubbard model, U=4}}{%
\begin{tabular}
[c]{|l|l|l|l|l|l|}\hline
method & \multicolumn{2}{c}{QSE} &  
\multicolumn{2}{c}{first order correction} &power method\\\hline
states & 2000 & 4000 & 2000 & 4000 & 1200 \\\hline
E & -5.374 & -5.387 & -7.620 & -7.621(5) & -9.74(5) \\\hline
\end{tabular}
}%
\]

A brief look at the summary of results for the ground energy shows
that despite the apparent convergence of the first-order correction,
the answer is still far from correct. Computations for distance
cutoffs of 6 and 10 will allow a determination of the precision of the 
power method result presented here.

Another interesting extension will be the calculation of ground state
energy for other fillings.  Exploring the fillings near the one
presented here will allow determination of the energy to add or remove
an electron from the lattice.

\def\umlaut{\"}

\nocite{}
{
\ssp 

\bibliography{mybibs}
}

\appendix



\chapter{Degenerate Perturbation Theory}

Let $|1\rangle$, $|2\rangle$ \dots $|n\rangle$ \dots be eigenstates of $H_0$
with energies $E_n$.
Assume $E_1$ through $E_m$ are near each other but are all
sufficiently far from $E_{m+1}$,  $E_{m+2}$  etc.
And let $\pket$ be an eigenstate of $(H_0 + V)$ with eigenvalue $W$ where
$W$ is close to the $E_1$ through $E_m$.

By the eigenvalue equation
$$
(H_0 + V)\pket = W\pket
$$
\noindent
which if we expand $\pket$ in terms of $\ket{n}$ becomes
$$
\sum_n (H_0 + V) \ket{n}\br{n}\pket = W\sum_n \ket{n}\br{n}\pket
$$
So, taking a contraction with $\bra{i}$ on the left we get
\begin{equation}
  \label{W0}
   E_i\br{i}\pket + \sum_{n} \bra{i}V\ket{n}\br{n}\pket = W \br{i}\pket
\end{equation}

This is just an (infinite dimensional) eigenvalue matrix equation.
\begin{equation}
  \label{W1}
  \left [
  \begin{array}{cccc}
    E_1 + \bra{1}V\ket{1} & \bra{1}V\ket{2} &  \bra{1}V\ket{3} & ... \\
    \bra{2}V\ket{1} & E_2 + \bra{2}V\ket{2} &  \bra{2}V\ket{3} & ... \\
    \bra{3}V\ket{1} & \bra{3}V\ket{2} &  E_3 + \bra{2}V\ket{3} & ... \\
    \vdots & \vdots & \vdots & \ddots \\
  \end{array}
  \right ]
  \left [
  \begin{array}{c}
    \br{1}\pket \\
    \br{2}\pket \\
    \br{3}\pket \\
    \vdots \\
  \end{array}
  \right ]
  =
  W
  \left [
  \begin{array}{c}
    \br{1}\pket \\
    \br{2}\pket \\
    \br{3}\pket \\
    \vdots \\
  \end{array}
  \right ]
\end{equation}
If we temporarily fix the value of $W$ this is just a homogeneous first-order
equation in $ \br{n}\pket $.  Eqn (\ref{W1}) will only have a solution for
special values
of $W$, but we can use it to determine
$     \br{m+1}\pket $,    $ \br{m+2}\pket $, etc. in terms of 
$     \br{1}\pket $ through   $ \br{m}\pket $.

If we treat  $ \br{i}\pket $ as fixed for $ i \le m$ we get the equation
$$
  \left [
    S
  \right ]
  \left [
  \begin{array}{c}
    \br{m+1}\pket \\
    \br{m+2}\pket \\
    \vdots \\
  \end{array}
  \right ]
  =
  \left [
  \begin{array}{c}
    \sum_{i \le m}    \bra{m+1}V\ket{i}\br{i}\pket \\
    \sum_{i \le m}    \bra{m+2}V\ket{i}\br{i}\pket \\
    \vdots \\
  \end{array}
  \right ]
$$
\noindent
where
$$
  \left [
    S
  \right ]
  =
  \left [
  \begin{array}{ccc}
    E_{m+1} + \bra{m+1}V\ket{m+1} - W & \bra{m+1}V\ket{m+2} & ... \\
    \bra{m+2}V\ket{m+1} & E_{m+2} + \bra{m+2}V\ket{m+2} - W & ... \\
    \vdots & \vdots & \ddots \\
  \end{array}
  \right ]
$$

\noindent
Since $W$ is not near $E_n$ for $n\ge m+1$ and $V$ is small, $[S]$ must have
non-zero determinant and be invertible.  Since the right hand side 
is linear in $\br{i}\pket$ for $i \le m$ we can ( by multipling by $[S^{-1}]$)
express  $\br{n}\pket$ for $n \ge m+1$ as a linear combination of
$\br{i}\pket$ .

For $n \ge m+1$ we therefore write
$$
  \br{n}\pket = \sum_{i \le m}A_{ni} \br{i}\pket
$$
Plugging this into (\ref{W0}) we get 
$$
  (W-E_n-\bra{n}V\ket{n})\sum_{i \le m} A_{ni} \br{i}\pket
  =
  \bigl(\sum_{i \le m} \bra{n}V\ket{i}\br{i}\pket\bigr)
  +
  \sum_{
    \begin{array}{c}
      n^\prime \neq n \\
      n^\prime \ge m+1 \\
    \end{array}
    }
  \sum_{i \le m}   \bra{n}V\ket{n^\prime}
  A_{n^\prime i} \br{i}\pket
$$
which can be solved for $A_{ni}$ by fixing coefficients of $\br{i}\pket$
\begin{equation}
  \label{W2}
  A_{ni} 
  =
  {1\over W-E_n-\bra{n}V\ket{n} }
  \biggl(
  \bra{n}V\ket{i}
  +
  \sum_{
    \begin{array}{c}
      n^\prime \neq n \\
      n^\prime \ge m+1 \\
    \end{array}
    }
  \bra{n}V\ket{n^\prime}
  A_{n^\prime i}
  \biggr)
\end{equation}

Although this equation has $A_{n^\prime i}$ on the right hand side it could be
solved iteratively since the  $A_{n^\prime i}$ comes with an extra factor of
$V$ relative to the  $A_{n i}$ on the left hand side.  However, remember that
$A_{n i}$ is really a function of $W$ so we are not ready to solve it until
we find $W$.

Now we are ready to turn to the eigenvalue problem in the
$\ket{1}$ through $\ket{m}$ subspace.  Once again using eqn (\ref{W0})
$$
  W \br{i}\pket
  =
  (E_i+\bra{i}V\ket{i}) \br{i}\pket
  +
  (\sum_{
    \begin{array}{c}
      j \neq i \\
      j \le m \\
    \end{array}
    }
  \bra{i}V\ket{j}\br{j}\pket)
  +
  \sum_{
    \begin{array}{c}
      n\ge m+1 \\
    \end{array}
    }
  \sum_{j \le m}   \bra{i}V\ket{n}
  A_{n j} \br{j}\pket
$$
\begin{equation}
  \begin{array}{rl}
    =
  \biggl(&E_i+\bra{i}V\ket{i}
  +
  \sum_{
    \begin{array}{c}
      n\ge m+1 \\
    \end{array}
    }
  \bra{i}V\ket{n}
  A_{ni}
  \biggr) \br{i}\pket \\
  &+
  \sum_{
    \begin{array}{c}
      j \neq i \\
      j \le m \\
    \end{array}
    }
  \biggl(
    \label{W3}
    \bra{i}V\ket{j}
    +
    \sum_{
      \begin{array}{c}
        n\ge m+1 \\
      \end{array}
      }
    \bra{i}V\ket{n}
    A_{nj}
    \biggr)
    \br{j}\pket  
  \end{array}
\end{equation}

Equation (\ref{W3}) is an eigenvalue equation, $([M]-W)[\br{i}\pket] = 0$,
for $W$ with the $m \times m$ matrix $[M]$ given by
$$
M_{ii} = 
  \biggl(E_i+\bra{i}V\ket{i}
  +
  \sum_{
    \begin{array}{c}
      n\ge m+1 \\
    \end{array}
    }
  \bra{i}V\ket{n}
  A_{ni}
  \biggr)
$$
and
$$
M_{ij} = 
  \biggl(
  \bra{i}V\ket{j}
  +
  \sum_{
    \begin{array}{c}
      n\ge m+1 \\
    \end{array}
    }
  \bra{i}V\ket{n}
  A_{nj}
  \biggr)
$$
The only problem is that $[M]$ is itself a function of $W$ through $A_{ni}$.
Again, however, this can be solved iteratively since the $A_{ni}$ terms all
come in with an extra factor of $V$.

So, the general procedure is to solve the eigenvalue equation to first order
in $V$ (ignoring all the terms with $A_{ni}$ since $A_{ni}$ itself has order 
1) and then to use this $W$ to solve for $A_{ni}$ to first order.  (We can
ignore the terms with $A_{n^\prime i}$ on the right hand side since they are
second order in $V$.)  Now we can continue by substituting the first order
value of $A_{ni}$ into $[M]$ and now calculating $W$ to second order.
If we substitute this and the first order values for $A_{n^\prime i}$ into
(\ref{W2}) we can solve for $A_{ni}$ to second order.

At order $k$ we solve the eigenvalue equation for $W$ using $A_{ni}$ to order
$(k-1)$ and then use $W$ to order $k$ and $A_{n^\prime i}$ to order $(k-1)$ to
solve for $A_{ni}$ to order $k$.  Of course, there are $m$ solutions for $W$
at order $k$ and we have to match the solution used for order $(k-1)$.
\cite{hsalwen}

\end{document}